\documentclass[
	reprint,
	superscriptaddress,
	nofootinbib,
	showkeys,
	aps,
	pra,
	longbibliography
]{revtex4-2}

\usepackage[utf8]{inputenc}

\usepackage[hyperref]{xcolor}
	\definecolor{goethe-blau}{cmyk}{1.0,0.2,0.0,0.4}
	\definecolor{hellgrau}{cmyk}{0.04,0.04,0.05,0.02}
	\definecolor{sandgrau}{cmyk}{0.12,0.09,0.13,0.0}
	\definecolor{dunkelgrau}{cmyk}{0.25,0.25,0.30,0.75}
	\definecolor{purple}{cmyk}{0.08,1.0,0.3,0.36}
	\definecolor{emo-rot}{cmyk}{0.04,1.0,0.8,0.07}
	\definecolor{senfgelb}{cmyk}{0.01,0.25,1.0,0.05}
	\definecolor{gruen}{cmyk}{0.62,0.4,0.87,0.09}
	\definecolor{magenta}{cmyk}{0.08,0.86,0.12,0.12}
	\definecolor{orange}{cmyk}{0.0,0.7,1.0,0.04}
	\definecolor{sonnengelb}{cmyk}{0.0,0.12,0.95,0.0}
	\definecolor{helles-gruen}{cmyk}{0.4,0.17,0.81,0.07}
	\definecolor{lichtblau}{cmyk}{0.8,0.0,0.06,0.04}

\usepackage[
	colorlinks,
	pdfpagelabels,
	breaklinks,
	pdfstartview=FitH,
	bookmarksopen=true,
	bookmarksnumbered=true,
	bookmarksopenlevel=2,
	plainpages=false,
	hypertexnames=false,
	citecolor=emo-rot,
	linkcolor=goethe-blau,
	urlcolor=purple,
	pdftitle={Solving the Lindblad equation with methods from computational fluid dynamics},		
	pdfauthor={Jan Rais, Adrian Koenigstein, Niklas Zorbach, Carsten Greiner},	
	pdfkeywords={Lindblad equation, open quantum systems, computational fluid dynamics}				
]{hyperref}

\usepackage{graphicx}
\usepackage{dcolumn}
\usepackage{braket}
\usepackage[tbtags]{amsmath}
\usepackage{amssymb}
\usepackage{epstopdf}
\usepackage{xspace}
\usepackage[capitalize]{cleveref}
\usepackage{bm}
\usepackage{tabularx}
\usepackage{lipsum}
\usepackage{multirow}
\usepackage[bottom]{footmisc}
\usepackage{orcidlink}
\usepackage{upgreek}
\usepackage{comment}

\usepackage[acronym]{glossaries-extra}
\glssetcategoryattribute{acronym}{nohyperfirst}{true}
\setabbreviationstyle[acronym]{long-short}

\newacronym{kt}{KT}{Kurganov-Tadmor}
\newacronym{fv}{FV}{Finite Volume}
\newacronym{muscl}{MUSCL}{Monotonic Upstream-centered Scheme for Conservation Laws}
\newacronym{pde}{PDE}{partial differential equation}
\newacronym{cfl}{CFL}{Courant-Friedrichs-Lewy}

\newacronym{lhs}{l.h.s.}{left hand side}
\newacronym{rhs}{r.h.s.}{right hand side}

\newcommand{\reff}{Ref.~}
\newcommand{\reffs}{Refs.~}

\newcommand{\vdistance}{\vphantom{\bigg(\bigg)}}
\newcommand{\Vdistance}{\vphantom{\Bigg(\Bigg)}}

\newcommand{\dd}{{\rm d}}
\newcommand{\ii}{{\rm i}}
\newcommand{\ee}{{\rm e}}

\newcommand{\kBoltzmann}{k_{\mathrm{B}}}

\allowdisplaybreaks

\begin{document}
	
	\title{Solving the Lindblad equation with methods from computational fluid dynamics}
	
	\author{Jan Rais \orcidlink{0000-0001-8691-6930}}
	\email{rais@itp.uni-frankfurt.de}
	\affiliation{Institut f\"ur Theoretische Physik, Johann Wolfgang Goethe-Universit\"at, Max-von-Laue-Strasse 1, 60438 Frankfurt am Main, Germany}
	
	\author{Adrian Koenigstein \orcidlink{0000-0001-7482-2195}}
	\email{adrian.koenigstein@uni-jena.de}
	\affiliation{Theoretisch-Physikalisches Institut, Friedrich-Schiller-Universit\"at, Fröbelstieg 1, 07743 Jena, Germany}
	
	\author{Niklas Zorbach \orcidlink{0000-0002-8434-5641}}
	\email{niklas.zorbach@tu-darmstadt.de}
	\affiliation{Institut f\"ur Kernphysik, Technische Universit\"at, Schlossgartenstraße 2, 64289 Darmstadt, Germany}
	
	\author{Carsten Greiner \orcidlink{0000-0001-8933-1321}}
	\email{carsten.greiner@itp.uni-frankfurt.de}
	\affiliation{Institut f\"ur Theoretische Physik, Johann Wolfgang Goethe-Universit\"at, Max-von-Laue-Strasse 1, 60438 Frankfurt am Main, Germany}
	
	\date{\today}
	
	\begin{abstract}
		Liouvillian dynamics describes the evolution of a density operator in closed quantum systems. 
		One extension towards open quantum systems is provided by the Lindblad equation.
		It is applied to various systems and energy regimes in solid state physics as well as also in nuclear physics.
		A main challenge is that analytical solutions for the Lindblad equation are only obtained for harmonic system potentials or two-level systems.
		For other setups one has to rely on numerical methods.

		In this work, we propose to use a method from computational fluid dynamics, the Kurganov-Tadmor central (finite volume) scheme, to numerically solve the Lindblad equation in position-space representation. 
		We will argue, that this method is advantageous in terms of the efficiency concerning initial conditions, discretization, and stability.
		
		On the one hand, we study, the applicability of this scheme by performing benchmark tests.
		Thereby we compare numerical results to analytic solutions and discuss aspects like boundary conditions, initial values, conserved quantities, and computational efficiency.
		On the other hand, we also comment on new qualitative insights to the Lindblad equation from its reformulation in terms of an advection-diffusion equation with source/sink terms.
	\end{abstract}
	

	\maketitle
	
\section{Introduction}

	Oftentimes the idea of ``leveraging synergies'' between different fields of research is proclaimed in project proposals.
	Putting this into practice, however, is not always straightforward.
	Within this work, we hope to convince the reader that the authors indeed managed to successfully combine methods from different fields of research -- computational fluid dynamics and open quantum systems.

\subsection{Contextualization: Approaches to open quantum systems}
\label{sec:oqa}

	Quantum mechanics, that describe closed quantum systems usually have unitary time evolution, because they obey reversible dynamics and conservation of energy. 
	However, realistic physical systems are oftentimes not isolated and therefore coupled to some environment.
	This breaks the time symmetry and leads to dissipation.

	One way to describe statistical systems is in terms of distribution functions (or density operators), which is for example done in the classical regime with the Liouville equation and for quantum systems with the Liouville-von-Neumann equation.
	This approach is invariant under time-reversal.
	
	To construct a time irreversible formulation including dissipation one uses phenomenological equations, such as the Langevin or Fokker-Planck equation, system-plus-reservoir approaches or one can even think of modifications of the equations of motion that lead to non-canonical/non-unitarity transformations and non-linear evolution equations \cite{Schuch:2018fvh}. 
	
\subsubsection{System-plus-reservoir approaches}

	The system-plus-reservoir approaches suggest to couple the system of interest to an environment, such that the environment together with the system can be considered to be a closed Hamiltonian system. 
	One couples the system to (infinitely many) bath degrees of freedom, usually, as it is done in the Caldeira-Leggett model \cite{Caldeira:1982iu}, linearly, assuming harmonic oscillators as bath particles.
	Then, the path integral of the whole system is used to propagate the density matrix to derive a master equation for the system only, by tracing out the bath degrees of freedom.
	However, the procedure of deriving such a master equation is usually based on additional assumptions:
	Traditionally, one uses trotterization, which in turn requires the system to be Markovian.
	Moreover, one assumes, that the dampings, which can be motivated from the dissipation-fluctuation theorem, respectively the coupling to the bath, have to be small while the temperature has to be large compared to typical system scales.
	We will comment on this in \cref{sec:reinterpretation} \cite{Caldeira:1982iu,DIOSI1993517}. 
	Another basic requirement is, that the bath and the environment are uncorrelated at $t = 0$.
	Based on these assumptions, Caldeira and Leggett derived a Fokker-Planck master equation \cite{Caldeira:1982iu}, which, however, does not  a priori preserve positivity and norm conservation of the diagonal matrix elements  of the density matrix. 

\subsubsection{From the Caldeira-Leggett to the Lindblad equation}

	An improvement of the Caldeira-Leggett master equation, which guarantees the aforementioned basic requirements, is the Gorini-Kossakowski-Sudarshan-Lindblad equation\footnote{For convenience and in accordance with most of the literature, we will use only the wording Lindblad equation.} \cite{KOSSAKOWSKI1972247,Lindblad:1975ef}, which is Markovian, positivity and norm-preserving by definition.
	This equation enables to describe various types of non-equilibrium systems. There are several ways to express the Lindblad equation; one general form is given by \cite{BRE02, Isar:1996ri}
		\begin{align}
			\dot{\hat{\rho}} ( t ) = \, & - \ii \big[ \hat{H}, \hat{\rho} ( t ) \big]_- +	\vdistance
			\\
			& + \tfrac{1}{2} \sum_{j} \, \big( \big[ \hat{V}_j \, \hat{\rho} ( t ), \hat{V}_j^\dagger \big]_- + \big[ \hat{V}_j, \hat{\rho} ( t ) \, \hat{V}_j^\dagger \big]_- \big) \, .	\vdistance	\nonumber
		\end{align}
	Typically such Lindblad-type equations are used in solid state physics to describe quantum dots or two-state systems \cite{BRE02, gardiner00, LEI20121408, Jin_2010}.
	Mostly, these equations are formulated in the Fock space \cite{Manzano:2020yyw}, where the ladder operators can be both, bosonic and/or fermionic. 
	The so-called Lindblad operators $\hat{V}_j$ then are just the jump operators, that describe the quantum transitions from one state to another. 
	
	As long as the energy eigenvalues of the considered systems are equidistant, the operators $\hat{V}_j$ can be easily derived. 
	However, if the systems are more complicated, theoretically every transition in between two arbitrary states has to be described by a different  jump operator.
	As a consequence, it is sometimes preferable to solve the Lindblad equation in position space, which is discussed in detail within the next paragraphs.
	
\subsubsection{The Lindblad equation in position space}
\label{sec:coord_space}

	Next, we list some aspects and advantages of computations of Lindblad dynamics in position space:
	\begin{enumerate}
	
		\item	The position-space formulation allows to stay as close as possible to the original Caldeira-Leggett model for general system potentials by introducing only the most necessary additives, which satisfy norm conservation and positivity, as it is done by equations in Lindblad-form \cite{Gorini:1975nb}. 
		The reason, we want to stay as close as possible to the Caldeira-Leggett master equation is, that this equation can be derived from the path integral for arbitrary system potentials and therefore is most convenient in order  to discuss the general framework and its approximations. 
		Referring to the first point, usually the path-integral formulation is done in position space.
		\item Because of item (1.), for a description in position space, it is not required to (re)formulate/generate Lindblad operators.
		This is advantageous, as there is no general technique to derive them systematically.
		This is for example discussed in \reffs\cite{Koide:2023awf,Gao:1997}.
		\item	We want to perform ``box''-calculations and calculations for background potentials.
		Thus, the density matrix should not spread into the entire position space during the temporal evolution, which is crucial for norm conservation.
		If we would compute the temporal evolution of the Lindblad equation in the energy eigenspace, problems concerning the truncation of the Hilbert space, caused by the structure of the Lindblad equation and the density matrix, which in energy representation has an infinite number of entries, would be unavoidable.
		In spatial representation, the truncation is dictated by the boundaries/boundary conditions of the box/potential.

		\item	Another aspect is due to the findings discussed in \reffs\cite{Bernad2018,Homa2019}:
		The modes of the wave functions, that are considered to determine the initial system are not constant during the temporal evolution and the interaction with the bath can under certain conditions lead to shifts of the original energy spectrum of the system. 
		Therefore, if the system is not described by a harmonic potential, every state gets modified differently during the time evolution. 
		This can also be seen in another open quantum approach discussed in \reff\cite{Neidig:2023kid} in the context of Kadanoff-Baym equations.

		However, one has to keep in mind, that even if the Lindblad dynamics gets stationary after some temporal evolution, thermalization -- in the strict sense that the system particle gets the temperature of the bath -- is not necessarily satisfied.
		By diagonalizing the stationary density matrix the wave functions of the thermalized system can be extracted, which can be further analysed to get insights into the mode shift during the thermalization process as well as quantities like the spectral function. 
		This, however, will be discussed in a separate publication \cite{Rais2025} based on the methods developed in this work.
	\end{enumerate}

\subsection{Numerical approaches to solve the Lindblad equation}

	There are various powerful numerical methods to solve the Lindblad equation.

	For example, it can be advantageous to solve the Lindblad equation in Fourier space, which in the case of the harmonic oscillator is a method that even provides an analytic solution to the problem \cite{Homa2019, Bernad2018, Anastopoulos:1994tz}. 
	However, also some numerical procedures use a Fourier space representation and then apply some time stepping \cite{Gao:1997}. 
	Time stepping for Lindblad equations is for example done with Runge-Kutta and Taylor-series methods, Pad\'e approximations, operator splitting-based methods or the Kraus representation approximation methods, where some of them are trace preserving, $\mathrm{tr} \, \rho = 1$, some are not \cite{Cao:2021fja}.
	Mostly, these methods are not applied to the Lindblad equation in position space representation, which is, as we motivated above, a very useful approach for certain systems.
	
	However, using these methods in position space, it turns out that most of the schemes are not unconditionally stable.
	For example, for the widely used Crank-Nicolson scheme \cite{Crank_Nicolson_1947} it is known that for some initial conditions spurious oscillations can occur, which lead to wrong results and oftentimes even to a breakdown of the numerical computation \cite{10.1093/comjnl/9.1.110, 10.1093/comjnl/7.2.163, 8b370aba-ebed-340f-8ce2-87c0149f028b}.
	It has been observed that, in order to describe heavy quarkonia, numerical schemes can also lead to norm violations of the density matrix, which has to be separately treated \cite{Escobedo:2019gzn, Akamatsu:2020ypb, Blaizot:2018oev}.
	Simply speaking, this problem can arise for \glspl{pde}, whose dynamics is dominated by the first derivative terms (advection terms) and the second derivative terms (diffusion terms) are small.
	This depends on the coefficients of the \glspl{pde} themselves as well as on the intitial conditions and is quantified by the \gls{cfl} condition (a measure for the local speed of information/fluid propagation) \cite{CFL}.
	For the Lindblad equation, these problems seem to be present for initial conditions with large gradients (excited states), non-smooth initial conditions, and for special values of the Lindblad coefficients.

	Hence, it would be highly desirable to have a numerical method, which is stable for various kinds of initial conditions and Lindblad coefficients, which can be easily implemented and used as a black-box solver.
	The research field of computational fluid dynamics has developed a variety of numerical methods for solving all types of \glspl{pde}.
	The method we propose in this work, the famous \gls{kt} central scheme, is such a method.
	It is a semi-discrete\footnote{It is discrete in space and continuous in time, which enables the use of arbitrary time stepping methods, which can ensure the \gls{cfl} condition by automatically adjusting the size of the time steps $\Delta t$.} high-order\footnote{It is second order accurate in space, such that errors should decrease quadratically with the grid spacing $\Delta x$.
	In practice, there are usually small deviations from an exact $\Delta x^2$ scaling.} scheme, which is able to handle discontinuities and is based on the finite-volume method \cite{EYMARD2000713}.
	Finite-volume methods are in turn based on the subdivision of the computational domain into volume cells as well as the formulation of the \gls{pde} in terms of conservation laws.
	The latter can be shown for the Lindblad equation, as we will discuss in \cref{sec:reinterpretation}, which even allows to reinterpret the Lindblad equation in terms of diffusion and advection fluxes and source/sink terms for the density matrix and allows general new insights into Lindblad dynamics.
	Furthermore, norm conservation is theoretically guaranteed by the method, which is a crucial requirement for the Lindblad equation.
	We use this as a benchmark test in this work.
	To our knowledge, this method has not been used to solve Lindblad-type equations yet, and we hope that it might turn out as a useful tool for the community.

\subsection{Goal of this work}

	The goal of this work is to demonstrate that finite volume schemes, like the \gls{kt} central scheme, are powerful methods for solving the Lindblad equation in position space.
	In particular, we show that this scheme is a suitable numerical approach to solve the Lindblad equation for various Lindblad coefficients and various kinds of initial conditions including higher excited states.
	Furthermore, we discuss the Lindblad equation in terms of diffusion and advection fluxes and source/sink terms, which is a new perspective on the Lindblad equation and provides new insights into the dynamics of open quantum systems, the time evolution of the density matrix, and the thermalization process. 

\subsection{Structure}

	This paper is structured as follows: first, we rewrite the before introduced Lindblad equation, \cref{sec:lindblad},  given in \cref{eq:lindblad_spatial} into an advection-diffusion equation in conservative form. 
	This allows us to discuss the terms appearing in the Lindblad equation on the level of advection- and diffusion-, source- and sink-terms, cf.\ \cref{sec:reinterpretation}. 
	In the next step, in order to discuss the numerical method, \cref{sec:numerics}, we solve the one-dimensional von-Neumann equation in order to discuss some minimal test cases (free particle in the square well potential with different initial conditions), \cref{sec:minimal}.
	Thereafter, we solve the full dissipative Lindblad equation for the one-dimensional particle in a square well potential and the one-dimensional harmonic oscillator, \cref{sec:non-trivial}.
	This will pave the way to discuss more physically motivated systems in terms of thermalization and relaxation time, as it will be discussed in \reff\cite{Rais2025} and is motivated in \reff\cite{Rais:2022gfg}. 
	
\subsection{Conventions}
\label{sec:conventions}

	Without loss of generality, we work on nuclear scales and set the mass to $m = m_d = 470$ MeV, the reduced mass of a deuteron, cf. \cite{Rais:2022gfg}, and $\hbar = \kBoltzmann = 1$. The computational domain is set to $L \times L = 40 \text{ fm} \times 40$ fm, if not stated explicitly otherwise\footnote{Even though we are considering a one dimensional quantum system, cf.\ \reff\cite{Rais:2022gfg}, the density matrix formalism leads to a two dimensional problem in coordinate space, $\rho(x,y,t)$. }.
	However, our findings are independent of the chosen units and can be easily scaled to other units.

\section{Open Quantum System Approaches and the Lindblad master Equation}\label{sec:lindblad}

	In this section, we provide a more formal though brief recapitulation of open quantum systems and the Lindblad master equation.
	The position space representation of the Lindblad equation serves as a starting point for the discussions.

	Open quantum systems are used to describe a system of interest, which can be a single particle, or even a chain of interacting particles, surrounded by a heat bath.
	This can be most generally written in terms of the Hamiltonian
		\begin{align}
			\hat{H} = \hat{H}_{\mathrm{S}} + \hat{H}_{\mathrm{B}} + \hat{H}_{\mathrm{SB}} \, ,
		\end{align}	
	which was introduced by Feynman and Vernon in \reff\cite{Feynman:1963fq}, and analysed in depth by Caldeira and Leggett in \reff\cite{Caldeira:1982iu}, where the famous Caldeira-Leggett master equation was originally proposed.
	
	In this framework, $ \hat{H}_{\mathrm{S}}, \hat{H}_{\mathrm{B}}$, and $\hat{H}_{\mathrm{SB}}$ (here S refers to the system particle, B to the environment/bath and SB to the interaction between the bath and the system) are given by 
		\begin{align}\label{eq:hamiltonian}
			\hat{H}_{\mathrm{S}} = \, & \tfrac{1}{2 M} \, \hat{p}^2 + V ( \hat{x} ) \, ,	\vdistance
			\\
			\hat{H}_{\mathrm{B}} + \hat{H}_{\mathrm{SB}} = \, & \sum_{\alpha = 1}^{N} \Big[ \tfrac{1}{2 M_\alpha} \, \hat{p}_\alpha^2 + \tfrac{1}{2} \, M_\alpha \, \Omega_\alpha^2 \, \big( \hat{q}_\alpha - \tfrac{c_\alpha}{M_\alpha \, \Omega_\alpha^2} \, \hat{x} \big)^2 \Big] \, ,	\vdistance	\nonumber
		\end{align}
	where $N$ is the number of bath particles, and  $V ( \hat{x} )$ is the system potential, $V ( \hat{x} ) = \frac{1}{2} \, M \, \omega^2 \, \hat{x}^2$ in the case of a harmonic potential.
	In this notation, $\hat{x}$ represents the system's spatial coordinate and $\hat{q}_\alpha$ the ones of the bath, $\Omega_\alpha$ are the bath oscillator frequencies and $c_\alpha$ the coupling constants. 
	Generally, one can also use other system potentials $V ( \hat{x} )$ to describe the characteristics of the system of interest. In solid state physics, usually quantum dots and qubits are described successfully by Markovian approaches, cf. \reffs\cite{PhysRevB.78.235311,Jin2010,LEI20121408,Zhang:2018pda}, optical traps or even dimers and polymers coupled linearly among each other \cite{Bode:2023vqw,Chikako}. 
	From our point of view, the harmonic oscillator is a desirable benchmark to apply numerical methods in order to investigate the thermalization of the system particle in the Lindblad master equation framework, because the long time behaviour can be fully described analytically, as already mentioned above.
	
	Without detailing the exact derivation of the Lindblad equation from the path integral, cf.\ \reffs\cite{BRE02,DIOSI1993517,Ingold,Haenggi,Elze:1998ew}, and assuming Markovian behaviour, considering an Ohmic heat bath for the environment and satisfying norm conservation and positivity of the density matrix, the most general form of the Lindblad-type Caldeira-Leggett master equation is given by \cite{BRE02,gardiner00}
		\begin{align}\label{eq:lindblad}
			\dot{\hat{\rho}}_{\mathrm{S}} = \, & - \ii \big[ \hat{H}_{\mathrm{S}}, \hat{\rho}_{\mathrm{S}} ( t ) \big]_- - \ii \gamma \, \big[ \hat{x}, \big\{ \hat{p}, \hat{\rho}_{\mathrm{S}} ( t ) \big\}_+ \big]_- 	\vdistance
			\\
			& - D_{p p} \, \big[ \hat{x},\big[ \hat{x}, \hat{\rho}_{\mathrm{S}} ( t ) \big]_- \big]_- + 2 D_{p x} \, \big[ \hat{x}, \big[ \hat{p}, \hat{\rho}_{\mathrm{S}} ( t ) \big]_- \big]_- 	\vdistance	\nonumber
			\\
			& - D_{x x} \, \big[ \hat{p}, \big[ \hat{p}, \hat{\rho}_{\mathrm{S}} ( t )\big]_- \big]_- \, ,	\vdistance	\nonumber
		\end{align} 
	which can be expressed in spatial representation \cite{Bernad2018}, $\bra{x}\hat{\rho} \ket{y}$, by 
		\begin{align}\label{eq:lindblad_spatial}
			& \ii \tfrac{\partial}{\partial t} \, \rho ( x, y, t ) =	\vdistance
			\\
			= \, & \Big[ \tfrac{1}{2 m} \, \big( \tfrac{\partial^2}{\partial y^2} - \tfrac{\partial^2}{\partial x^2} \big) + ( V ( x ) - V ( y ) ) - \ii D_{p p} \, ( x - y )^2 	\vdistance	\nonumber
			\\
			& - \ii \gamma \, ( x - y ) \, \big( \tfrac{\partial}{\partial x} - \tfrac{\partial}{\partial y} \big) - 2 D_{p x} \, ( x - y ) \, \big( \tfrac{\partial}{\partial x} + \tfrac{\partial}{\partial y} \big) 	\vdistance	\nonumber
			\\
			& + \ii D_{x x} \, \big( \tfrac{\partial}{\partial x} + \tfrac{\partial}{\partial y} \big)^2 \Big] \, \rho ( x, y, t ) \, .	\vdistance	\nonumber
		\end{align} 
	Here, the constant $\gamma$-dependent\footnote{Note, that in general, $\gamma$ can also be time dependent, $\gamma = \gamma(t)$. We want to emphasize here, that this does not cause numerical difficulties.} coefficients $D_{p p}$, $D_{p x}$ and $D_{x x}$ were derived in \reff \cite{BRE02,DEKKER198467,DIOSI1993517, Sandulescu:1985wv, PhysRevA.16.2126} and have to satisfy certain conditions, in order to make \cref{eq:lindblad_spatial} physically reasonable, cf. \cref{sec:harmonic_osci}. 
	These conditions are derived and discussed in \reffs\cite{DEKKER198467,Homa2019,Roy:1999,Ramazanoglu:2009} and later used especially in the case of the harmonic potential.
	The ``dissipation coefficient" $\gamma$ is generally motivated by the Ohmic heat bath spectrum and satisfies $\gamma = \eta/2M$, where $\eta$ expresses the friction related to the force-force correlator
		\begin{align}\label{eq:forceforce}
			\braket{F(\tau)F(s)} = 2\eta T \delta(\tau-s)\, ,
		\end{align}
	cf.\ \reff\cite{Caldeira:1982iu,Lindenberg:1984zz,Chen:2023pgx}.
	This is of course valid for all temperatures of the classical non-interacting fluid. 
	The Lindblad equation is valid only for high (or medium) temperatures\footnote{In the fluctuation part of the path integral the kernel is given by \cite{Lindenberg:1984zz}
		\begin{align}\label{eq:diss-fluc}
			A_b(t,t') = \int_{-\infty}^{\infty} \dd \omega^\prime \, \rho_{\mathrm{B}} ( \omega^\prime ) \cosh \big( \tfrac{\omega^\prime}{2 T} \big) \cos ( \omega^\prime \, ( t - t^\prime ) ) \, ,
		\end{align}
	where $\rho_{\mathrm{B}}(\omega)$ is the Ohmic bath spectrum. 
	Therefore, to derive a master equation from this, it is necessary to Taylor expand the hyperbolic cosine function, which can be done for second or even to third order, and therefore explains the high-temperature limit.} and the influence functional, see \reff\cite{Feynman:1963fq}, as well as the system functional are proportional to the spatial variable, which makes it hard to derive the dissipation-fluctuation theorem without further assumptions.\footnote{To derive the master equation, the path integral influence functional is given by \reffs\cite{Feynman:1963fq,kleinert}. 
	To derive the dissipation-fluctuation theorem, \cref{eq:diss-fluc}, out of the influence functional one has to assume, that \cref{eq:diss-fluc},
		\begin{align}
			A_b ( t, t^\prime ) = \tfrac{1}{2} \big\langle \big\{ \tilde{F} ( t ) , \tilde{F} ( s ) \big\} \big\rangle - \big\langle \tilde{F} ( t ) \big\rangle \big\langle  \tilde{F} ( s ) \big\rangle \, ,
		\end{align}
	to recover \cref{eq:forceforce}.
	Hence, one has to assume that 
		\begin{align}
			\rho_B ( \omega ) = \chi_{\text{FF}}^{\prime \prime} ( \omega ) = \sum_{\alpha} \tfrac{\uppi C_\alpha^2}{2 m_\alpha \Omega_\alpha} \delta( \omega - \omega_\alpha ) \, .
		\end{align}}
	
	Indeed, \cref{eq:lindblad_spatial} satisfies norm conservation,  which means, that all terms has to vanish in order to satisfy $\partial_t \int \dd x \, \rho ( x, x, t ) = 0$.
	This has to be explicitly shown only for the last term of \cref{eq:lindblad_spatial}, because all the other term satisfy this condition trivially.
	However, the last term vanishes, if and only if the boundaries of the integration domain go from $-\infty$ to $\infty$, cf. \cref{sec:normconservation}.
	
	Therefore, we want to point out here, that for computations in a finite box, it is crucial to neglect the $D_{x x}$-proportional term, if one demands norm conservation.
	
	Within this work, we also use the Lindblad equation in terms of \cref{eq:lindblad_spatial} \cite{BRE02, Homa2019, Bernad2018, Gao:1997}.

\section{Reinterpretation of the Lindblad equation as an advection-diffusion equation with sources and sinks}
\label{sec:reinterpretation}

	In this section, we introduce a new perspective on the interpretation of the Lindblad equation and discuss the equation as an advection-diffusion equation with sources and sinks.
	
	However, let us start the discussion by reminding the reader that it was and is in general hard to physically interpret \cref{eq:lindblad} or \cref{eq:lindblad_spatial} term by term, since the contributions are not easily related to the physical quantities of the system and phenomenological parameters.
	Generally, the entire part, which does not belong to the pure von-Neumann equation is referred to be the ``dissipator".
	Usually, parts of the terms were therefore absorbed in an effective non-hermitian Hamiltonian, to extract a dissipative part, added to the von-Neumann equation with mixtures of the (anti-)commutators \cite{Gao:1997, Manzano:2020yyw, Ohlsson:2020gxx}.
	Here, we want to exemplify this with the $D$-constants, which are oftentimes denoted as diffusion constants: $D_{x x}$ as spatial diffusion, $D_{p x}$ cross correlated diffusion between space and momentum and $D_{pp}$ the momentum diffusion.
	However, its interpretation on the \gls{pde} level partially deviates, as we will demonstrate below.

	Next, let us provide another interpretation of the Lindblad equation, namely in terms of an advection-diffusion equation with sinks and/or sources in conservative form.
	In fact, the novel formulation implies that we do not need to introduce an effective Hamiltonian, nor interpret term by term proportional to the (anti-)commutators, cf.\ \cref{eq:lindblad}, but are in the position to give exact mathematical meaning to the very kind of differential equation, which also allows us physical interpretations.
	\subsection{Reformulation of the Lindblad equation}
	Splitting the density matrix into real and imaginary parts, rearranging the terms and performing integrations by parts, see \cref{sec:kt_form} for details, we can rewrite \cref{eq:lindblad_spatial} as follows,
		\begin{align}\label{eq:dgl}
			& \partial_t \vec{u} + \partial_x \vec{f}^{\, x} [ \vec{x}, \vec{u} \, ] + \partial_y \vec{f}^{\, y} [ \vec{x}, \vec{u} \, ] =	\vdistance
			\\
			= \, & \partial_x \vec{Q}^{\, x} [ \partial_x \vec{u}, \partial_y \vec{u} \, ] + \partial_y \vec{Q}^{\, y} [ \partial_x \vec{u}, \partial_y \vec{u} \, ] + \vec{S} [ t, \vec{x}, \vec{u} \, ] \, .	\vdistance	\nonumber
		\end{align} 
	Here, $\vec{u} = \vec{u} ( \vec{x}, t ) = ( \rho_I ( x, y, t ), \rho_R ( x, y, t ) )^T$ is a vector that contains the imaginary- and real part of the density matrix, where $\vec{f}^{x/y}, \vec{Q}^{x/y}$ and $\vec{S}$ are explicitly  given in \cref{sec:kt_form}.
	Note that the equation is now formulated entirely in terms of real quantities.
	
	To interpret \cref{eq:dgl}, we compare the equation with the general form of an advection-diffusion equation with sources and sinks,
		\begin{align}\label{eq:diff-adv}
			\tfrac{\partial}{\partial t} \, \xi = \vec{\nabla} \cdot \big( D \, \vec{\nabla} \, \xi - \vec{v} \, \xi \big) + S ( \xi ) \, .
		\end{align}
	Here, $\xi$ is usually some concentration or temperature, $D$ the diffusion coefficient, $\vec{v}$ the field velocity, and $S$ the source term related to sources or sinks of $\xi$. 
	In our case, $\xi = \vec{u}$, the vector of the real and imaginary parts of the probabilistic density (matrix).
	Thus, the comparison to the terms in \cref{eq:diff-adv} allows the following interpretation:
	This \gls{pde} has contributions of parabolic (diffusion) terms as well as hyperbolic (advection) terms.
	Explicitly, the diffusion term, which is of second order in derivatives, is
		\begin{align}\label{eq:diff_flux}
			& \vec{\nabla} \cdot \big( D \, \vec{\nabla} \xi \big) =	\Vdistance
			\\
			= \, & \partial_x \vec{Q}^{\, x} [ \partial_x \vec{u}, \partial_y \vec{u} \, ] + \partial_y \vec{Q}^{\, y} [ \partial_x \vec{u}, \partial_y \vec{u} \, ] =	\Vdistance	\nonumber
			\\
			= \, & \bigg[ \partial_x
			\begin{pmatrix}
				D_{x x} \, ( \partial_x + \partial_y )	&	\frac{1}{2 m} \, \partial_x
				\\
				- \frac{1}{2 m} \, \partial_x	&	D_{x x} \, ( \partial_x + \partial_y )
			\end{pmatrix} +
			\Vdistance	\nonumber
			\\
			& + \partial_y
			\begin{pmatrix}
				D_{x x} \, ( \partial_x + \partial_y )	&	- \frac{1}{2 m} \partial_y
				\\
				\frac{1}{2 m} \, \partial_y	&	D_{x x} \, ( \partial_x + \partial_y )
			\end{pmatrix}
			\bigg] \vec{u} =	\Vdistance	\nonumber
			\\
			= \, & \bigg[
			\begin{pmatrix}
				D_{x x}	&	\frac{1}{2 m}
				\\
				- \frac{1}{2 m}	&	D_{x x}
			\end{pmatrix}
			\partial_x^2 +
			\begin{pmatrix}
				D_{x x}	&	- \frac{1}{2 m}
				\\
				\frac{1}{2 m}	&	D_{x x} 
			\end{pmatrix}
			\partial_y^2 +	\Vdistance	\nonumber
			\\
			& + 2
			\begin{pmatrix}
				D_{x x}	&	0
				\\
				0	&	D_{x x}
			\end{pmatrix}
			\partial_y \partial_x \bigg] \vec{u} \, ,	\Vdistance	\nonumber
		\end{align}
	cf. \cref{eq:fqs3}  and \cref{eq:fqs4}, where the mixing term, proportional to $\partial_y\partial_x$, arises from the geometry of the computational domain, in which the Lindblad equation is solved.
	As its name suggests, the (spatial) diffusion term is in general responsible for the spreading of the density matrix in space.
	Interestingly, due to the factor $\ii$ on the \gls{lhs}\ of \cref{eq:lindblad_spatial} the actual diffusive terms are those proportional to $D_{x x}$ on the diagonal of the matrices in \cref{eq:diff_flux}, which do not mix the real and imaginary parts of the density matrix like the von-Neumann terms $\propto \frac{1}{2 m}$.
	Therefore, this $D_{x x}$-term is one main contributor to the time symmetry breaking and actually deserves the name ``spatial diffusion'' coefficient.

	The so called drift, or advection term (sometimes convection term) of \cref{eq:diff-adv} can be written, comparing to \cref{eq:fqs1} and \cref{eq:fqs2} as
		\begin{align}\label{eq:drift}
			& \vec{\nabla} \cdot \big( \vec{v} \, \xi \big) =	\Vdistance
			\\
			= \, & \partial_x \vec{f}^{\, x} [ \vec{x}, \vec{u} \, ] + \partial_y \vec{f}^{\, y} [ \vec{x}, \vec{u} \, ] = \Vdistance	\nonumber
			\\
			= \, & \partial_x
			\begin{pmatrix}
				- 2 D_{p x} \, ( x - y ) \, \rho_R + \gamma \, ( x - y ) \, \rho_I
				\\
				+ 2 D_{p x} \, ( x - y ) \, \rho_I + \gamma \, ( x - y ) \, \rho_R
			\end{pmatrix} +	\Vdistance	\nonumber
			\\
			& + \partial_y
			\begin{pmatrix}
				- 2 D_{p x} \, ( x - y ) \, \rho_R - \gamma \, ( x - y ) \, \rho_I
				\\
				+ 2 D_{p x} \, ( x - y ) \, \rho_I - \gamma \, ( x - y ) \, \rho_R
			\end{pmatrix} =	\Vdistance	\nonumber
			\\
			= \, & \bigg[
			\begin{pmatrix}
				\gamma	&	- 2 D_{p x}
				\\
				2 D_{p x}	&	\gamma
			\end{pmatrix} \partial_x -
			\begin{pmatrix}
				\gamma	&	2 D_{p x}
				\\
				- 2 D_{p x}	&	\gamma
			\end{pmatrix} \partial_y
			\bigg] ( x - y ) \, \vec{u} \, .	\Vdistance	\nonumber
		\end{align}
	The advection term is in general responsible for the directed motion/evolution of  a fluid, here the density matrix, in space.
	Interestingly, we find that the advection fluxes $\vec{f}^{\, x}$ and $\vec{f}^{\, y}$ depend on $x - y$, the distance from the diagonal of the density matrix as well as on the real and imaginary parts of the density matrix itself.
	(The explicit position dependence already implies that they contain some source/sink contribution in their conservative formulation.)
	Therefore, the advection term is not only responsible for the directed movement of the density matrix in space, but also for the interaction of the real and imaginary parts of the density matrix via the terms proportional to $D_{p x}$.
	Hence, in the fluid-dynamical formulation the $D_{p x}$-term is not diffusive but advective.
	It is only dissipative and breaks time symmetry of the system for situations, where the solution \gls{pde} produces non-analyticities.
	Let us note that $\gamma$ mainly determines the ``fluid velocity", here the propagation velocity of $\vec{u}$, orthogonal to the diagonal of the density matrix.
	Both aspects will be further discussed in the next section, where we switch to different coordinates.
	
	In general, we can conclude that the advection is large for excited states of the system, where the density matrix is not diagonal and has high gradients, while it is small for small gradients and close to the diagonal of the density matrix, which also explains the fast equilibration of these systems.

	Finally, the source/sink term of \cref{eq:diff-adv} is given by
		\begin{align}\label{eq:source_matrix}
			& S ( \xi ) = \vec{S} ( \vec{u} \, ) =	\Vdistance
			\\
			= \, & \begin{pmatrix}
				2 \gamma - D_{p p} \, ( x - y )^2	&	V ( y ) - V(x)
				\\
				V ( x ) - V ( y )	&	2 \gamma - D_{p p} \, ( x - y )^2 
			\end{pmatrix} \vec{u} \, ,	\Vdistance	\nonumber
		\end{align}
	cf. \cref{eq:fqs5}. In fact, we find that the external potential appears as a source/sink term in the \gls{pde}, which is, however, proportional to the real and imaginary parts of the density matrix.
	Similar to the diffusive term, the contributions from the von-Neumann equation couple real and imaginary parts of the density matrix via the \gls{pde}.
	The actual source/sink contributions are given by the diagonal entries of the source matrix \labelcref{eq:source_matrix}.
	Here, we again find the coefficient $\gamma$, but also the coefficient $D_{p p}$.
	In fact, the dependence on $\gamma$ is artificial, since the advection fluxes are position dependent and therefore contain a source contribution, which exactly cancels the $\gamma$ term, which is discussed in the next section.
	It follows that the real source is the $D_{p p}$ part, that contributes only to the off-diagonal elements of the density matrix.
	
\subsection{Interpretation in relative and centre-of-mass coordinates}

	To further elucidate the interpretation of the terms, let us switch to relative and centre of mass coordinates $r = \frac{1}{2} \, ( x - y )$ and $q = \frac{1}{2} \, ( x + y )$ and $\tilde{\vec{u}} ( r, q, t )$.
	Hence, $r$ basically measures the distance from the diagonal of the density matrix, while the $q$-coordinate is oriented along the diagonal.
	Now, starting with the source term, we find
		\begin{align}\label{eq:relS}
			& \vec{S} ( \tilde{\vec{u}} \, ) =	\Vdistance
			\\
			= \, &
			\begin{pmatrix}
				2 \gamma - 4 D_{p p} \, r^2	&	V ( r - q ) - V ( r + q )
				\\
				V ( r + q ) - V ( r - q )	&	2 \gamma - 4 D_{p p} \, r^2 
			\end{pmatrix} \tilde{\vec{u}} \, .	\Vdistance	\nonumber
		\end{align}
	Ignoring the $2 \gamma$ entries, which cancel with a contribution from the advection flux, cf. \cref{eq:drift2}, we find that for $r = 0$ the actual source contributions completely vanish and one is left with the von-Neumann terms -- the off-diagonal elements.
	However, to better understand the role of $- 4 D_{p p} r^2$ for $r^2 > 0$, let us assume that all other contributions of the \gls{pde} are zero.
	Then, the \gls{pde} reduces to
		\begin{align}
			\partial_t \tilde{\vec{u}}  = - 4 D_{p p} \, r^2 \, \tilde{\vec{u}} 
		\end{align}
	and is solved by
		\begin{align}
			\tilde{\vec{u}}  ( r, t ) = \tilde{\vec{u}}_0 \, \exp ( - 4 D_{p p} \, r^2 \, t ) \, .	\label{eq:decay-off-diagonal}
		\end{align}
	Hence, we find that the source term actually leads to an exponential suppression of the elements of the density matrix, which are off the diagonal, with a rate proportional to $4 D_{p p} r^2$.
	Hence, this term is also responsible for the decay of the off-diagonal elements of the density matrix, which is a key feature of the Lindblad equation, while it does not produce dissipative effects.
	Using $D_{p p} = 2 m \gamma T$\footnote{\label{note1}The coefficients $D_{xx}, D_{px}$ and $D_{pp}$ will be introduced and motivated in \ref{sec:non-trivial}, \cref{eq:d-params} explicitly.}, which is discussed below, we find that large $m$ and $T$ lead to a faster decay of the off-diagonal elements of the density matrix, which in general decay faster for larger $r^2$.
	The same applies to large $\gamma$, which, however, additionally contributes to the other $D$-coefficients, as we will see below.
	
	Again, in centre of mass and relative coordinates, the diffusion term, \cref{eq:diff_flux}, is given by
		\begin{align}\label{eq:diffusion}
			 & \partial_q
			\begin{pmatrix}
				D_{x x} \, \partial_q	&	\frac{1}{2 m} \, \partial_r
				\\
				- \frac{1}{2 m} \, \partial_r	&	D_{x x} \, \partial_q
			\end{pmatrix}
			\tilde{\vec{u}} =	\Vdistance	
			\\
			= \, & D_{x x} \, \partial_q^2
			\begin{pmatrix}
				\rho_I ( r, q, t )
				\\
				\rho_R ( r, q, t )
			\end{pmatrix}
			+ \tfrac{1}{2m} \, \partial_q \partial_r
			\begin{pmatrix}
				\rho_R ( r, q, t )
				\\
				- \rho_I ( r, q, t )
			\end{pmatrix} \, .	\Vdistance	\nonumber
		\end{align}
	Here, one clearly sees that the first term proportional to $D_{x x}$ is a pure diffusion equation, which in fact diffuses the density matrix parallel to its diagonal, while the second term describes how the real and imaginary parts of the density drive each other in space.
	As we will discuss, for $D_{x x} = \frac{\gamma}{6 m T}$\textsuperscript{\ref{note1}} the diffusion is enhanced for large $\gamma$, while large $m$ and/or $T$ reduce spatial diffusion.
	This allows the interpretation, that the ``by hand" introduced term $D_{x x}$, which is justified by the minimal invasive principle to obtain Lindblad form \cite{BRE02, PhysRevA.16.2126}  actually causes the spatial diffusion and is therefore key to the thermalization process and the dissipative dynamics of the system.

	The so called drift, or advection term of \cref{eq:diff-adv}, \cref{eq:drift} can be rewritten with the coordinates introduced above as 
		\begin{align}\label{eq:drift2}
			& \vec{\nabla} \cdot \big( \vec{v} \,  \tilde{\vec{u}}  \big) =	\Vdistance
			\\
			= \, &
			\begin{pmatrix}
					2 \gamma	&0
					\\
					0	&	2 \gamma
			\end{pmatrix}  \tilde{\vec{u}}  + 2 r \bigg[
			\begin{pmatrix}
				0	& - 2 D_{p x}
				\\
				2 D_{p x}	& 0
			\end{pmatrix} \partial_q +
			\begin{pmatrix}
				\gamma	&	 0
				\\
				0	&	\gamma
			\end{pmatrix} \partial_r
			\bigg]  \tilde{\vec{u}} \, .	\Vdistance	\nonumber
		\end{align}
	As mentioned earlier, the first contribution which stems from the position dependence of the advection flux is actually a source term that cancels the $2 \gamma$ terms in \cref{eq:relS}.
	It is therefore ignored in the discussion.
	Inspecting the other $\gamma$-dependent contribution in \cref{eq:drift2} we clearly see that $2 r \gamma$ is the advection velocity of the advection that is orthogonal to the diagonal of the density matrix.
	Hence, the $\gamma$-term is responsible for the directed movement of the components of the density matrix away from its diagonal.
	On the other hand, the $D_{p x}$-term is responsible for the interaction of the real and imaginary parts of the density matrix.
	Interestingly, it solely depends on the gradients parallel to the diagonal of the density matrix, but is also proportional to the distance from the diagonal.
	Hence, we expect the term to be relevant for the formation of the profile of the density matrix in the $q$-direction, before the off-diagonal entries are too suppressed.
	Still, in medium distance to the diagonal its contribution will be significant and using $D_{p x} = - \frac{\gamma T}{\Omega}$\textsuperscript{\ref{note1}} we expect increasing importance of the term for larger $\gamma$ and $T$, see below.

	To summarize, the Lindblad equation can be rewritten in terms of a conservation law with advection-diffusion terms and sources/sinks.
	In fact, the terms with coefficients $D_{p p}$, $D_{x x}$ and $\gamma$ are the main contributors to the diffusion, advection and source/sink terms, respectively, while the terms, that are already part of the von-Neumann equation always feed the real into the imaginary part and vice versa.
	The role of $D_{p x}$ is special, because it also leads to an interaction of the real and imaginary parts, which does not seem to be dissipative, but is still part of the Lindblad coefficients and not contained in the von-Neumann equation.

	Before we turn to the numerical method, let us briefly comment on the computational domain and the boundary conditions.
	Even though it seems appealing to use the coordinates $r$ and $q$ to solve the Lindblad equation, it is more convenient to use the original coordinates $x$ and $y$.
	The reason is that the boundary conditions are easier to implement, because they can be deduced from the boundary conditions of the wave-functions in the initial condition.
	(The wave function and the density matrix have to vanish at the boundary of the computational domain.)
	Rotating the coordinates by $45^\circ$, as is done when using $r$ and $q$, would make an implementation of these boundary conditions more complicated.
	Especially, if the problem is restricted to a finite-sized domain (e.g.\ for a box) $x \in [ -\frac{L}{2}, \frac{L}{2} ]$, such that $y \in [ -\frac{L}{2}, \frac{L}{2} ]$, a rotation of the coordinates would lead to a diamond-shaped computational domain in the $r$-$q$-plane, which is hard to handle within a finite volume scheme that is based on a rectangular grid.

\section{Numerical method}
\label{sec:numerics}

	In this section we introduce the numerical methods we use to solve the Lindblad equation.
	We discuss the basic idea of a \gls{fv} method, which is the underlying concept for the actual numerical scheme, the \gls{kt} scheme. 
	
	Afterwards, we also briefly introduce the \gls{kt} scheme itself, including all formulae of the explicit discretization, which can be used in combination with a time-stepper of choice to implement the scheme as a semi-discrete black-box solver.
	Finally, we comment on the implementation of boundary conditions.

	We note, that the entire discussion in this section is basically a summary of \reff\cite{KURGANOV2000241}, which contains all the details.
	Furthermore, we encourage the reader to skip (parts of) this section or skim through it in a first reading and come back to it when the actual implementation is of interest.

\subsection{Finite-volume method}

	Having derived the conservative form of \cref{eq:lindblad_spatial}, in particular \cref{eq:dgl}, we identified the functions $\vec{f}^{x/y} [ t, \vec{x}, \vec{u} ( t, \vec{x} \, ) ]$ as (non-linear) advection fluxes and $\vec{Q}^{x/y} [ t, \partial_x \vec{u} ( t, \vec{x} \, ), \partial_y \vec{u} ( t, \vec{x} \, ) ]$ as diffusion (dissipation) fluxes. 
	The source term is given by $S [ t, \vec{x}, \vec{u} ( t, \vec{x} \, ) ]$.
	We have discussed potential interpretations for all the appearing terms in \cref{sec:reinterpretation}.
	
	Next, we are looking for the temporal evolution of $\vec{u} ( t, \vec{x} \, )$ from some initial time $t_0$ to $t_N > t_0$.
	To this end, let us define the finite computational domain $\Omega = \mathcal{V} \times [ t_0, t_N ]$, where $\mathcal{V} \subset \mathbb{R}^2$ is the spatial volume and $t_{0/N}$ is the initial/final time with the initial condition $\vec{u} ( t_0, \vec{x} \, )$ and Dirichlet (Neumann) boundary conditions specifying $(\partial_{x,y}) \vec{u} (t, \vec{x} \, ) \vert_{x,y \in \partial \mathcal{V}}.$ 
	\gls{fv} methods discretize the computational domain into spatial control volumes $\mathcal{V}_i$ where the set of spatial control volumes, or how we call it in this work, the number of cells with cell centre $\vec{x}_i$, covers the spatial computational domain $\mathcal{V}$.
	Let
		\begin{align}\label{eq:cellaverage}
			\vec{\bar{u}}_i ( t ) \equiv \tfrac{1}{\vert \mathcal{V}_i \vert} \int_{\mathcal{V}_i} \dd \xi_x \, \dd \xi_y \, \vec{u} ( t, \vec{\xi} \, )
		\end{align}
	be the sliding cell average, where $\mathcal{V}_i = \{ \vec{\xi}: \vert \xi_x - x_i \vert \leq \frac{\Delta x}{2}, \vert \xi_y - y_i \vert \leq \frac{\Delta y}{2} \}$.
	Then \cref{eq:dgl} can be integrated over a control volume centred at $\vec{x}_i$, using the divergence (Gauss-Ostrogradsky) theorem on the fluxes to derive an ``integral'' form of \cref{eq:dgl},
	\begin{widetext}
		\begin{align}
			\partial_t \vec{\bar{u}}_i \, & + \frac{1}{\Delta y} \int_{y_i - \frac{\Delta y}{2}}^{y_i + \frac{\Delta y}{2}} \mathrm{d} y \, \frac{\vec{f}^{\, x} [ t, x_{i} + \frac{\Delta x}{2}, y, \vec{u} ( t, x_{i} + \frac{\Delta x}{2}, y ) ] - \vec{f}^{\, x} [ t, x_{i} - \frac{\Delta x}{2}, y, \vec{u} ( t, x_{i} - \frac{\Delta x}{2}, y ) ]}{\Delta x} +	\Vdistance
			\\
			& + \frac{1}{\Delta x} \int_{x_i - \frac{\Delta x}{2}}^{x_i + \frac{\Delta x}{2}} \mathrm{d} x \, \frac{\vec{f}^{\, y} [ t, x, y_{i} + \frac{\Delta y}{2}, \vec{u} ( t, x, y_{i} + \frac{\Delta y}{2} ) ] - \vec{f}^{\, y} [ t, x, y_{i} - \frac{\Delta y}{2}, \vec{u} ( t, x, y_{i} - \frac{\Delta y}{2} ) ]}{\Delta y} = \ldots \, .	\Vdistance	\nonumber
		\end{align}
	\end{widetext}
	The central challenge of a \gls{fv} scheme is to evaluate these integrals in order to evolve the cell averages $\vec{\bar{u}}_j$ in time from $t_n$ to $t_{n + 1}$.
	To this end, the fluxes $\vec{f}^{x/y}$ and $\vec{Q}^{x/y}$ have to be calculated at the cell boundaries $x_{j \pm \frac{1}{2}} = x_j \pm \frac{\Delta x}{2}$ and $y_{k \pm \frac{1}{2}} = y_k \pm \frac{\Delta x}{2}$.
	This, however, requires some kind of reconstruction of $\vec{u}$ on the cell boundaries.
	In general, there are several \gls{fv} schemes.
	Some are based on Riemann solvers (e.g.\ the Roe \cite{ROE1981357} or the HLLE solver \cite{Harten:1997,Einfeldt:1988}), however, there are also some that do not require Riemann solvers as for example the \gls{kt} scheme (by A. Kurganov and E. Tadmor) \cite{KURGANOV2000241}, which uses a piecewise linear reconstruction of the cell averages and a slope limiter to avoid oscillations.
	The derivation/construction of these schemes usually also requires the discretization of the temporal direction with some $\Delta t$ for a single time step.
	As a consequence, some schemes have even to be used in this fully discrete form.
	An advantage of the \gls{kt} scheme is that there is a well-defined $\Delta t \to 0$ limit, while keeping the spatial directions discrete.
	This allows to combine the \gls{kt} scheme with an arbitrary time integrator, which is especially useful for stiff systems or systems that require adaption of the time step size to fulfil the \gls{cfl} condition \cite{CFL}.
	
\subsection{KT (central) scheme}

	Next, let us present and discuss the \gls{kt} (central) scheme in its semi-discrete form.
	For explicit derivations and details, we refer to the original \reff\cite{KURGANOV2000241}.

	We immediately start with the final result, \cite[Eq.~(4.16)]{KURGANOV2000241}, which is continuous in time and discrete in space and explain the single terms and contributions afterwards.
		\begin{align}\label{eq:kt_form}
			\partial_t \vec{\bar{u}}_{j, k} ( t ) = \, & - \frac{\vec{H}^{\, x}_{j + \frac{1}{2}, k} - \vec{H}^{\, x}_{j - \frac{1}{2}, k}}{\Delta x} - \frac{\vec{H}^{\, y}_{j, k + \frac{1}{2}} - \vec{H}^{\, y}_{j, k - \frac{1}{2}}}{\Delta y} +	\Vdistance	\nonumber
			\\
			& + \frac{\vec{P}^{\, x}_{j + \frac{1}{2}, k} - \vec{P}^{\, x}_{j - \frac{1}{2}, k}}{\Delta x} + \frac{\vec{P}^{\, y}_{j, k + \frac{1}{2}} - \vec{P}^{\, y}_{j, k - \frac{1}{2}}}{\Delta y} +	\Vdistance	\nonumber
			\\
			& + \vec{S}_{j, k} \, .	\Vdistance	\nonumber
		\end{align}
	Here, $\vec{\bar{u}}_{j, k} ( t )$ is the vector of the cell averages of the fluids (in this work, the real and imaginary part of the density matrix) for the cell with cell centre $(x_j, y_k)$ and cell boundaries at $x_{j \pm \frac{1}{2}} = x_j \pm \frac{\Delta x}{2}$ and $y_{k \pm \frac{1}{2}} = y_k \pm \frac{\Delta y}{2}$.
	The labels $j$ and $k$ are the indices of the cells in $x$- and $y$-direction, respectively, and allow for different cell sizes and numbers in $x$- and $y$-direction, which is however not used in this work.
	The numerical advection fluxes at the cell interfaces are $\vec{H}^{x}_{j \pm \frac{1}{2}, k}$ and $\vec{P}^{x}_{j \pm \frac{1}{2}, k}$ are the diffusion fluxes in $x$-direction, while $\vec{S}_{j, k}$ is the source term.
	The advection and diffusion fluxes in $y$-direction are named analogously with $y$ and $j, k \pm \frac{1}{2}$ instead of $x$ and $j \pm \frac{1}{2}, k$.

	The crucial point is, how these numerical fluxes are calculated from the cell averages of the previous time step, which defines the scheme.
	One ingredient to the fluxes are the values of $\vec{u}$ at the cell boundaries, which have to be reconstructed from the cell averages.

\subsubsection{Piecewise linear reconstruction}

	The first step in the \gls{kt} scheme is the reconstruction of the values of the fluid on the cell boundaries from the cell averages.
	To this end, the cell averages are interpolated by a piecewise linear function.
		\begin{align}
			\vec{u}^\pm_{j + \frac{1}{2}, k} = \, & \vec{\bar{u}}_{j + 1, k} \mp \tfrac{\Delta x}{2} \, ( \partial_x \vec{u} \, )_{j + \frac{1}{2} \pm \frac{1}{2}, k} \, ,	\vdistance
			\\
			\vec{u}^\pm_{j, k + \frac{1}{2}} = \, & \vec{\bar{u}}_{j, k + 1} \mp \tfrac{\Delta y}{2} \, ( \partial_y \vec{u} \, )_{j, k + \frac{1}{2} \pm \frac{1}{2}} \, .	\vdistance
		\end{align}
	Hereby, one needs to estimate the spatial derivatives of the fluid, $( \partial_{x/y} \vec{u} \, )_{j, k}$ in order to avoid over or underestimation of the slopes.
	The latter would lead to spurious oscillations in the solution.
	In practice, this is done by so-called slope limiters,
		\begin{align}
			& ( \partial_x u^\alpha )_{j, k} =	\Vdistance
			\\
			= \, & f_\mathrm{limiter} \bigg( \frac{\bar{u}^\alpha_{j + 1, k} - \bar{u}^\alpha_{j, k}}{\Delta x}, \frac{\bar{u}^\alpha_{j, k} - \bar{u}^\alpha_{j - 1, k}}{\Delta x} \bigg) \, ,	\Vdistance	\nonumber
			\\
			& ( \partial_y u^\alpha )_{j, k} =	\Vdistance
			\\
			= \, & f_\mathrm{limiter} \bigg( \frac{\bar{u}^\alpha_{j, k + 1} - \bar{u}^\alpha_{j, k}}{\Delta y}, \frac{\bar{u}^\alpha_{j, k} - \bar{u}^\alpha_{j, k - 1}}{\Delta y} \bigg) \, ,	\Vdistance	\nonumber
		\end{align}
	where $f_\mathrm{limiter}$ is the slope limiter function.
	There are various limiters that can be used for this purpose.
	One of the most popular limiters is the MinMod limiter, which is given by
		\begin{align}
			f_\mathrm{MinMod} ( a, b ) =
			\begin{cases}
				\min ( | a |, | b | ),	& \text{if } a \cdot b > 0 \, ,	\vdistance
				\\
				0 \, ,	& \text{otherwise} \, ,
			\end{cases}
		\end{align}
	which is also used in this work.

\subsubsection{Advection}

	Having reconstructed the values of the fluid on the cell boundaries, the advection fluxes can be calculated.
\begin{widetext}
	\begin{align}
		\vec{H}^{\, x}_{j + \frac{1}{2}, k} = \, & \frac{\vec{f}^{\, x} \big[ \vec{u}^{\, +}_{j + \frac{1}{2}, k} \big] + \vec{f}^{\, x} \big[ \vec{u}^{\, -}_{j + \frac{1}{2}, k} \big]}{2} - \frac{a^{\, x}_{j + \frac{1}{2}, k}}{2} \, \Big( \vec{u}^{\, +}_{j + \frac{1}{2}, k} - \vec{u}^{\, -}_{j + \frac{1}{2}, k} \Big) \, ,	\Vdistance
		\\
		\vec{H}^{\, y}_{j, k + \frac{1}{2}} = \, & \frac{\vec{f}^{\, y} \big[ \vec{u}^{\, +}_{j, k + \frac{1}{2}} \big] + \vec{f}^{\, y} \big[ \vec{u}^{\, -}_{j, k + \frac{1}{2}} \big]}{2} - \frac{a^{\, y}_{j, k + \frac{1}{2}} }{2} \, \Big( \vec{u}^{\, +}_{j, k + \frac{1}{2}} - \vec{u}^{\, -}_{j, k + \frac{1}{2}} \Big) \, .	\Vdistance
	\end{align}
\end{widetext}
	As can be seen, the first ingredient is the average of the physical fluxes at the cell boundaries evaluated for the left- and right-reconstructed values of the fluid.
	The second part is a correction term, which is proportional to the local wave speed $a^{\, x/y}_{j (\pm \frac{1}{2}), k (\pm \frac{1}{2})}$ on the cell boundaries.
	These wave speeds are given by the maximum of the eigenvalues of the Jacobian of the fluid velocities,
		\begin{align}
			& a^{\, x}_{j + \frac{1}{2}, k} =	\Vdistance
			\\
			= \, & \mathrm{max} \bigg[ \rho \bigg( \frac{\partial \vec{f}^{\, x}}{\partial \vec{u}} \big[ \vec{u}^{\, +}_{j + \frac{1}{2}, k} \big] \bigg), \rho \bigg( \frac{\partial \vec{f}^{\, x}}{\partial \vec{u}} \big[ \vec{u}^{\, -}_{j + \frac{1}{2}, k} \big] \bigg) \bigg] \, ,	\Vdistance	\nonumber
			\\
			& a^{\, y}_{j, k + \frac{1}{2}} =	\Vdistance
			\\
			= \, & \mathrm{max} \bigg[ \rho \bigg( \frac{\partial \vec{f}^{\, y}}{\partial \vec{u}} \big[ \vec{u}^{\, +}_{j, k + \frac{1}{2}} \big] \bigg), \rho \bigg( \frac{\partial \vec{f}^{\, y}}{\partial \vec{u}} \big[ \vec{u}^{\, -}_{j, k + \frac{1}{2}} \big] \bigg) \bigg] \, ,	\Vdistance	\nonumber
		\end{align}
	where $\rho$ is the spectral radius of the Jacobian of the fluxes, $\rho ( A ) = \max\{ | \lambda_\alpha |, \ldots, | \lambda_\omega | \}$, where $A$ is a matrix and $\lambda_k$ are its eigenvalues.
	Hence, apart from the advection fluxes, which can be directly read off the \gls{pde}, see \cref{eq:final_form} for the Lindblad equation, the \gls{kt} scheme requires the calculation of the Jacobian of the fluxes and its eigenvalues, which is a straightforward task for the Lindblad equation, see \cref{app:eigenvalues}.

\subsubsection{Diffusion}

	Next, we turn to the discretization of the diffusion fluxes,
\begin{widetext}
	\begin{align}
		\vec{P}^{\, x}_{j + \frac{1}{2}, k} = \, & \frac{\vec{Q}^{\, x} \big[ \vec{\bar{u}}_{j, k}, \frac{\vec{\bar{u}}_{j + 1, k} - \vec{\bar{u}}_{j, k}}{\Delta x}, ( \partial_y \vec{u} \, )_{j, k} \big] + \vec{Q}^{\, x} \big[ \vec{\bar{u}}_{j + 1, k}, \frac{\vec{\bar{u}}_{j + 1, k} - \vec{\bar{u}}_{j, k}}{\Delta x}, ( \partial_y \vec{u} \, )_{j + 1, k} \big]}{2} \, ,	\Vdistance
		\\
		\vec{P}^{\, y}_{j, k + \frac{1}{2}} = \, & \frac{\vec{Q}^{\, y} \big[ \vec{\bar{u}}_{j, k}, ( \partial_x \vec{u} \, )_{j, k}, \frac{\vec{\bar{u}}_{j, k + 1} - \vec{\bar{u}}_{j, k}}{\Delta y} \big] + \vec{Q}^{\, y} \big[ \vec{\bar{u}}_{j, k + 1}, ( \partial_x \vec{u} \, )_{j, k + 1}, \frac{\vec{\bar{u}}_{j, k + 1} - \vec{\bar{u}}_{j, k}}{\Delta y} \big]}{2} \, .	\Vdistance
	\end{align}
\end{widetext}
	Here, the diffusion fluxes are calculated as the averages of the physical diffusion fluxes evaluated for the cell averages and finite difference derivatives in flux direction.
	Derivatives perpendicular to the flux direction enter as limited derivatives, which are calculated as described above.
	In total, in \cref{eq:kt_form} the diffusion term enters like a central finite difference of the fluxes and actually reduces to the second-order central finite difference scheme for the one-dimensional heat equation.

\subsubsection{Sources and sinks}

	Lastly, we provide our discretization of the source/sink term,
		\begin{align}
			\vec{S}_{j, k} = \, & \vec{S} \big[ \vec{\bar{u}}_{j, k} \big] \, .
		\end{align}
	Here, the source term is evaluated at the cell center and is directly added to the \gls{rhs} of the \gls{pde}.
	Of course, this constitutes an approximation, since in a \gls{fv} scheme the source term should be integrated over the cell volume.
	However, since this is not analytically possible, we decided to use this approximation, which is certainly valid in the limit of vanishing cell sizes and turned out as a good approximation in other applications of the scheme.

\subsection{Implementation of the boundary conditions}
\label{sec:boundary_conditions}
	
	In this section, we want to discuss the boundary conditions and their discretized implementation on the level of the cell averages.

	Let us start with the analytical perspective.
	In general, the density matrix is defined on a subset of $\mathbb{R}^2$, because we are considering a one dimensional problem in coordinate representation, $\bra{x} \hat{\rho} \ket{y}$.
	If there is no external potential and no spatial box, the density matrix is defined on the entire $\mathbb{R}^2$.
	Theoretically, there are no spatial boundary conditions for this scenario and the \gls{pde} problem is well-posed with an appropriate initial condition.
	The same holds true in the presence of an external potential (as long as the potential is not restricting the spatial domain, e.g.\ by a box potential).
	Here, the density matrix usually falls off towards spatial infinity, while it is still defined on the entire $\mathbb{R}^2$.
	However, if the density matrix is confined to a finite domain subset of $\mathbb{R}^2$, e.g.\ by a box potential, the density matrix has to vanish at the boundaries of this computational domain.

	Let us turn to the numerical perspective.
	The \gls{kt} scheme is of the type of a \gls{muscl} \cite{VANLEER1979101}.
	Hence, the piecewise linear reconstruction and time evolution of a cell average $\vec{\bar{u}}_{j, k}$ requires in total the knowledge of the neighbouring cells averages $\{ \vec{\bar{u}}_{j - 2, k}, \vec{\bar{u}}_{j - 1, k}, \vec{\bar{u}}_{j + 1, k}, \vec{\bar{u}}_{j + 2, k} \}$ as well as $\{ \vec{\bar{u}}_{j, k - 2}, \vec{\bar{u}}_{j, k - 1}, \vec{\bar{u}}_{j, k + 1}, \vec{\bar{u}}_{j, k + 1} \}$, which is a so-called 5-point stencil in each direction (including the cell itself).
	For the cells at the computational-domain boundaries this implies that some additional cells are required for the scheme.
	These additional cells are called ghost cells and are used to implement the spatial boundary conditions.

	For the situations for the Lindblad equation, which we described above, this implies the following.
	Without any potential and restrictions of the physical domain, one sill has to restrict the computational domain to a finite sized domain.
	This can be done via some mapping or by simply choosing a sufficiently large computational domain with sufficiently localized initial conditions.
	As long as the density matrix does not spread to the boundaries of the computational domain, the boundary conditions are not relevant and we can simply choose the cell averages in the ghost cells to be zero.
	However, as soon as the density matrix spreads to the boundaries, one will experience severe errors.

	The situation for a potential $V ( x )$ is similar.
	However, here, the density matrix is usually confined to a finite domain of $\mathbb{R}^2$ by the potential and falls off towards spatial infinity.
	Hence, as long as the computational domain is chosen much larger than the spreading of the density matrix, the boundary conditions are not relevant and one can choose the cell averages in the ghost cells to be zero.

	For a box potential, the situation is different.
	Here, the density matrix has to vanish outside the box.
	If we align the cell interfaces of the cells at the boundary of the computational domain with the boundary of the box, there are basically two options to choose the cell averages in the ghost cells.
	The first option is to again set the ghost cell averages to zero.
	Here, however, the reconstruction of $\vec{u}$ on the boundary of the computational domain will not vanish exactly.
	This may be considered to be a sub-leading error.
	Another choice is to mirror the cell averages of the last physical cells to the ghost cells, also mirroring the sign.
	This ensures that the reconstructed values of $\vec{u}$ vanish at the boundary of the computational domain, which corresponds to nodes in the wave function.
	Also this choice is not exact and may lead to minor errors, because one might experience a minimal outflow of the density matrix at the boundaries.

	We tested both approaches and found that the second approach is more accurate, while the difference for a sufficiently large number of cells is negligible.

	To be specific, for the confining box, we implemented the ghost cells as follows,
		\begin{align}
			&	\vec{\bar{u}}_{0, k} = -\vec{\bar{u}}_{1, k} \, ,	&&	\vec{\bar{u}}_{-1, k} = - \vec{\bar{u}}_{2, k} \, ,	\vdistance
			\\
			&	\vec{\bar{u}}_{N + 1, k} = -\vec{\bar{u}}_{N, k} \, ,	&&	\vec{\bar{u}}_{N + 2, k} = - \vec{\bar{u}}_{N - 1, k} \, ,	\vdistance
		\end{align}
	where the first ghost cell is at $j = 0$ and the last physical cell is at $j = N$ and analogously for the $y$-direction.
	Here, $N$ is the number of physical cells in $x$-direction and the first physical cell in $x$ direct is at $j = 1$.

\subsection{Time integration}\label{sec:timeintegration}

	For the numerical calculations, we use the \gls{kt} scheme implemented in {\it Python 3} \cite{10.5555/1593511}. We utilize several packages, including {\it numpy} \cite{harris2020array}, {\it scipy} \cite{2020SciPy-NMeth}, and {\it matplotlib} \cite{Hunter:2007}. For the time integration of the semi-discrete scheme introduced in the previous section, we employed the {\it solve\_ivp} function from {\it scipy}, using the explicit Runge-Kutta method of order 5(4) ({\it RK45}) as the time-stepping method. For all calculations, we used the same numerical precision: $r_{\mathrm{tol}} = 10^{-8}$ for the relative tolerance and $a_{\mathrm{tol}} = 10^{-8}$ for the absolute tolerance, cf. \reff \cite{2020SciPy-NMeth}.
	
 \section{Minimal test case: von-Neuman equation in a box}
 \label{sec:minimal}
 
	In this section, we present our first results.
	As a minimal though non-trivial test case, we solve the von-Neumann equation for a free particle in a box potential.
	Hence, all Lindblad coefficients $D_{p p}$, $D_{p x}$, $D_{x x}$, and $\gamma$ are set to zero.
	This implies, that the advection and the source term vanish exactly and the diagonal contributions (the true diffusion) in \cref{eq:diff_flux} are zero as well.
	The only non-vanishing contributions are the off-diagonal contributions in \cref{eq:diff_flux}, which are the ``diffusive'' terms that mix the real and imaginary part of the density matrix.

	For initial conditions, that are stationary solutions of a quantum mechanical system, the solution of this equation has to be constant.
	Violations of this constancy are due to numerical errors and serve as a measure for the numerical accuracy of the scheme and the correct implementation of the boundary conditions.

	On the other hand, for arbitrary initial conditions we expect some evolution of the density matrix and some oscillatory behaviour due to the reflection at the boundaries of the box potential.
	Even though there is no analytic solution for this case, we can still use the violation of the norm of the density matrix as a measure for the numerical accuracy as well as the violation of the symmetry of the problem.
	The parameter we investigate is the amount of cells covering the computational domain.
		\begin{figure*}
			\begin{center}
				\includegraphics[width=\linewidth]{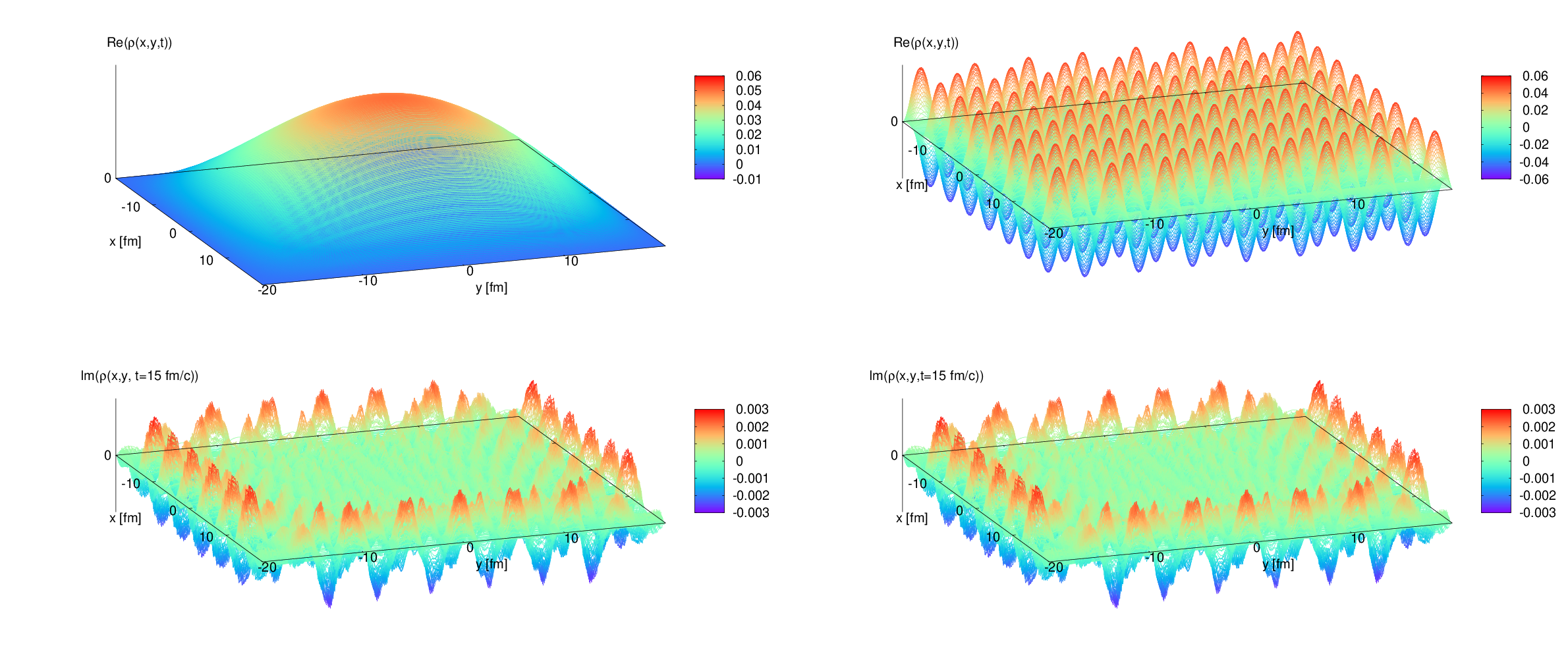}
				\caption{%
					The density matrix of a stationary state of a free particle in a square well potential in position space, as described by \cref{eq:free_part_wave}.
					The left panel depicts $n = 1$ and the right panel $n=15$. The upper row shows the real part of $\rho(x,y,t)$ for an arbitrary time.
					The lower row figures show the respective imaginary part for $t = 15$ fm/c.
					The number of computational cells is $500 \times 500$.%
				}
				\label{fig:rho3d}
			\end{center}
		\end{figure*}

\subsection{Free, one-dimensional particle in a square well potential}

	To begin the discussion, we consider the potential
		\begin{align}
			V ( x ) =
			\begin{cases}
				0 \, ,	&	\text{for} \, x \in [ - L/2, L/2 ] \, ,
				\\
				\infty \, ,	&	\text{otherwise,}
			\end{cases}
		\end{align}
	where $L = 40$ fm. The wave functions that build a complete set of orthogonal normalized eigenfunctions of $\hat{H}_\text{S}$ are given by
		\begin{align}\label{eq:free_part_wave}
			\psi_n ( x ) = \braket{x \vert \psi_n} =
			\begin{cases}
				\sqrt{\frac{2}{L}} \sin ( k_n x ) \, ,	&	\text{if} \, n \, \text{is even,}
				\\
				\sqrt{\frac{2}{L}} \cos ( k_n x ) \, ,	&	\text{if} \, n \, \text{is odd,}
			\end{cases}
		\end{align}
	where $k_n = \frac{n \uppi}{L}$.
	The density matrix is therefore given by
		\begin{align}\label{eq:density_matrix}
			\rho ( x, y, t ) = \sum_{n, m} \rho_{n, m} ( t ) \, \braket{x \vert \psi_n} \braket{\psi_m \vert y} \, ,
		\end{align} 
	where
		\begin{align}\label{eq:rho-mn}
			\rho_{n, m} ( t ) = \rho_{n, m}( 0 ) \, \exp \big(- \tfrac{\ii}{\hbar} \, ( E_n - E_m ) \, t \big) \, ,
		\end{align}
	is a solution for the ``free" von Neumann-equation,
		\begin{align}
			\partial_t \rho ( x, y, t ) = - \ii \bra{x} \big[ \hat{H}, \hat{\rho} \big]_- \ket{y} \, ,
		\end{align}
	which yields, if we decompose real and imaginary part of the density matrix,
		\begin{align}\label{eq:final_form_1}
			\partial_t
			\begin{pmatrix}
				\rho_I
				\\
				\rho_R
			\end{pmatrix}
			= \, & \tfrac{1}{2 m} \, \bigg[
			\partial_x^2
			\begin{pmatrix}
				\rho_R
				\\
				- \rho_I
			\end{pmatrix}
			- \partial_y^2
			\begin{pmatrix}
				\rho_R
				\\
				- \rho_I
			\end{pmatrix}
			\bigg] \, .
		\end{align}
	In \cref{eq:rho-mn}, $\rho_{n, m} ( 0 )$ corresponds to the initially populated state. 
	As one can also see already from \cref{eq:rho-mn}, only a state which is not an energy eigenstate of the system does change during a time evolution.
	In \cref{eq:final_form_1}, we clearly see that the real and imaginary part of the density matrix are coupled by the Laplacian.
	Even though it has the form of diffusion fluxes of the \gls{kt} scheme, the von-Neumann equation is not a normal diffusion equation, because $\partial_t \rho_{\mathrm{I}/\mathrm{R}}$ is not proportional to $\partial_{x/y}^2 \rho_{\mathrm{I}/\mathrm{R}}$ but to $\partial_{x/y}^2 \rho_{\mathrm{R}/\mathrm{I}}$.
	Still, the \gls{kt} is formally applicable, which calls for numerical tests.

\subsection{Stationary States}

	Let us first discuss the situation of stationary states.
	In \cref{fig:rho3d} we show two examples.
	The left column describes the density matrix $\rho ( x, y, t )$ for the case $n=1$, the right column the case $n = 15$, using \cref{eq:free_part_wave}. 
	Since the density matrix of an eigenstate remains constant during time evolution with the von-Neumann equation, we only show the initial density distribution on the $x$-$y$-plane. 
	Numerically, it is not as trivial, as \cref{eq:rho-mn} suggests. 
	Applying the \gls{kt} scheme \cref{eq:final_form_1} is solved numerically.
	As one can see in the lower row of \cref{fig:rho3d}, the imaginary part is not, as in the analytical solution, completely vanishing, which comes from the coupling between imaginary and real part and the accumulation of numerical errors.
	However, the imaginary part is very small.
	For reasons of better illustration we only show $\rho_{\mathrm{I}} ( x, y, t )$ at $t = 15$ fm/c.

\begin{figure}
			\begin{center}
				\includegraphics[width=1.0\columnwidth,clip=true]{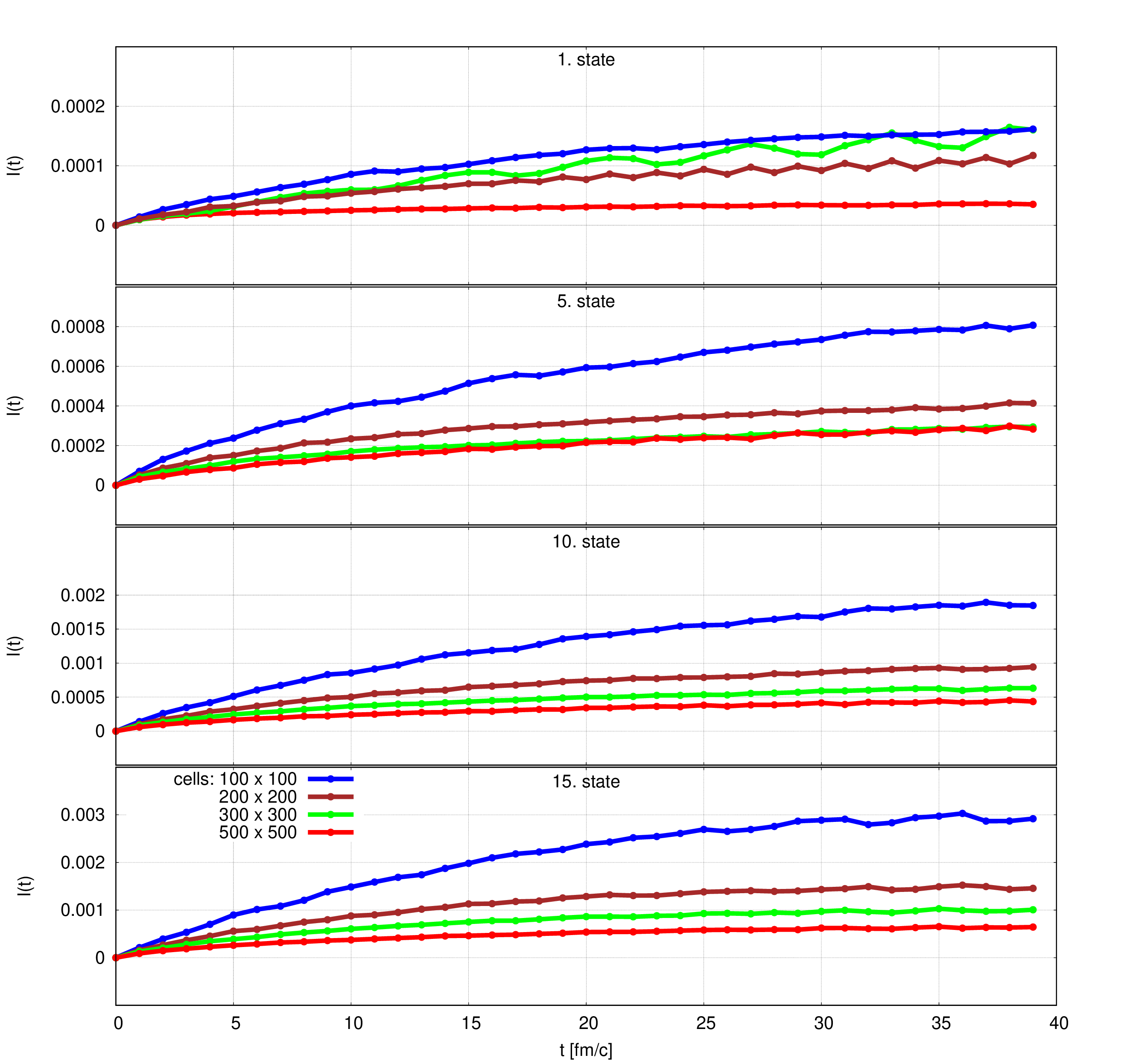}
				\caption{The cell-number-normalized violation of the (analyticaly vanishing) imaginary part of the density matrix, $I(t)$ as defined in \cref{eq:im_sum}, for the stationary state for different initial conditions, $n \in \{ 1, 5, 10, 15 \}$ as a function of time.
				The different colors correspond to different numbers of computational cells.
				}
				\label{fig:integral_im}
			\end{center}
		\end{figure}

	If one integrates the imaginary part and divides by the number of cells, 
		\begin{align}\label{eq:im_sum}
			I ( t ) \equiv \tfrac{1}{N_{\mathrm{cells}}} \sum_{\mathcal{V}_i} \big| \rho_I ( x, y, t ) \big|,
		\end{align}
	one can see in \cref{fig:integral_im}, that the accumulated error gets smaller, the more cells we implement, except for $n = 1$.
	Another result regarding the lower row of \cref{fig:rho3d} is that the imaginary parts of the density matrix have the largest values close to the boundary of the system. This effect stems from the imperfect boundary conditions and has to be accepted as a discretization artifact.

\begin{figure}
			\begin{center}
				\includegraphics[width=1.0\columnwidth,clip=true]{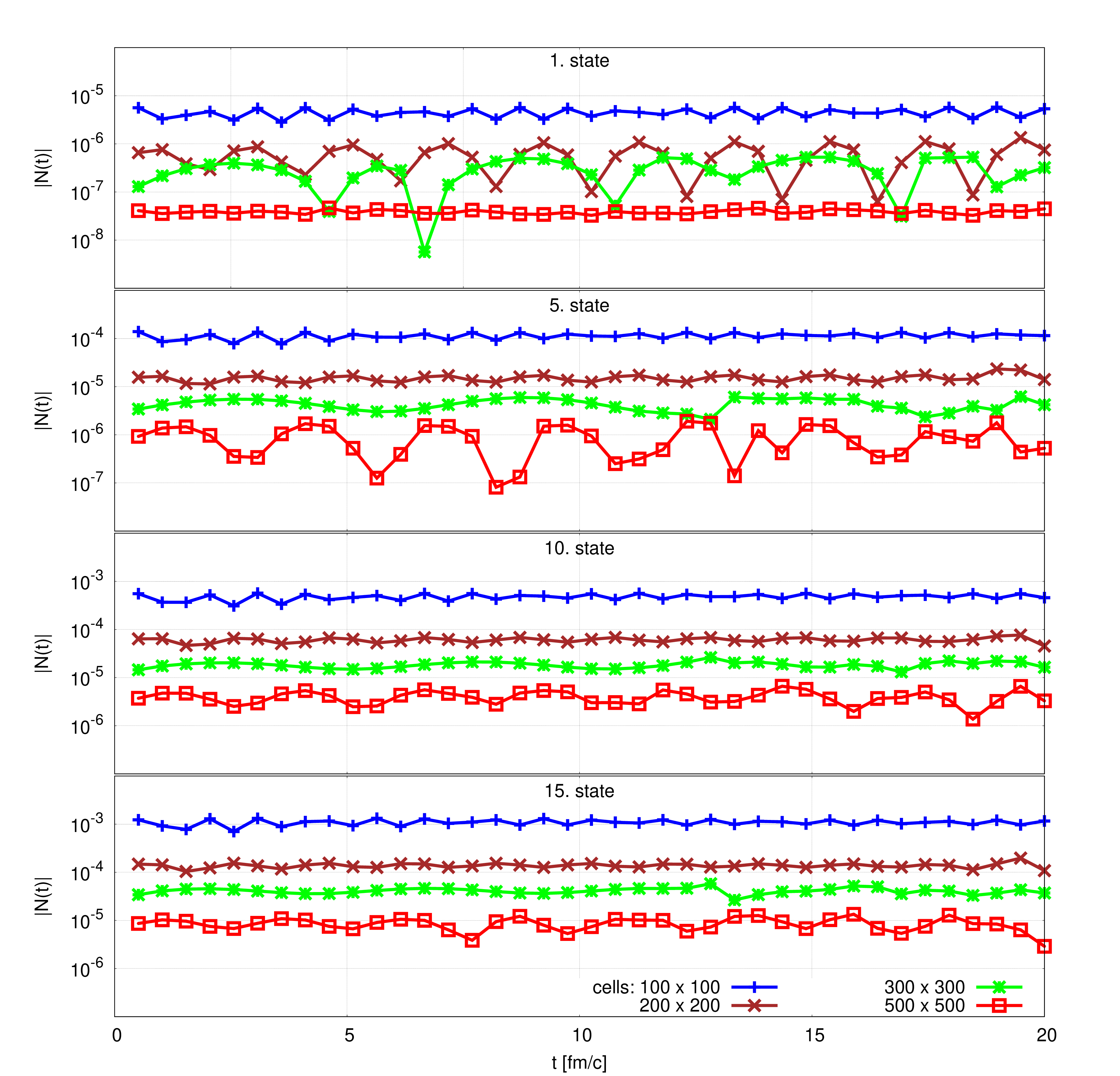}
				\caption{Deviation of the norm of the density matrix, $| N ( t ) |$ as defined in \cref{eq:norm}, from $t = 0$ to $t = 20$ fm/c for the stationary states $n \in \{ 1, 5, 10, 15 \}$.
				The different colours stand for different numbers of computational cells.
				}
				\label{fig:norm_von_neumann}
			\end{center}
		\end{figure}

	Another benchmark test of the reliability of the \gls{kt} scheme is to regard the conserved quantities.
	This is given by $\int \dd x \, \rho ( x, x, t ) = 1$. 
	For a stationary system, the density matrix should remain in its initially prepared state. 
	However, numerically fluctuations arise. 
	If we calculate 
		\begin{align}\label{eq:norm}
			N(t) =\frac{\int \dd x \, \rho ( x, x, t )}{\int \dd x \, \rho ( x, x, 0)} - 1 \, ,
		\end{align}
	we can quantify the deviation of the norm during the temporal evolution. 
	Let us mention here, that the norm of the initial discretized density matrix is not a priori exactly 1, even though the analytical input wave functions are normalized. 
	This has two origins. First the numerical summation is performed over the diagonal of the rectangular space, that defines the computational domain. Therefore, the geometry of the physical space is rotated by $\uppi/2$, which means, that one integrates over the diagonal of the cells. 
	Therefore,
		\begin{align}
			\lim_{\mathcal{V}_i \rightarrow 0} \int \dd x \, \rho ( x, x, t ) = 1 \, .
		\end{align}
	In \cref{fig:norm_von_neumann} we show how \cref{eq:norm} is applied to states with $n = 1, 5, 10, 15$.
	One can see, that the deviation decreases, if the number of cells is increased for all initial conditions.
	In all cases the deviation still is negligibly small and one can see, that for the cases $200 \times 200$, $300 \times 300$, and $500 \times 500$ cells also the initial condition does not influence the norm deviation during the temporal evolution, which is a strong advantage of the \gls{kt} scheme. In \cref{fig:integral_im} and \cref{fig:norm_von_neumann} one can also see, that the errors scale with approximately $(\Delta x)^2$, which explains, why the curves for $200\times 200$, $300\times 300$ and $500\times 500$ are much closer to each other on the linear scale then for the case, where we consider $100\times 100$ cells.

	Let us mention at this point, that the dots in \cref{fig:norm_von_neumann,fig:integral_im} simply denote the times, where we extract and store the data from the solution of the \gls{pde}.
	These are not the discrete time steps of the integrator.
	The lines are only drawn to guide the eye and to point out tendencies of the temporal behaviour of the norm and the imaginary part of the density matrix.
		\begin{figure}
			\begin{center}
				\includegraphics[width=1.0\columnwidth,clip=true]{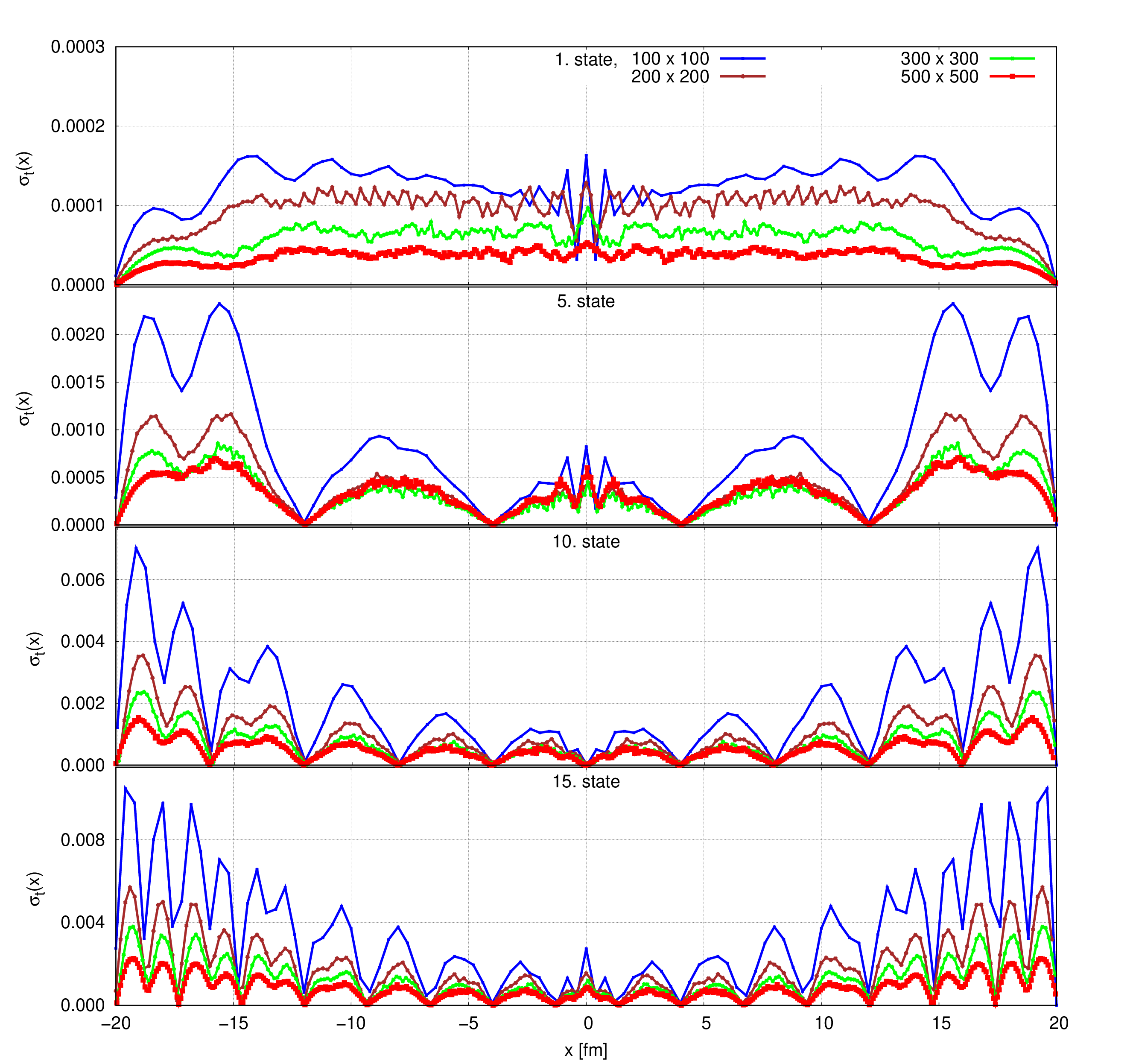}
				\caption{Deviation of the numerical solution form the analytical solution for $\rho ( x, x )$ measured in terms of $\sigma_t(x)$, \cref{eq:sigmat}.
				From top to bottom results for $n = 1, 5, 10, 15$ for the full lengths of the square well potential, averaged over time from times $t_0 = 0$ to $t_N=20$ fm/c are presented, cf. \cref{eq:sigmat}.
				The different colors denote different numbers of computational cells.
				}
				\label{fig:stationary_states}
			\end{center}
		\end{figure}
		\begin{figure*}
			\begin{center}
				\includegraphics[width=\linewidth]{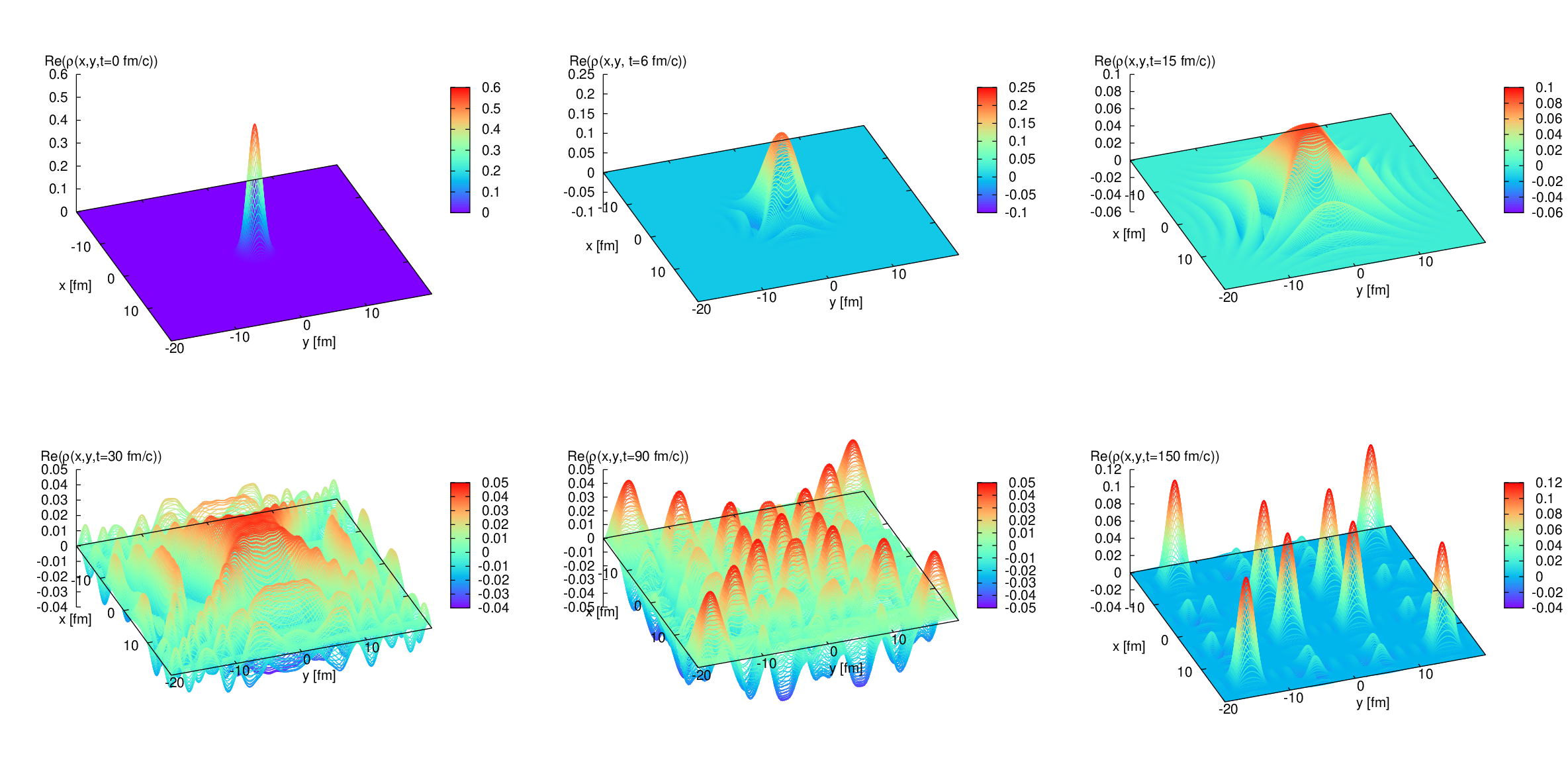}
				\includegraphics[width=\linewidth]{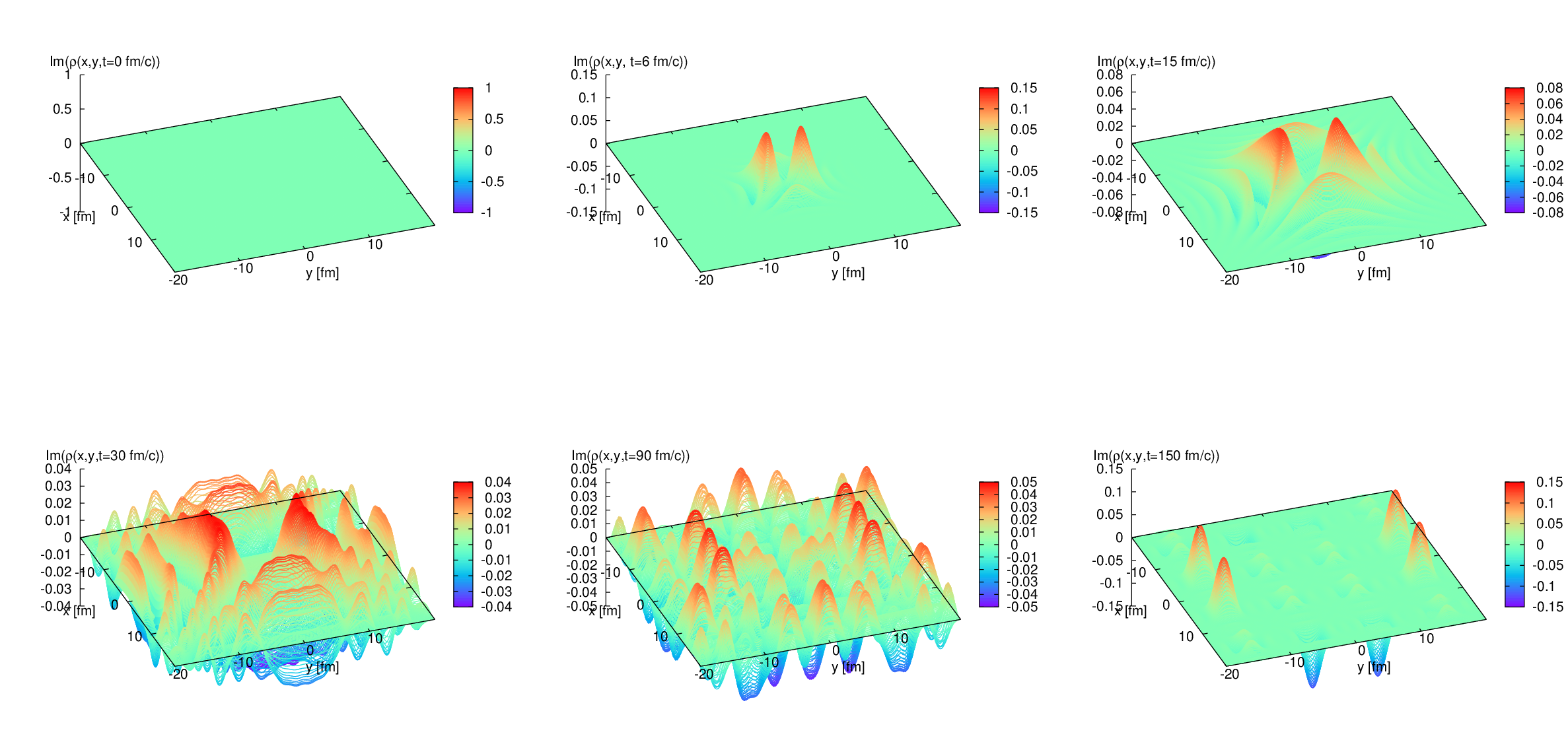}
				\caption{Temporal evolution of the real part (upper two rows) and of the imaginary part (lower two rows) of the density matrix of a Gaussian wave packet, \cref{eq:gauss}, with $a = 1$ fm$^{-2}$ for times $t = 0$ to $t = 150$ fm/c.
				The number of computational cells is $300 \times 300$.
				}
				\label{fig:rho3d_gauss}
			\end{center}
		\end{figure*}
		\begin{figure*}
			\begin{center}
				\includegraphics[width=\linewidth]{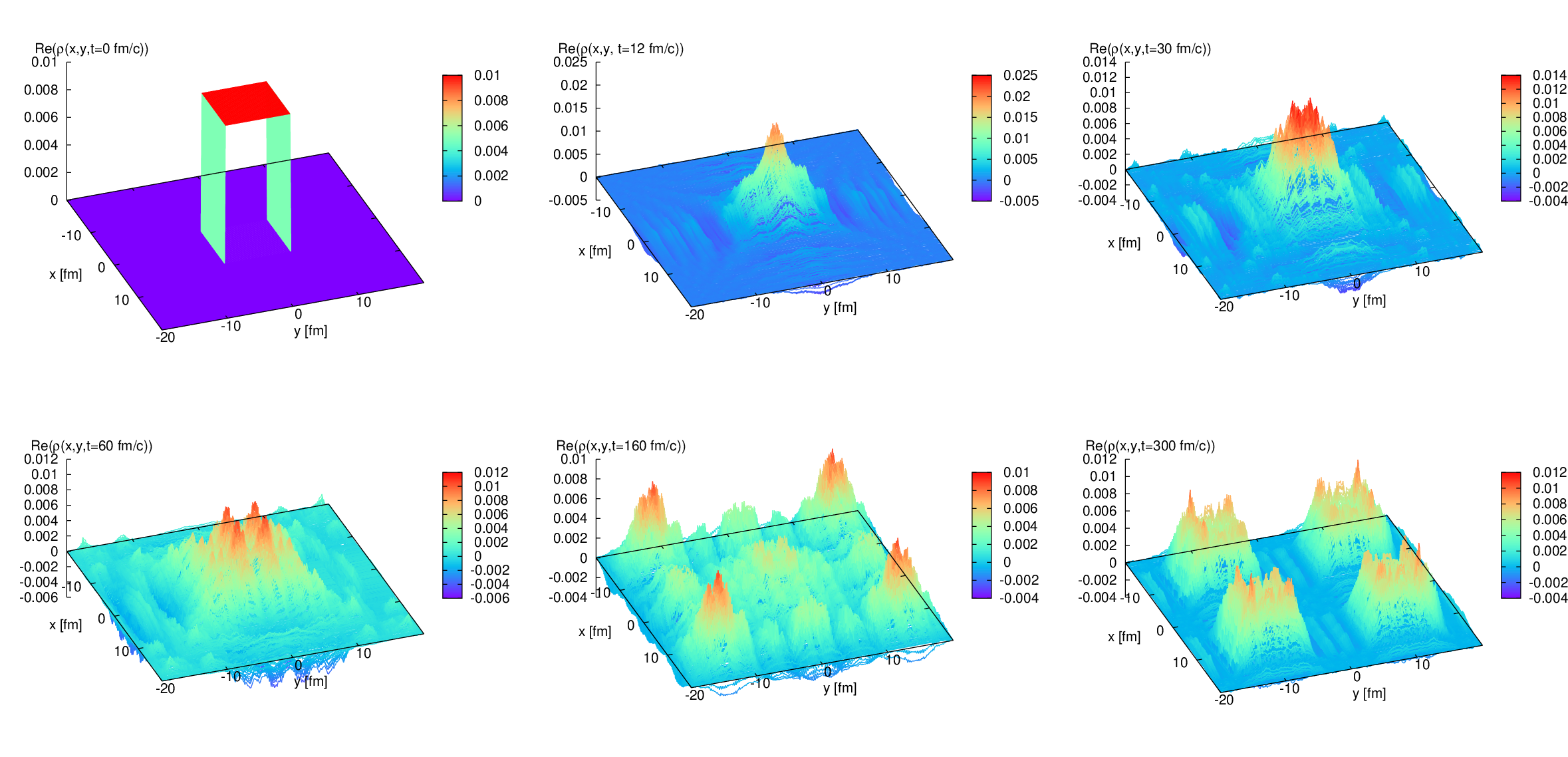}
				\includegraphics[width=\linewidth]{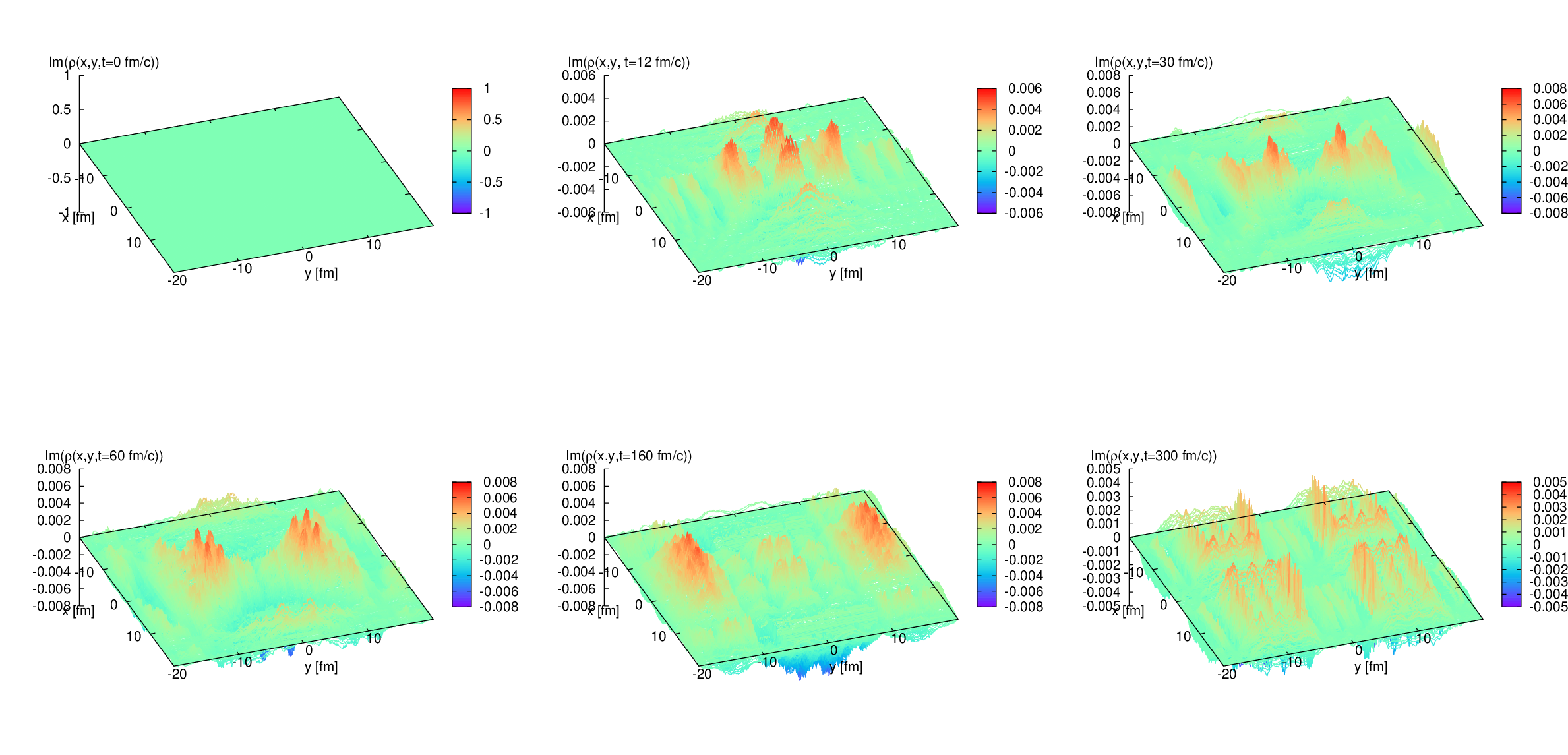}
				\caption{Temporal evolution of the real part (upper two rows) and the imaginary part (lower two rows) of the density matrix of a initially prepared rectangular wave function, \cref{eq:box}, where $b = 5$ fm for times $t = 0$ to $t = 300$ fm/c.
				The number of computational cells is $300 \times 300$.
				}
				\label{fig:rho3d_box}
			\end{center}
		\end{figure*}
	As a next step, we calculate the standard deviation
		\begin{align}\label{eq:sigmat}
			\sigma_t ( x ) = \sqrt{\tfrac{1}{N} \sum_{n = 0}^N \big( \rho( x, x, t_n ) - \rho ( x, x, t_0 ) )^2} \, ,
		\end{align} 
	where $t_n$ is the $n$-th evaluation time step,
	to show the fluctuations around the analytic result.
	Again, \cref{fig:stationary_states} shows $\sigma_t ( x )$ for the cases $n = 1, 5, 10, 15$.
	As expected, the fluctuation increase for higher $n$.
	For the cases $n = 5, 10, 15$ the fluctuations are highest close to the boundaries, which is due to boundary effect and also explains the imaginary part of the density matrix in \cref{fig:rho3d}, which is non-negligible near the boundaries.
	For the first state, the boundary effects do not impact the fluctuations in a crucial way, because the density matrix is centred. 
	It is worth to mention, that the fluctuations, which are not caused by boundary effects do not increase with a smaller amount of cells. 
	If one compares the order of the maximum value of $\sigma_t ( x )$ for $x \in [ - l, + l ]$, for some $l \ll L$, such that boundary effects are negligible, one can see, that the fluctuations for different numbers of cells are of the same order. 
	The deviations can even increase for higher amounts of cells, as can be seen in the first row of \cref{fig:stationary_states}, due to small numerical deviations. 
	Let us mention, that the accuracy of the Runge-Kutta solver used, is chosen to be constant at $r_{\text{tol}} = a_{\text{tol}} = 10^{- 8}$, cf. \cref{sec:timeintegration}, which might already have influence on the results.
	
\subsection{Arbitrary initial conditions}

	Having discussed the stationary problem, we want to solve \cref{eq:final_form_1} for two arbitrary prepared initial conditions, namely a Gaussian wave packet and a rectangular box-like wave function, which are given by
		\begin{align}\label{eq:gauss}
			\psi_{0, \text{Gauss}} = \, & \big( \tfrac{a}{\uppi} \big)^{\frac{1}{4}} \, \ee^{- \frac{a}{2} \, x^2} \, ,
		\end{align}
	where $a \gg \frac{1}{L^2}$ and
		\begin{align}\label{eq:box}
			\psi_{0, \text{Box}} = \, & \tfrac{1}{2 b} \, [ \theta ( x - b ) - \theta ( x + b ) ] \, ,
		\end{align}
	which is illustrated in \cref{fig:rho3d_gauss,fig:rho3d_box} (upper left panels).
	In the following, we discuss the numerical results and the patterns, which can be seen in the figures.
		\begin{figure}[t]
			\begin{center}
				\includegraphics[width=1.0\columnwidth,clip=true]{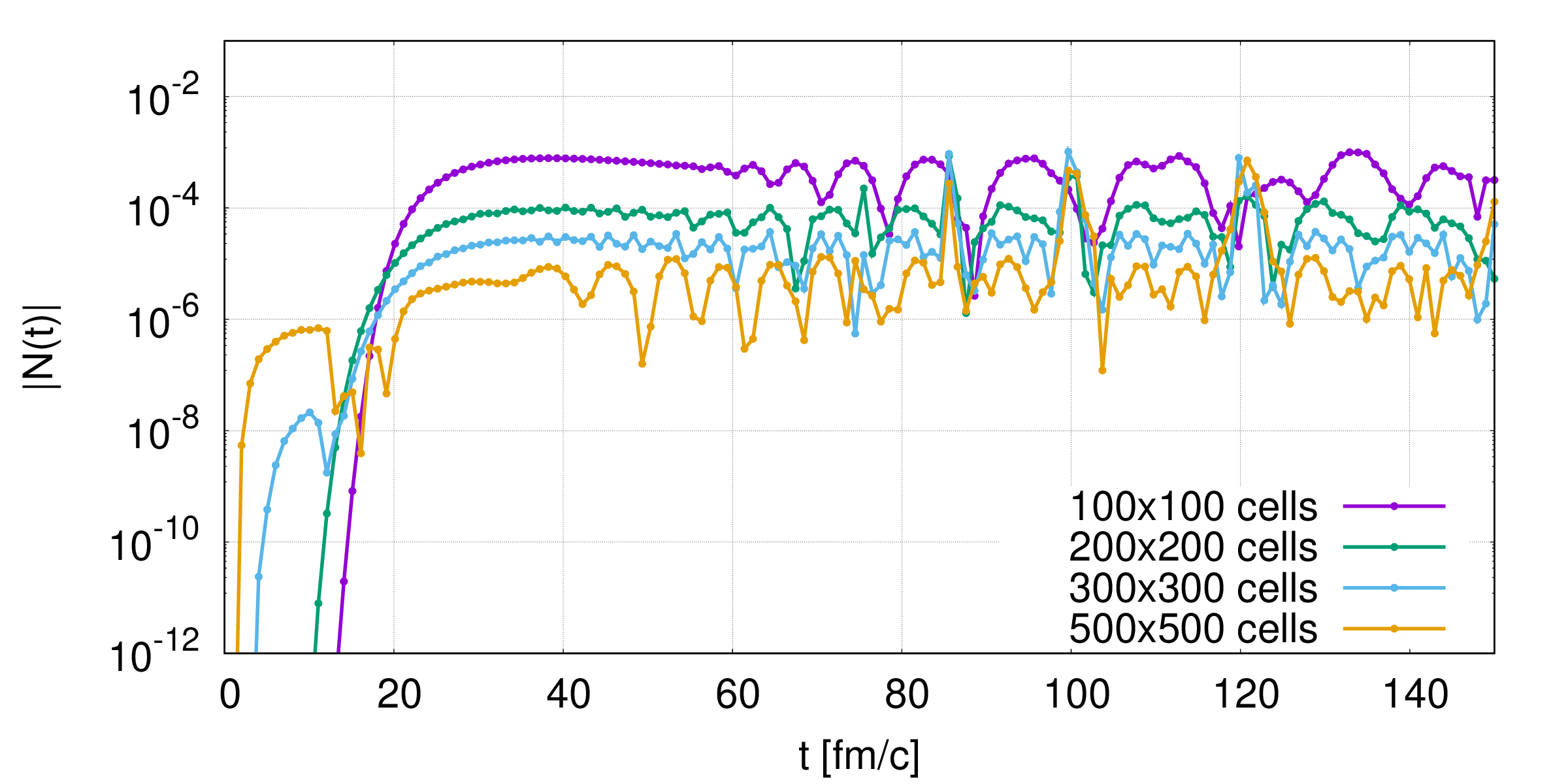}
				\caption{Deviation of the norm of the density matrix from one, $| N ( t ) |$ as defined in \cref{eq:norm}, for the Gaussian initial condition \labelcref{eq:gauss}, see also \cref{fig:rho3d_gauss}.
				Different colors correspond to different numbers of computational cells. 
				}
				\label{fig:norm_gauss}
			\end{center}
		\end{figure}
	Since both functions are real valued, the initial density matrix is simply given by
		\begin{align}\label{eq:psipsi}
			\rho ( x, y, t =0) = \psi_0 ( x ) \, \psi_0 ( y ) \, ,
		\end{align}
	and a dynamical propagation of these initial conditions is expected. 
	Even though there are no analytical solutions for these initial conditions, we can validate the quality of the simulation calculating the deviation of the norm for these two test cases. 
	We also want to point out the patterns that are obtained during the simulation.
		\begin{figure}
			\begin{center}
				\includegraphics[width=1.0\columnwidth,clip=true]{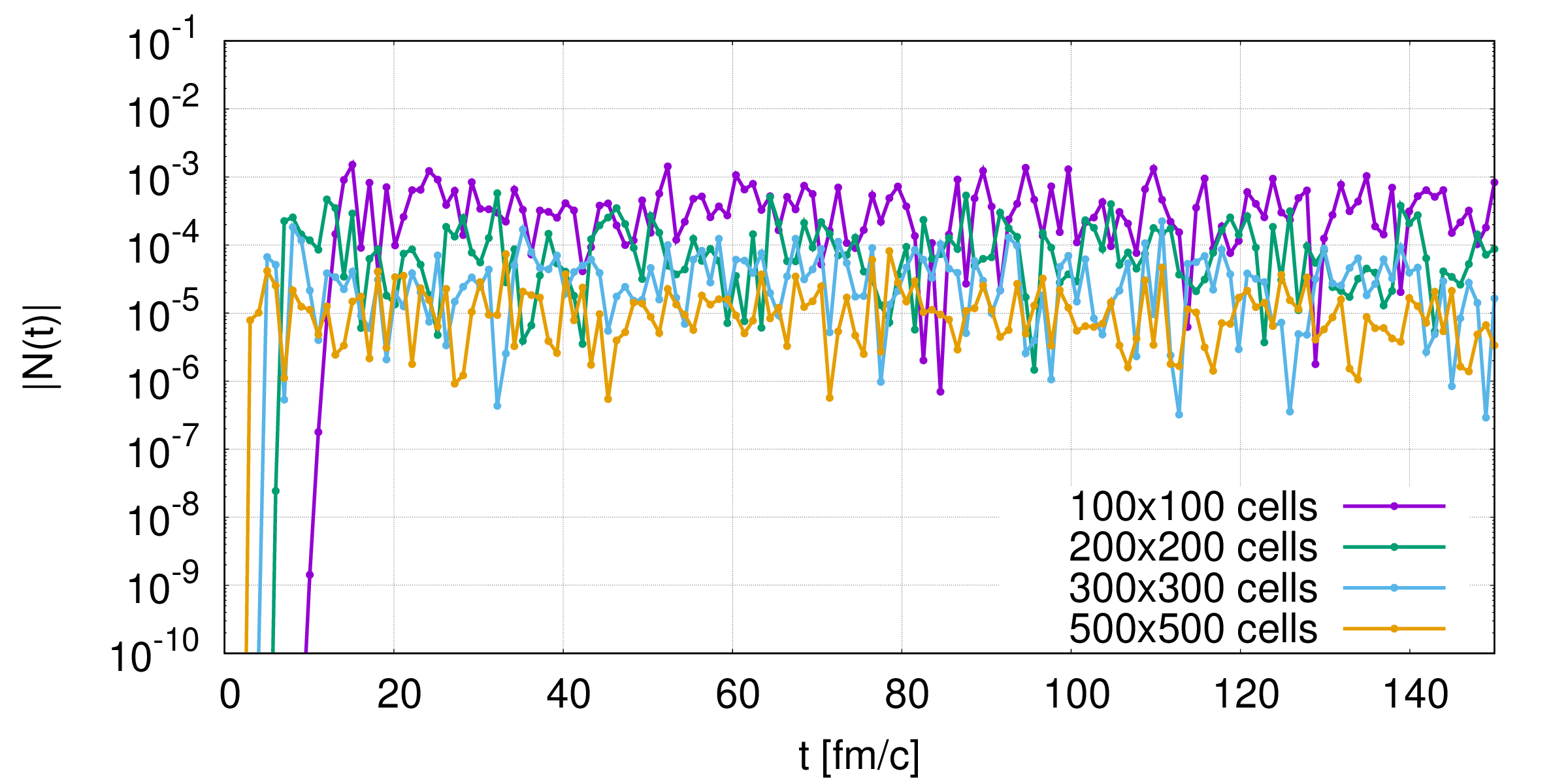}
				\caption{Deviation of the norm of the density matrix from one, $| N ( t ) |$ as defined in \cref{eq:norm}, for the ``box'' initial condition \labelcref{eq:box}, see also \cref{fig:rho3d_box}.
				Different colors correspond to different numbers of computational cells.}
				\label{fig:norm_box}
			\end{center}
		\end{figure}

\subsubsection{Gaussian wave packet}

	Therefore, inspecting \cref{fig:rho3d_gauss}, an initially prepared Gaussian broadens in time until it reaches the boundaries, which reflect the wave, such the it starts to interfere with itself. 
	The result is highly non-trivial interference pattern, which develops two orthogonal symmetry axes, dictated by the boundaries of the computational domain.
	The imaginary part of the propagating Gaussian is illustrated in \cref{fig:rho3d_gauss} (lower two rows).
	One can see, that the density matrix reaches a ``quasi" stationary state, where oscillations arise only symmetrically.

	To close the discussion of the Gaussian initial wave function, we want to discuss \cref{fig:norm_gauss}, which again shows the function $N ( t )$, \cref{eq:norm}, for different numbers of cells. 
	As the Gaussian wave packet is popular for its smooth behaviour, the deviation $N ( t )$ is small already for a rather small number of cells. 
	If one compares \cref{fig:norm_gauss} with \cref{fig:rho3d_gauss}, one can see, that for $t = 20$ fm/c, the norm is perfectly conserved for all cases, until the density matrix reaches the boundaries of the computational domain. 
	Therefore, \cref{fig:norm_gauss} also shows at which time $\rho(x,y,t)$ reaches the boundaries (here approximately $20$ fm/c) and also shows, that errors due to the discretization of the boundary condition are unavoidable. 
	Still these effects are very small.
	Again one can see, that the error scale is approximately $(\Delta x)^2$.

\subsubsection{Box-like wave function}

	For the second initial condition, which is given by \cref{eq:box} and \cref{eq:psipsi}, the temporal evolution, until a quasi steady state is reached takes longer than for the case of the Gaussian initial condition, see \cref{fig:rho3d_box}. 
	Even though the symmetry axes are the same, the interference pattern is different, mostly distributed along the off-diagonal entries of the density matrix. 
	Therefore, caused by the geometry of the system, there are four symmetric blocks, which build sub-blocks of the density matrix.
	Without further analysing this pattern, we want to remark, that this result is not too surprising because the initial condition itself mirrors the geometry of the computational domain and leads to a multiple reflection pattern. 

	The imaginary part of the density matrix in this case is small all over the temporal evolution and obeys the same symmetry axes. 
	Even though it is not completely understandable nor analytically describable how interference patterns build up for this case, (otherwise one would need to calculate an infinite number of coefficients, to describe this initial condition in the eigenbasis of the double square well,) the numerical evaluation is highly reliable. 
	This can also be seen, regarding $N ( t )$ in terms of \cref{eq:norm} for different numbers of cells, which is depicted in \cref{fig:norm_box} where again the deviation is less than $0.1$ percent for the given amount of cells and therefore the simulation turns out to be highly accurate. 
	Again, $N ( t )$ is extracted from the \gls{pde} solution at discrete times (the dots), the lines are introduced to guide the eye.
	One can see, that a simulation with this initial condition has highly oscillating behaviour in $N(t)$.
	This can be easily explained with the high amount of discontinuities, which emerge from the step function and additionally scatter off the boundaries of the computational domain. 
	However, an ordinary Crank-Nicolson solver is not expected to handle this task at all, \reffs \cite{10.1093/comjnl/9.1.110,10.1093/comjnl/7.2.163,8b370aba-ebed-340f-8ce2-87c0149f028b}.

\section{Nontrivial Test Cases: full Lindblad dynamics of a particle in a square well potential and the harmonic oscillator}
\label{sec:non-trivial}

	In this section, we discuss the full Lindblad dynamical evolution of the density matrix calculated numerically with the \gls{kt} scheme.
	Still, these calculations serve as a  testing ground for the numerical scheme, which in the case of the harmonic oscillator can also be compared to analytic results. 
	We also want to discuss the physical interpretation and comment on thermalization, because from a physics perspective, for all systems the stationary (long time limit) state has to be thermal.
		\begin{figure}
			\begin{center}
				\includegraphics[width=1.0\columnwidth,clip=true]{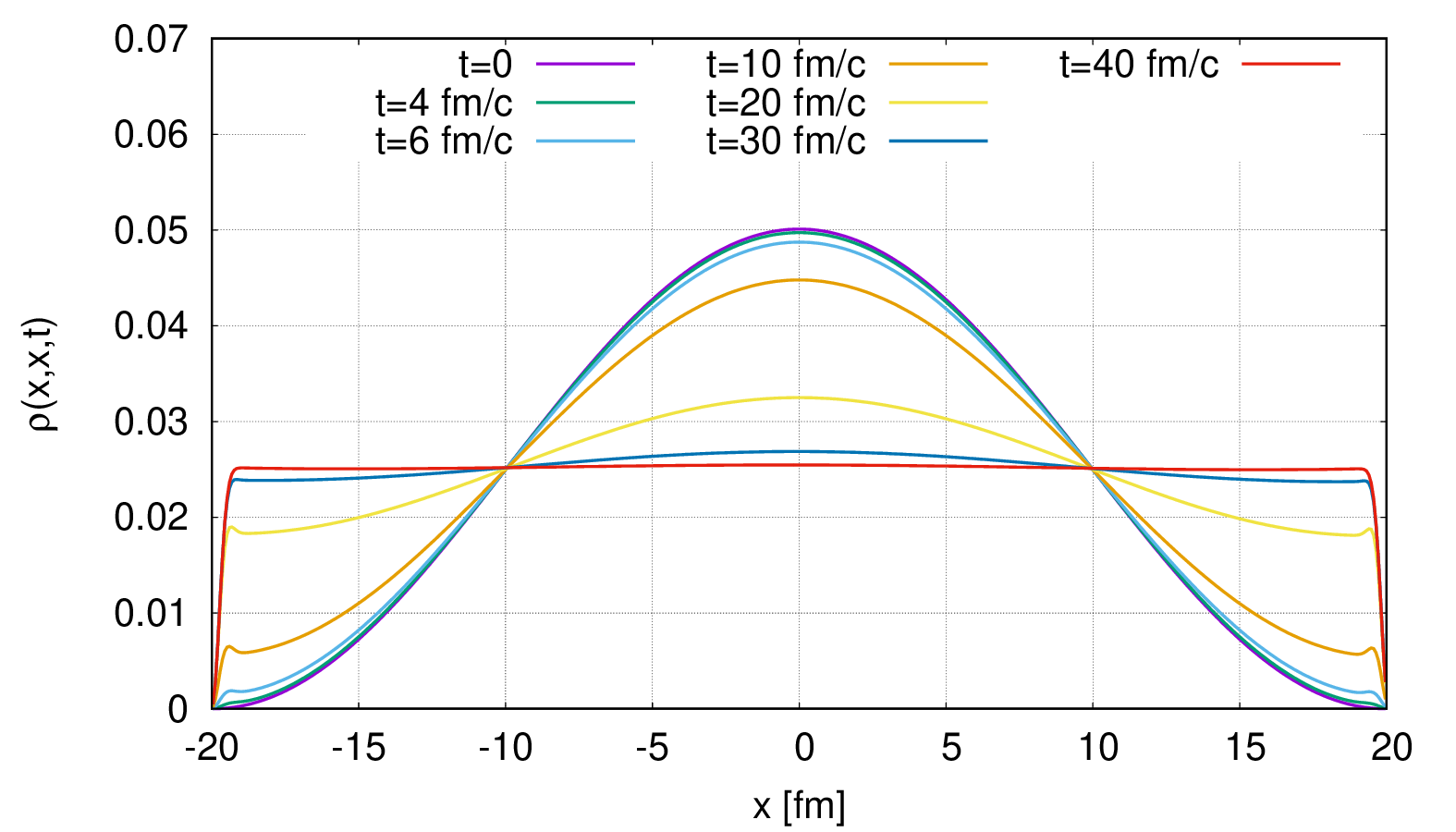}
				\caption{Temporal evolution of $\rho(x,x,t)$ for different times $t$ for the free particle following Lindblad dynamics.
					Parameters are $T = 300$ MeV for the temperature, $\gamma = 0.5$ c/fm for the damping, and $\Omega = 4T$ for the cutoff frequency.
					The initially populated state is $n = 1$ and the number of computational cells is $500 \times 500$.}
				\label{fig:rho_xx_init_0}
			\end{center}
		\end{figure}
		\begin{figure}
			\begin{center}
				\includegraphics[width=1.0\columnwidth,clip=true]{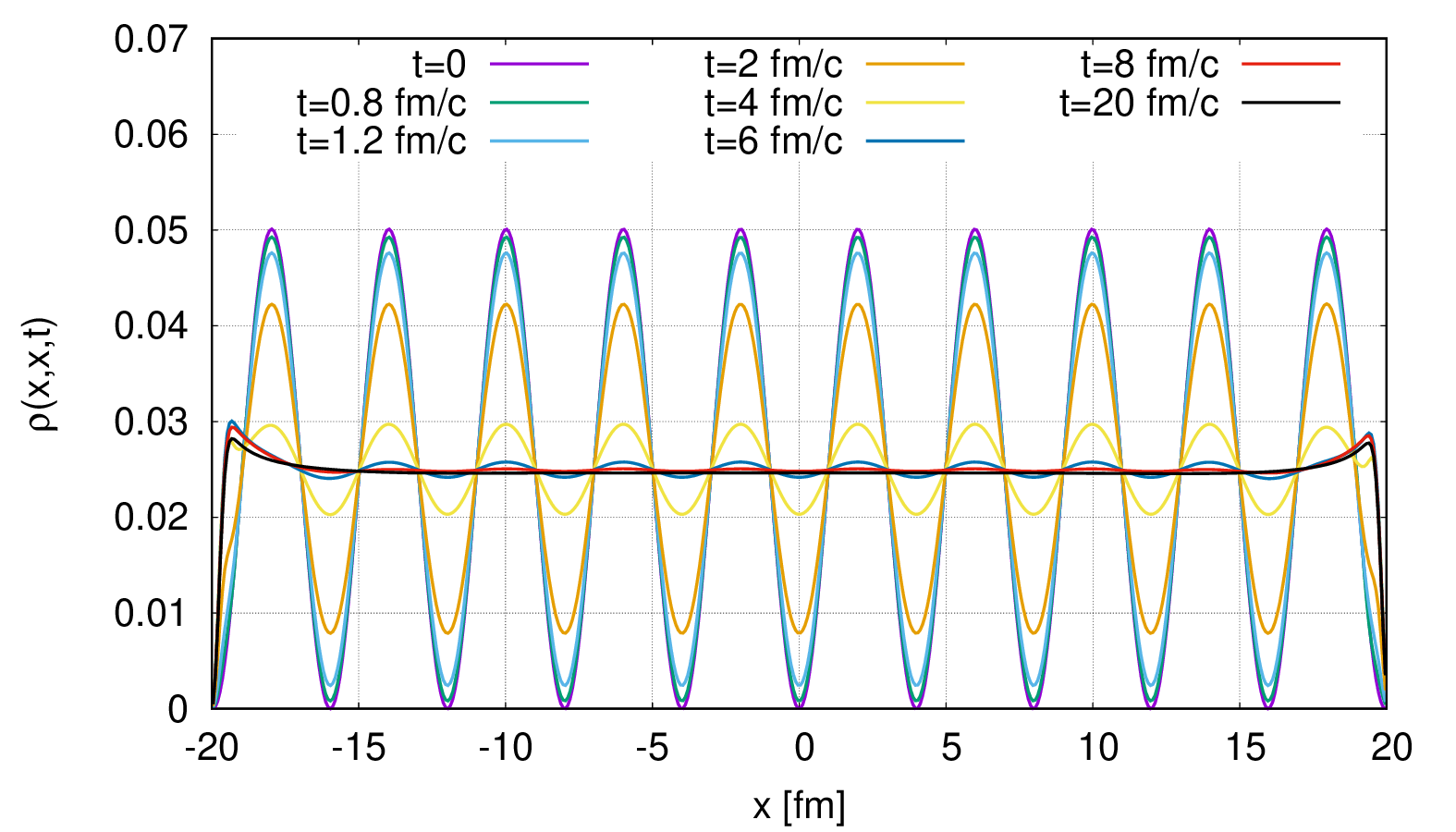}
				\caption{Temporal evolution of $\rho(x,x,t)$ for different times $t$ for the free particle following Lindblad dynamics.
					Computations are performed for the same choices of parameters as in \cref{fig:rho_xx_init_0}, while $n = 10$.}
				\label{fig:rho_xx_init_10}
			\end{center}
		\end{figure}
		\begin{figure}
			\begin{center}
				\includegraphics[width=1.0\columnwidth,clip=true]{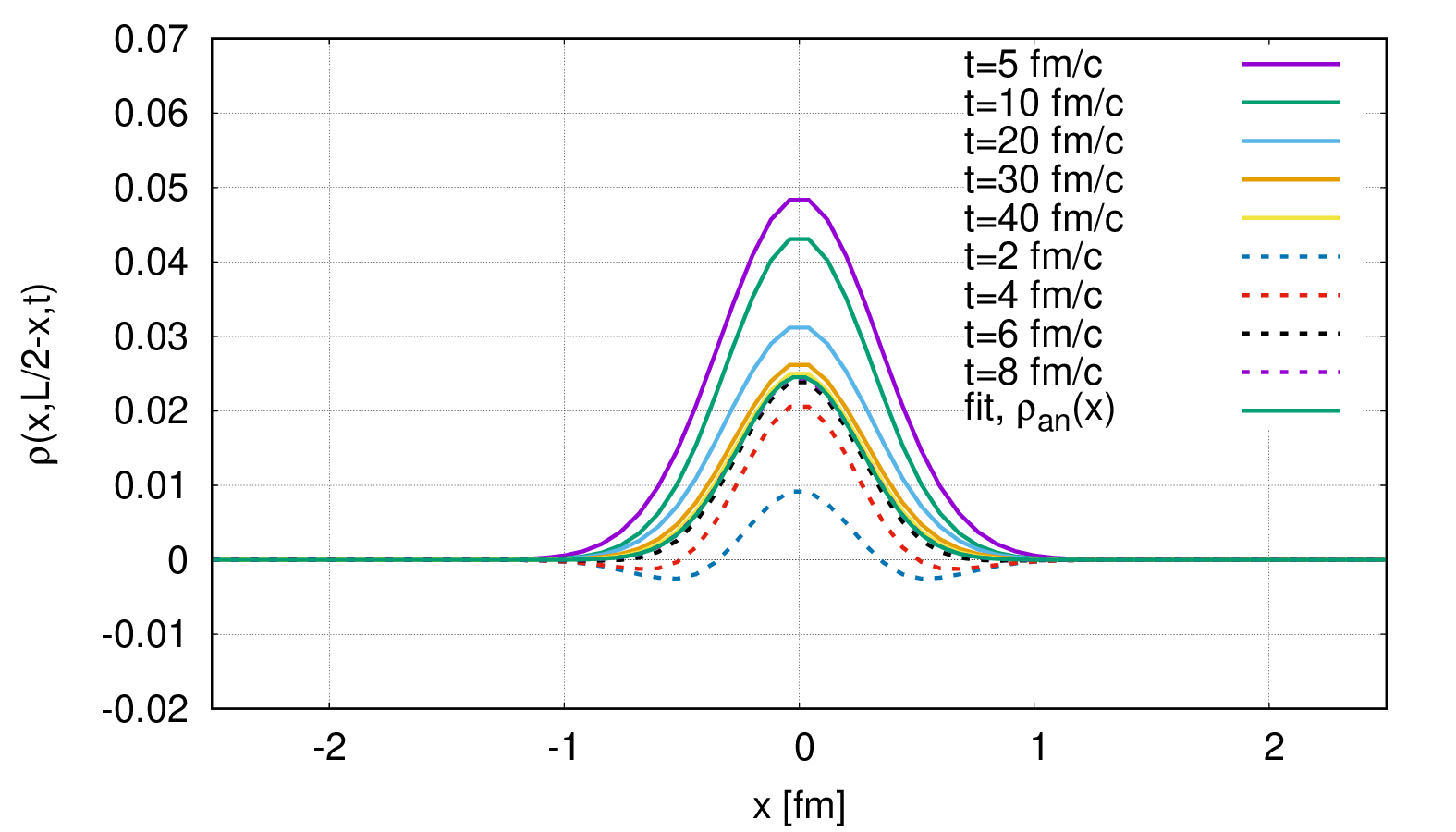}
				\caption{Temporal evolution of $\rho(x,L/2-x,t)$ for different times $t$ for the free particle with initial condition $n = 1$ (solid line) and $n = 10$ (dashed line) towards equilibration following Lindblad dynamics.
				Computations are performed for the same choices of parameters as in \cref{fig:rho_xx_init_0,fig:rho_xx_init_10}.}
				\label{fig:rho_xx_cross}
			\end{center}
		\end{figure}
		\begin{figure*}
			\begin{center}
				\includegraphics[width=\linewidth]{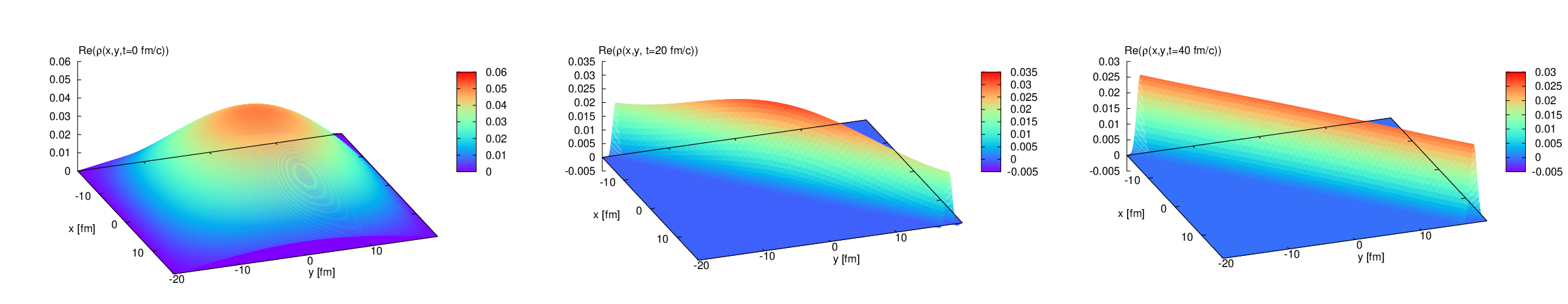}
				\includegraphics[width=\linewidth]{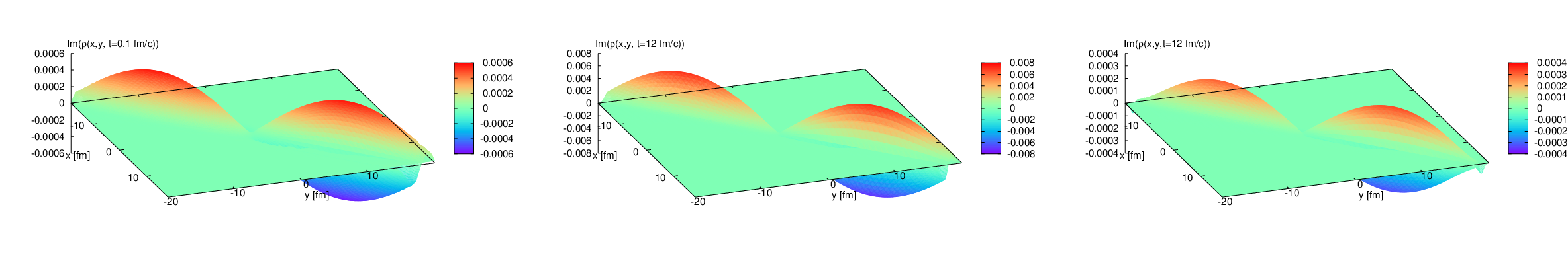}
				\caption{ Temporal evolution of $\rho(x,y,t)$ for different times $t$ for the free particle in a square well potential following Lindblad dynamics.
					Parameters are chosen as in \cref{fig:rho_xx_init_0,fig:rho_xx_init_10} for $n = 1$.
					The first two rows illustrate the real part, the second two rows illustrate the imaginary part. The number of cells is $500 \times 500$.}
				\label{fig:rho3d_free_init0}
			\end{center}
		\end{figure*}
		\begin{figure*}
			\begin{center}
				\includegraphics[width=\linewidth]{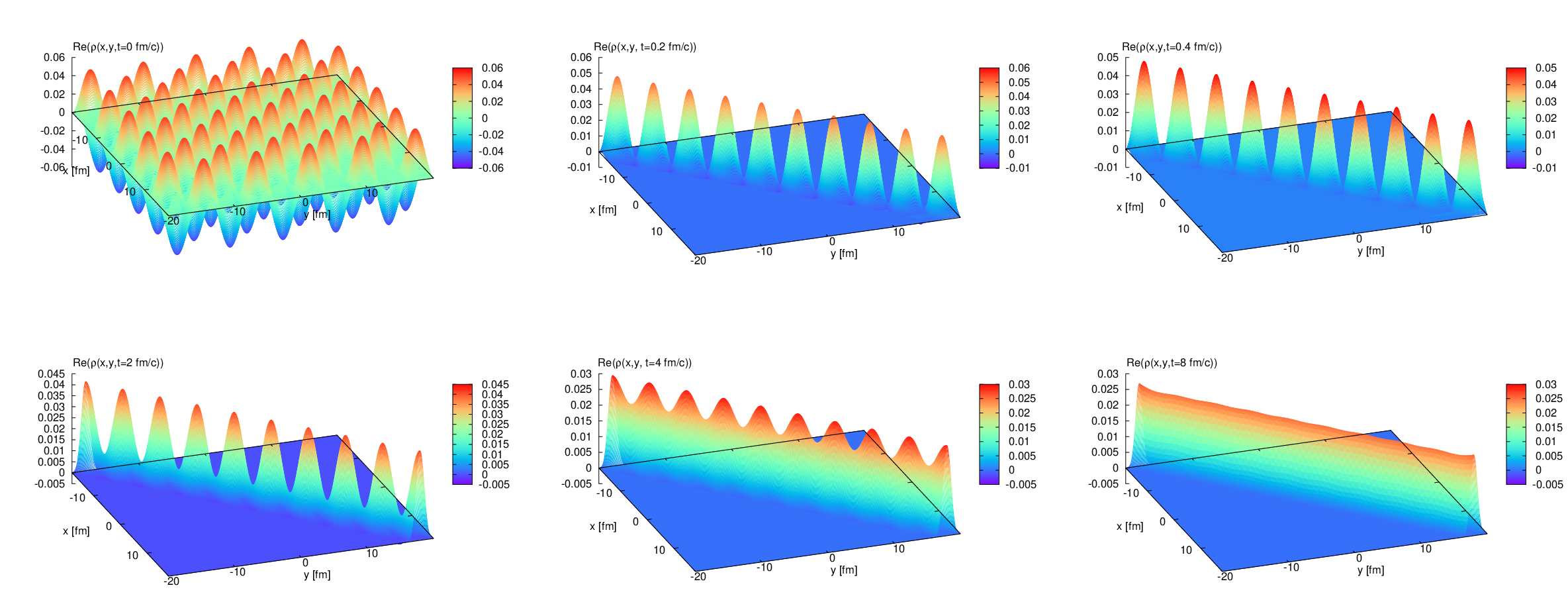}
				\includegraphics[width=\linewidth]{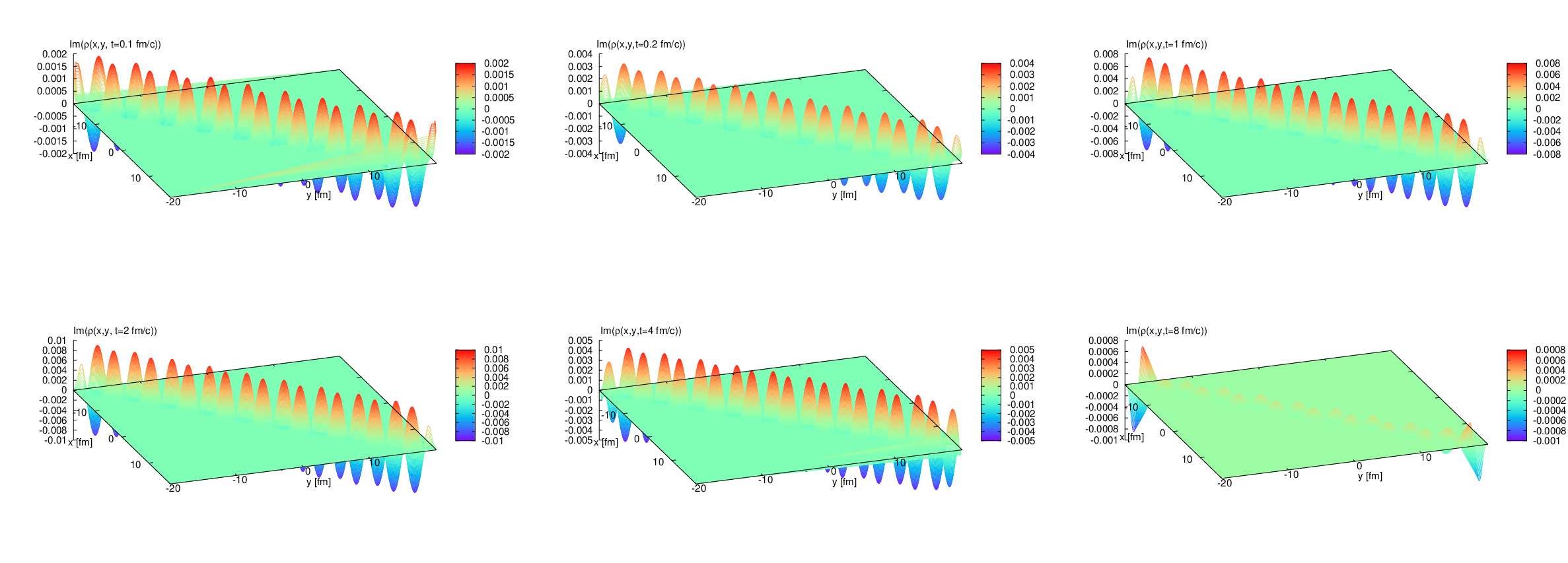}
				\caption{ Temporal evolution of $\rho(x,y,t)$ for different times $t$ for the free particle in a square well potential following Lindblad dynamics.
					Parameters are chosen as in \cref{fig:rho_xx_init_0,fig:rho3d_free_init0,fig:rho_xx_init_10} for $n = 10$.
					The first two rows illustrate the real part, the last two rows illustrate the imaginary part. The number of cells is $500 \times 500$.
				}
				\label{fig:rho3d_free}
			\end{center}
		\end{figure*}
	We consider two cases: the free particle in a square-well potential and the harmonic oscillator.
	Firstly, let us comment on the choice of these two cases.
	The infinite square well potential is numerically interesting because of the boundary effects that inevitably emerge during the temporal evolution. 
	Therefore, this is the case, where boundary effects can be studied best.
	On the other hand, the harmonic oscillator is the case, where boundary effects do not have to be taken into account, if the harmonic potential is chosen to be sufficiently large, see \cref{sec:boundary_conditions}.
	This allows to study the validity of the numerical scheme without boundary effects and therefore also without interference patterns that emerge by reflection, as was shown in \cref{sec:minimal}.

	Combining both findings will allow to perform ``box simulations" of Lindblad dynamics for arbitrary setups determined by repulsive or attractive potentials with highly excited initial conditions, cf. \reff\cite{Rais:2022gfg}. 
	This problem, which is more phenomenology driven, is treated in an upcoming publication \cite{Rais2025}, where a special focus will be set on the question how thermalization is potentially achieved, or not, within a Lindbladian time evolution and how relaxation times and thermalization can be studied within the  density matrix formalism. 
	Therefore, a special interest is given to the formation of bound states, which is also motivated in \reff\cite{Rais:2022gfg}. 
	
	Secondly, we will focus the discussion on cases, where $D_{x x} = 0$ and only briefly discuss the case, where $D_{x x} \neq 0$, since we have shown in \cref{sec:normconservation}, that a term  $D_{x x} \neq 0$ does not conserve the norm of the density matrix, which is reasonable from the viewpoint of a diffusion equation, which is also discussed in \cref{sec:normconservation}.
	
	Another reason for choosing $D_{x x} = 0$ is that the result, which we are using, describing the thermal state of the harmonic oscillator is derived analytically only for this case. 
	Furthermore, the choice of $D_{x x}$ itself is subtle and system-potential dependent \cite{Bernad2018}.

	Returning to our testing setup, there are basically two main ingredients which determine the validity of the simulation.
	First, there is the number of cells included, which is a purely numerical question (it will turn out, that full Lindblad dynamics needs more cells than Liouvillian dynamics without effects from dissipation, advection, and sources/sinks in order to satisfy the norm conservation condition).
	Second, there are the physically motivated parameters $\gamma$, which is called ``damping" coefficient, the temperature of the system $T$, and the cutoff frequency $\Omega$ of the environmental Ohmic bath.
	Those parameters underlie certain constraints and have to be chosen such, that these constraints are fulfilled. 
	We use the parameters $D_{p p}$, $D_{p x}$, and $D_{x x}$, cf.\ \cref{eq:lindblad_spatial}, found in \reffs\cite{BRE02,DEKKER198467,DIOSI1993517,Homa2019} given by
		\begin{align}\label{eq:d-params}
			&	D_{p p} = 2 \gamma m T \, ,
			&&	D_{x x} =
			\begin{cases}
				0
				\\
				\tfrac{\gamma}{6mT} 
			\end{cases} \, ,
			&&	D_{p x} = - \tfrac{\gamma T}{\Omega} \, ,
		\end{align}
	which, if $D_{xx} \neq 0$, have to fulfil the Dekker inequality \cite{DEKKER198467},
		\begin{align}\label{eq:condition_2}
			D_{p p} \, D_{x x} - D_{p x}^2 \geq \tfrac{\gamma^2}{4} \, .
		\end{align}
	$D_{p p}$ was found by Caldeira and Leggett in \reffs\cite{Caldeira:1982iu,BRE02}. 
	In the original Caldeira-Leggett model is $D_{p x} =0$. 
	For Lindblad form $D_{p x}$ can be chosen, as we have done it, in the high temperature limit \cite{BRE02}; for medium temperatures $D_{p x } = \Omega \gamma/(6\uppi T)$ \cite{DIOSI1993517}. Other definitions for different regimes are found in \reff\cite{DEKKER19811}.
	
	Physically we expect, that the system equilibrates into the stationary state,
	\begin{align*}
		\rho_{\mathrm{eq}} = \tfrac{1}{Z} \, \ee^{- \beta H_S}  \, .
	\end{align*}
	For a classical non-interacting gas, this density distribution is described by a fully, equally populated, purely diagonal density matrix.
	Therefore, we expect, that the quantum density matrix of the free particle in a box will diagonalize, too.
	However, because of its quantum nature, it will not diagonalize sharply, but gets smeared along the diagonal axis.
	This can be also understood in terms of decoherence, \cite{BRE02,Unruh:1989dd}. 
	The  orthogonal axis to the diagonal $\rho(x,x,t)$, $\rho(x, -x, t)$ allows to study thermalization in the strict sense of statistical thermodynamics, cf.\ \reff\cite{Rais2025}.
	The final distribution of the free, one-dimensional particle in a square-well potential is derived in \cref{app:distribution}, given by
		\begin{align}\label{eq:cross_rho}
			\rho_{\mathrm{free}} ( x, y,  t \rightarrow \infty ) = \tfrac{1}{L} \ee^{- \frac{m T}{2} ( x - y )^2}
		\end{align}
	and acts as a benchmark test, to investigate, whether the right ``thermal distribution" is reached in the stationary case.
	This has to be the case because a cooling down/or heating of the bath is excluded by the infinite number of bath degrees of freedom.

\subsection{Particle in a square well potential}\label{sec:full_lindblad}

	We turn to the first non-trivial example and solve \cref{eq:lindblad_spatial} for a particle in a square well potential. 
	
	At this point and in this subsection, to keep the focus on the numerical evaluation.
	Furthermore, we will comment on thermalization and decoherence during the discussion.
	 
	Let us first have a look on the diagonal of $\rho(x,y,t)$, namely $\rho(x,x,t)$. 
	This is shown in \cref{fig:rho_xx_init_0,fig:rho_xx_init_10} for different initial conditions \labelcref{eq:free_part_wave}, where $n = 1$ and $n = 10$. 
	Indeed, the diagonal is approximately equally populated, except of the ``forbidden domain" boundary, where the wave function has to vanish for both initials and boundary effects due to the finite computational domain.
	Another important result is, that both initial conditions, $n = 1$ and $n = 10$ evolve to the same stationary state (up to boundary effects).
	Also the axis orthogonal to the diagonal,  $\rho(x, - x, t)$, depicted for both initial conditions in \cref{fig:rho_xx_cross} tends towards the same distribution. 
	Here, the solid lines depict $n = 1$ and clearly show the diagonalization during the time evolution.
	The dashed lines represent the case, where $n = 10$, and show how the diagonal gets populated at the before unpopulated axis, due to the even initial condition.
	Using \cref{eq:cross_rho} as a fit function, setting $y=-x$, one can extract a temperature $T_{\text{fit}} = 302.4$ MeV, which nicely agrees with the bath temperature of $T = 300$ MeV.
	Also the factor $\frac{1}{L}$ in \cref{eq:cross_rho} is  obtained in the fit, where $L_{\text{fit}}=  40.65$.
	We conclude that the the system indeed thermalizes and deviations are most likely due to boundary effects, but can also emerge due to the before mentioned mode shifts.
		\begin{figure}[t]
			\begin{center}
				\includegraphics[width=1.0\columnwidth,clip=true]{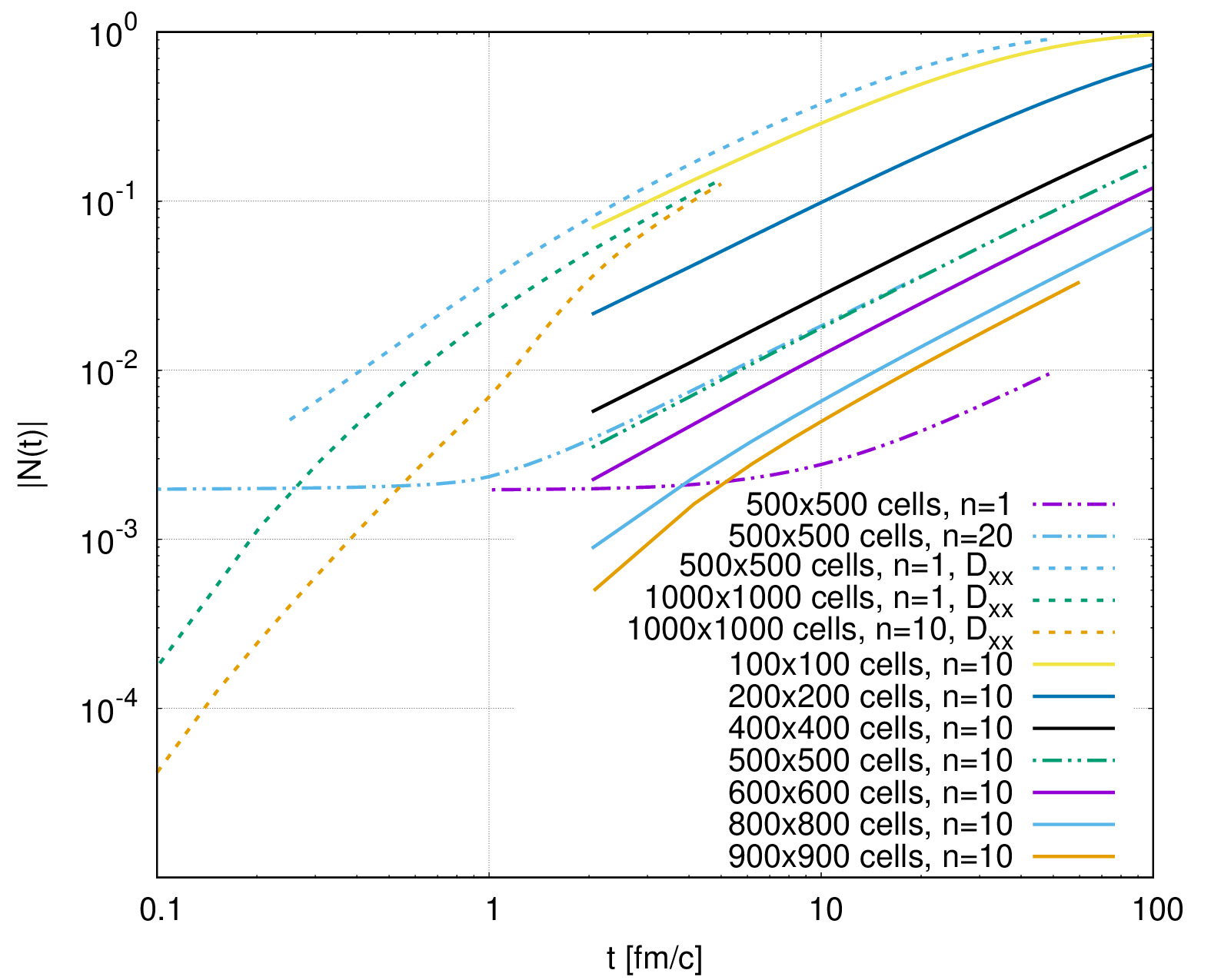}
				\caption{The norm of the density matrix for the parameter set given in \cref{fig:rho_xx_init_10,fig:rho_xx_init_0,fig:rho3d_free_init0,fig:rho3d_free} for different excited initial states $n$ for the initial condition $500 \times 500$ and for the initial condition $n=10$ and $n=20$ with different numbers of cells (dotted-dashed lines), $100 \times 100$ to $900 \times 900$ (solid lines). 
				The dotted lines illustrate the case where $D_{x x}$ is included to the simulation.
				The reason, that the curves start or stop at a certain values is because of the times we use to extract the data. 
				}
				\label{fig:norm_free}
			\end{center}
		\end{figure}
	
	Another important result is, that the system, for which a higher state is initially populated equilibrates much faster than the case, where the ground state is originally populated. 
	This can be seen in \cref{fig:rho_xx_init_0,fig:rho_xx_init_10}, where for $n = 1$ equilibration is achieved at times later than approximately $30$ fm, while for $n = 10$ the system equilibrates already roughly at $8$ fm/c.
	
	In the following, let us also have a look at the temporal evolution of the density matrix in the $x$-$y$-plane. 
	As already mentioned, the density matrix tends to diagonalize, which can be seen in \cref{fig:rho3d_free_init0,fig:rho3d_free}. 
	In these figures we show the density matrix again for an initial state with $n = 1$ and $n = 10$ to demonstrate, how the system gets stationary even for a more involved initial condition.
	To discuss the behaviour of the density matrix in detail, let us begin with the off-diagonal elements, and especially those far from the diagonal. 
	These get destructed/pushed towards the diagonal immediately, when the system is brought into contact with the thermal bath, which can be seen especially in \cref{fig:rho3d_free}, where the entries far from the diagonal are destructed already in between $0-0.2$ fm/c.
	From time $t\approx 1$ fm/c, only the oscillations on the diagonal parts remain unless they equally populate towards an equilibration.
	
	The imaginary part, depicted in the last row of \cref{fig:rho3d_free_init0} and the last two rows of \cref{fig:rho3d_free} is initially zero. 
	In the $n = 1$ case the original density matrix sharpens during the temporal evolution along the diagonal, to then vanish for later times (decoherence).
	Later on, the system obtains coherent dynamics and equilibrates to a purely real valued state \cite{gardiner00,BRE02}.
	For $n = 10$, the phase of decoherence takes place faster then in the case where $n = 1$.
	The imaginary contributions left and right from the diagonal, increase up to a certain maximal value, and afterwards start to decrease again. 
	What can be seen in the last two figures of the last row is a final state after decoherence, with still some very small imaginary parts, due to boundary effects and discretization artifacts.
	However, these are approximately two orders of magnitude smaller than the maximum amplitude of the imaginary part during the time evolution and at least one order of magnitude smaller than the real part.

	Finally, we want to discuss the norm of the density matrix, which is shown in \cref{fig:norm_free}.
	We again present $\vert N ( t )\vert $ from \cref{eq:norm} for different numbers of cells, from $100 \times 100$ to $900 \times 900$ cells, for times $t = 0$ to $t = 100$ which shows the impact of the number of cells on the validity of the result.
	First, we illustrate three different initial conditions $n = 1, 10, 20$ for the same number of cells, $500\times 500$, to discuss the impact of the initial condition, and then we pick one initial condition, $n=10$, to discuss the impact of the number of cells included. 
	It turns out, that the initial condition where $n=1$ conserves the norm best, the $n=10$ and $n=20$ cases lie on each other but still obey only small deviations, especially for time $t<10$ fm/c.
	
	However, it can be seen, that, due to the before mentioned geometry of the computational domain, which is rotated by $45^\circ$, for lower amounts of cells, the norm decreases dramatically towards 0.
	This is due to the fact, that real parts of the density matrix are coupled the imaginary and vice versa:
	The cells on the diagonal are simply too large to have exactly vanishing imaginary part and always pick up some imaginary contributions from off the diagonal.
	For $100\times 100$ cells the resolution to pick up the initial condition ($n=10$) is not high enough.
	Towards higher amounts of cells, at approximately $500\times 500$, which is also the amount of cells we use for the above calculations, the norm conservation is satisfying compared the time scale of the equilibration to the time scale, where the norm deviation starts to be crucial.
	For the case, where $D_{x x} \neq 0$ the norm is violated even more. 
	Here, there are two overlapping  effects: the norm violation due to the geometry of the computational domain and the norm violation due the the finite size of the box, which was derived in \cref{sec:normconservation}.
	Hence, effects of norm violation are visible in our setup, there is a rather controlled way to handle them by increasing the amount of cells, without performing artificial rescaling etc.

\subsection{Harmonic oscillator}
\label{sec:harmonic_osci}

		\begin{figure*}
			\begin{center}
				\includegraphics[width=\linewidth]{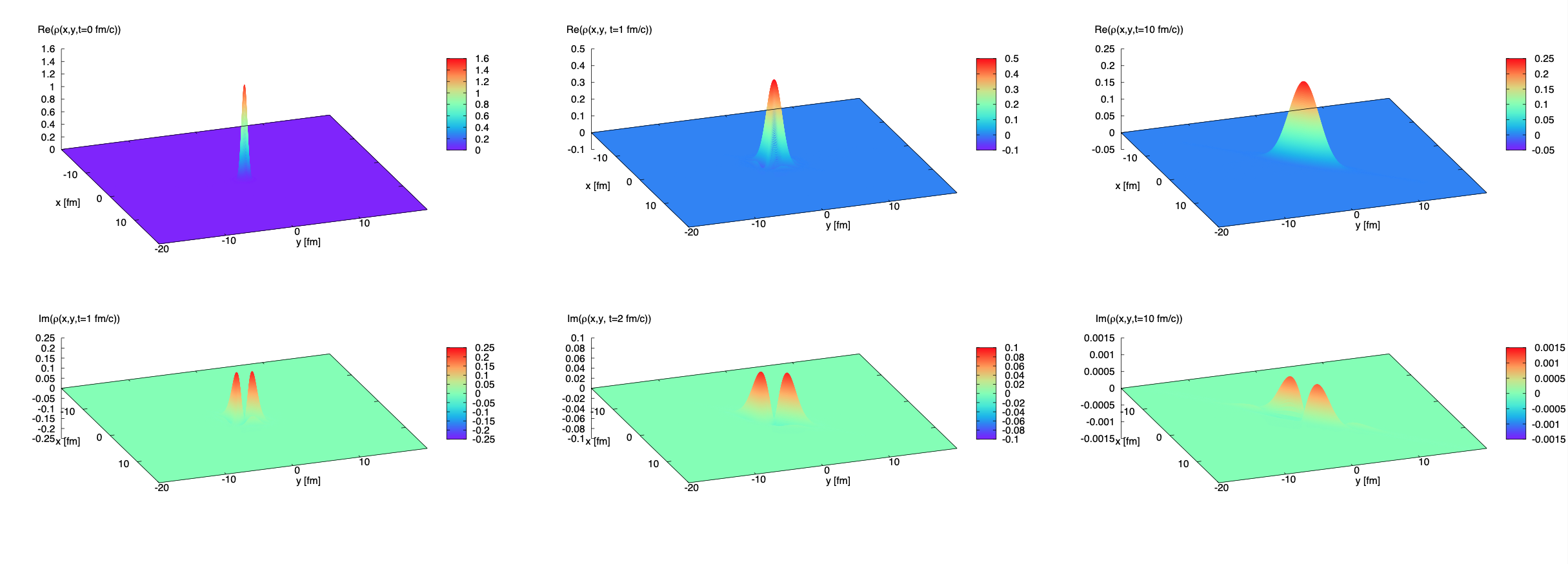}
				\caption{ Temporal evolution of $\rho ( x, y, t )$ at different times $t$ for the particle in a harmonic potential following Lindblad dynamics.
					The temperature is chosen to be $T = 300$ MeV, the damping $\gamma = 0.5$ c/fm, $\omega = 0.5$ c/fm and the cutoff frequency is chosen to be $\Omega = 4 T$.
					The initially populated state is $n = 1$ and the numbers of computational cells is $500 \times 500$.
					The first row illustrates the real part, the last row illustrates the imaginary part.}
				\label{fig:rho3d_ho_init1}
			\end{center}
		\end{figure*}
		\begin{figure}
			\begin{center}
				\includegraphics[width=1.0\columnwidth,clip=true]{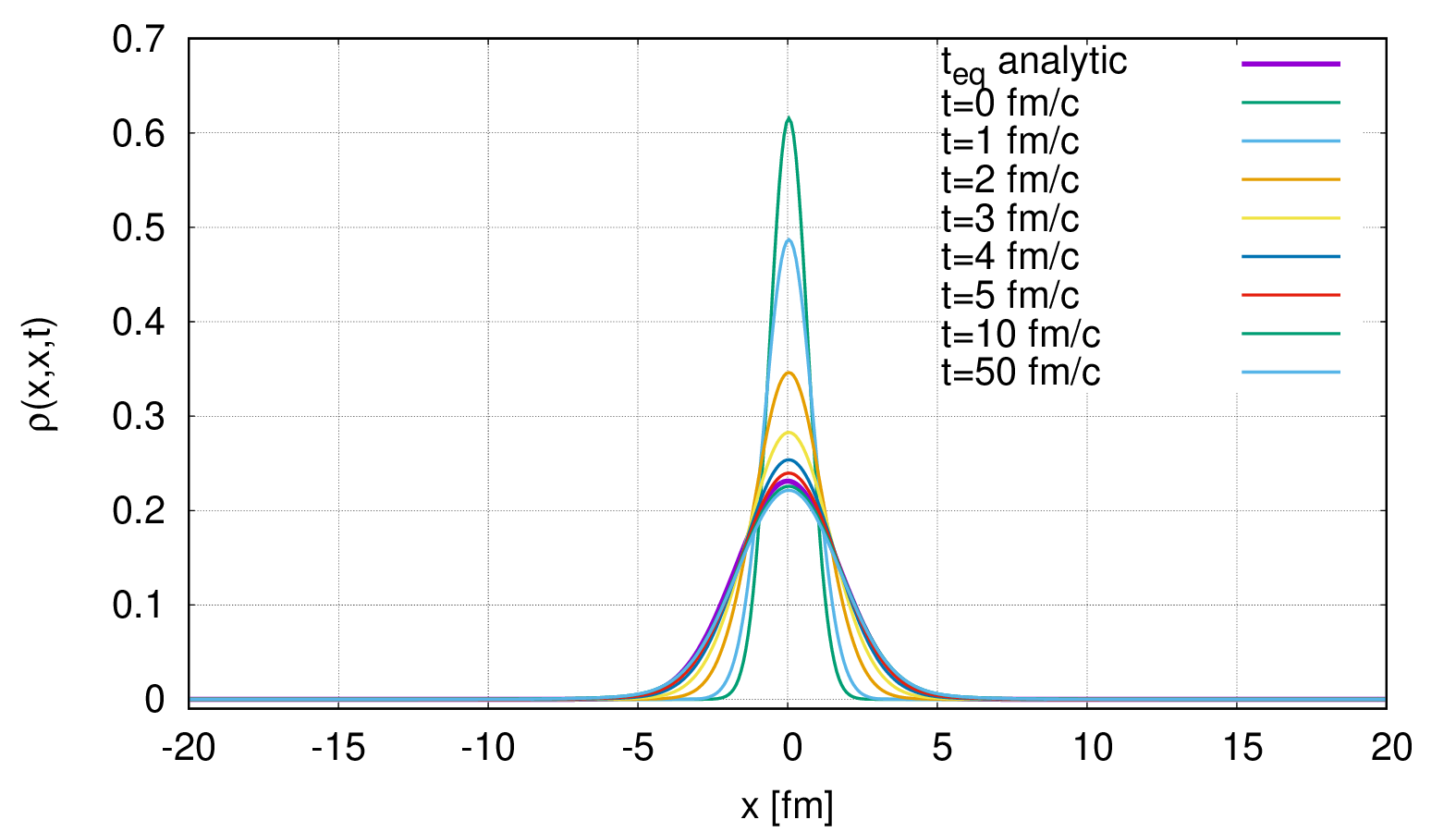}
				\caption{The diagonal of the density matrix $\rho ( x, x, t )$ in the harmonic oscillator potential at different times, until equilibrium is reached.
					Parameters are $n = 1$, $T = 300$ MeV, $\gamma = 0.5$ c/fm, $\omega = 0.5$ c/fm and the cutoff frequency $\Omega = 4 T$.
					We use $500 \times 500$ computational cells.
					The thick purple line corresponds to the analytical equilibrium solution.}
				\label{fig:rho_xx_ho_1}
			\end{center}
		\end{figure}
		\begin{figure}
			\begin{center}
				\includegraphics[width=1.0\columnwidth,clip=true]{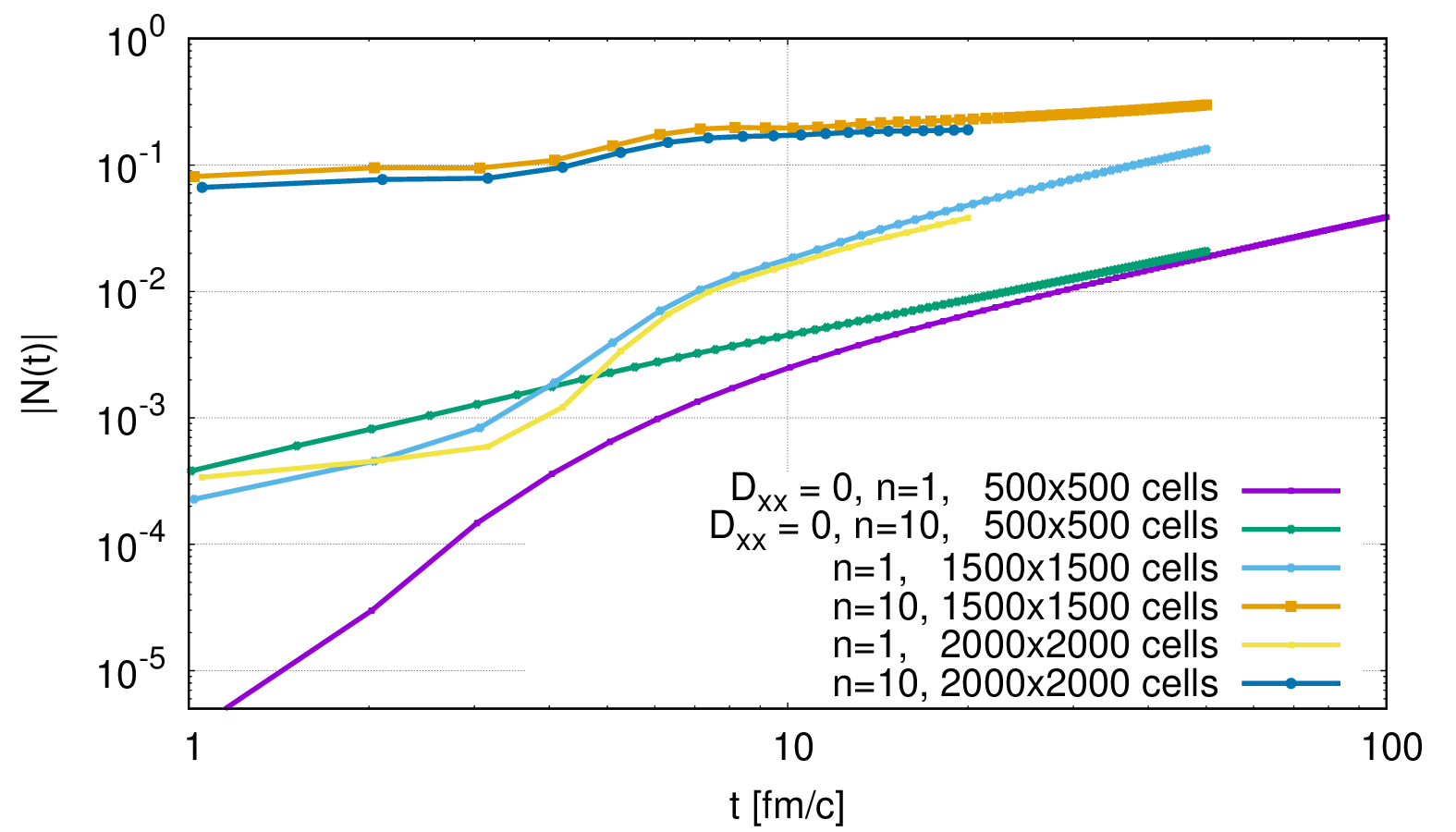}
				\caption{The time evolution of the norm $| N ( t ) |$ for parameters $T = 300$ MeV, $\gamma = 0.5$ c/fm, $\omega = 0.5$ c/fm and the cutoff frequency $\Omega = 4 T$ for different initial conditions, $n = 1$ and $n = 10$ for the case $D_{x x} = 0$ and $500 \times 500$ cells. For comparison, $N ( t )$  with $D_{x x} = \gamma/(6mT)$ are given for an amount of cells of  $1500 \times 1500$ and $2000 \times 2000$, with $\gamma=0.035$, to satisfy \cref{eq:condition_2}.  }
				\label{fig:norm_ho}
			\end{center}
		\end{figure} 
		\begin{figure*}
			\begin{center}
				\includegraphics[width=\linewidth]{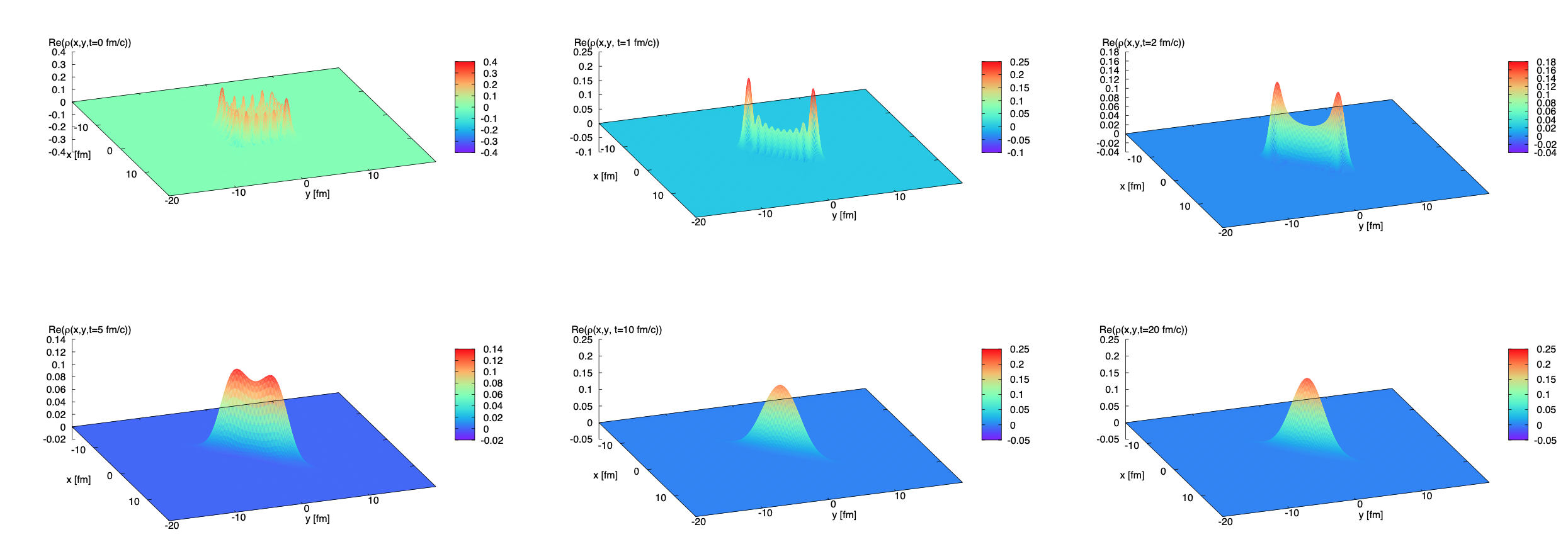}
				\includegraphics[width=\linewidth]{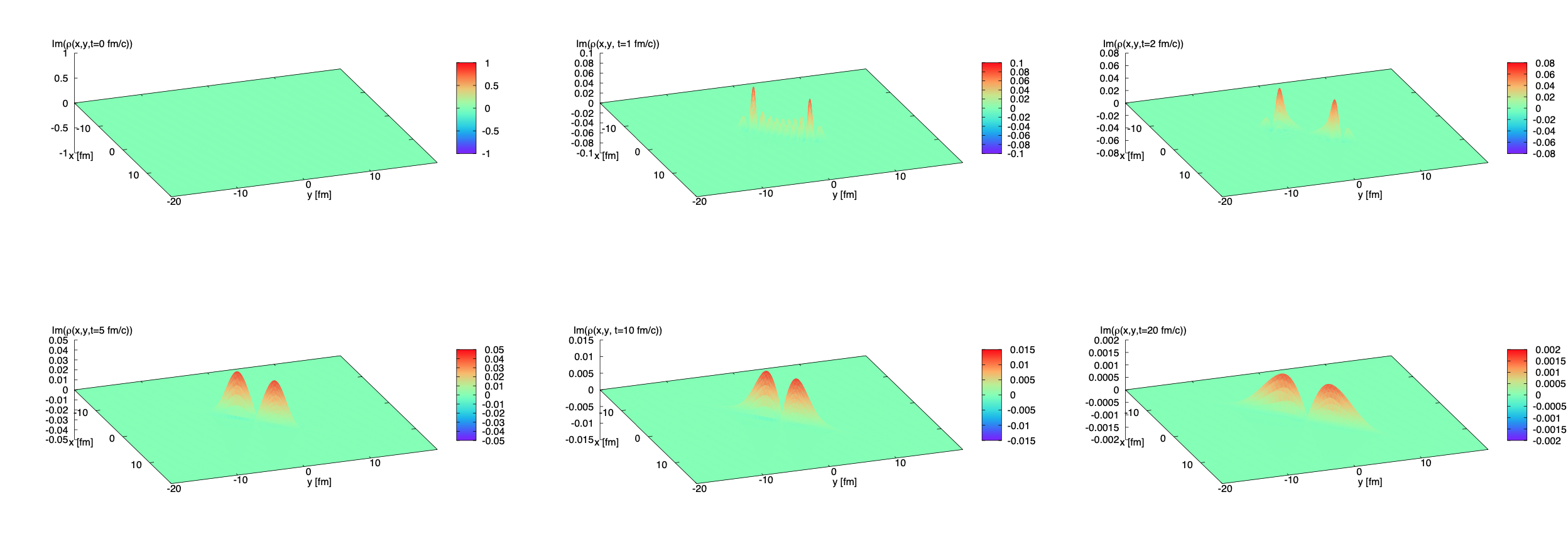}
				\caption{Temporal evolution of $\rho ( x, y, t )$ at different times $t$ for the particle in a harmonic potential following Lindblad dynamics.
				The parameters are the same as in \cref{fig:rho3d_ho_init1}, but for $n = 10$.
				The first two rows illustrate the real part, the last two rows illustrate the imaginary part.}
				\label{fig:rho3d_ho_init10}
			\end{center}
		\end{figure*}
	In this section, we turn to the Lindblad dynamics in an harmonic oscillator background potential.
	The reason for this is twofold:
	First, we know that the thermal state -- the $t \to \infty$ solution -- can be solved analytically, and therefore we are able to compare and benchmark the numerical results with this solution.
	Second, the harmonic potential confines the evolution of the density matrix to a finite domain without introducing a spatial box.
	Hence, keeping the computational domain sufficiently large, we can ignore the effect of boundary conditions, cf. \cref{sec:boundary_conditions}.

\begin{figure}
			\begin{center}
				\includegraphics[width=1.0\columnwidth,clip=true]{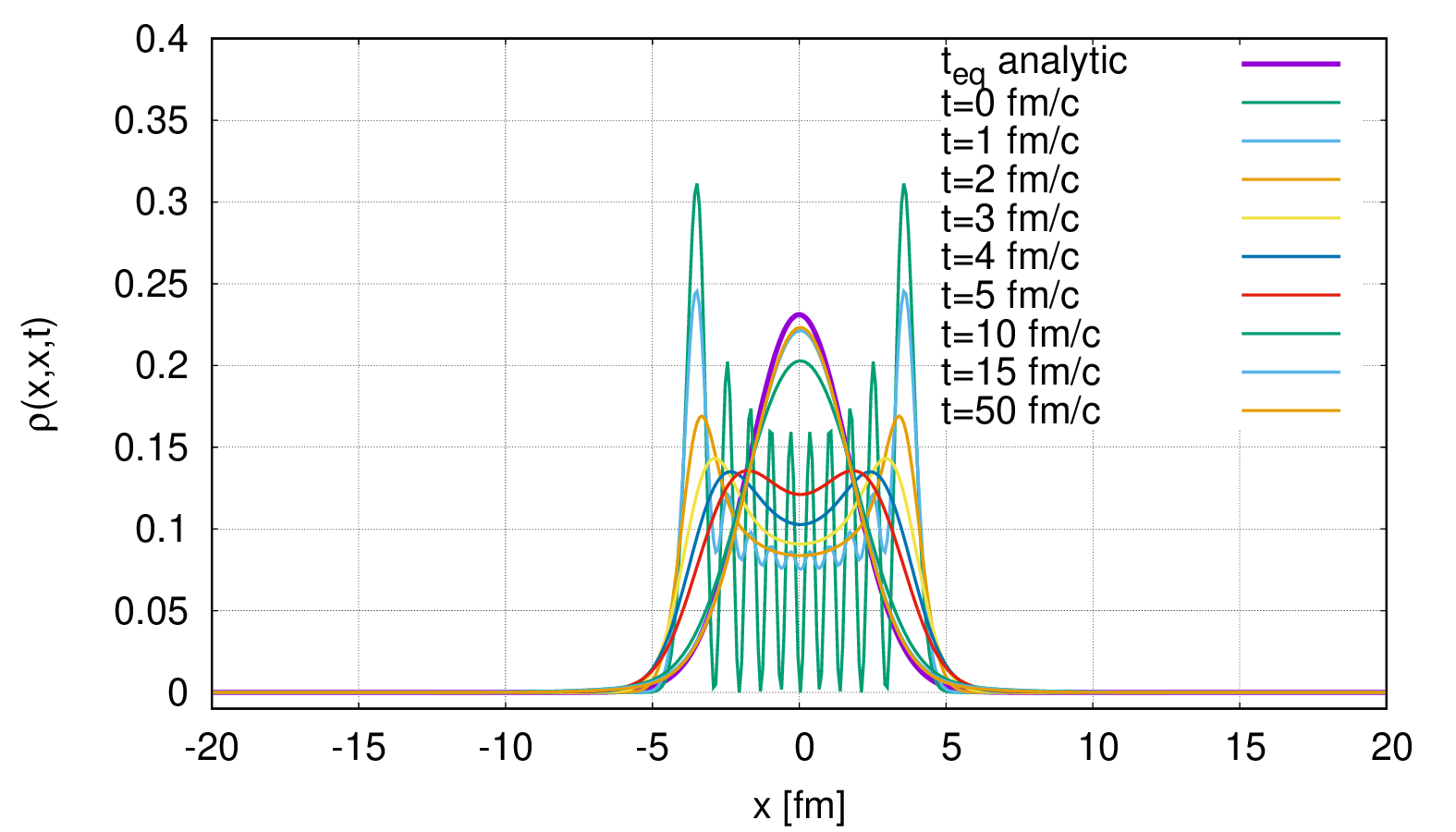}
				\caption{The diagonal of the density matrix $\rho ( x, x, t )$ in the harmonic oscillator potential at different times, until equilibrium is reached.
					Parameters are $n = 10$, $T = 300$ MeV, $\gamma = 0.5$ c/fm, $\omega = 0.5$ c/fm and the cutoff frequency $\Omega = 4 T$.
					We use $500 \times 500$ computational cells.
					The thick purple line corresponds to the analytical equilibrium solution.}
				\label{fig:rho_xx_ho_10}
			\end{center} 
		\end{figure}

	In \cref{fig:rho3d_ho_init1}, upper row, we initialized the simulation with a sharp (in comparison to the total size of the computational domain) Gaussian wave packet, as the ground state of the harmonic oscillator at $t = 0$, which is symmetrically distributed on the $x$-$y$-plane.
	The final (thermal) distribution is expected to be the same as in the previous case (up to the confining effects from the oscillator potential).
	Indeed, this is also seen in our simulations in \cref{fig:rho3d_ho_init10,fig:rho_xx_ho_10}.

	Therefore, let us briefly recapitulate the analytic results found in \reff\cite{Ramazanoglu:2009} and confirmed and generalized in \reffs\cite{Bernad2018,Homa2019}.
	Note, that these works provide a detailed interpretation of each of the diffusion coefficients to obtain the equilibrium solution for $t \to \infty$,
		\begin{align}\label{eq:density_equi}
			& \rho_{\text{analyt}} ( x, y, \infty ) =	\vdistance
			\\
			= \, & \tfrac{\sqrt{\gamma} m \omega}{\sqrt{\uppi ( D_{p p} - 4 \gamma m D_{p x} )}} \times \vdistance	\nonumber
			\\
			& \times \exp \Big( - \tfrac{\gamma [ m \omega ( x + y ) ]^2}{4 ( D_{p p} - 4 \gamma m D_{p x} )} - \tfrac{D_{p p} ( x - y )^2}{4 \gamma} \Big) \, .	\vdistance	\nonumber
		\end{align}
	Solving the eigenproblem of the steady state, 
		\begin{align}
			\int_{-\infty}^{\infty} \dd y \, \rho ( x, y, \infty ) \, \phi_n ( y ) = \epsilon_n \, \phi_n ( x ) \, ,
		\end{align}
	one obtains
		\begin{align}\label{eq:wave}
			\phi_n ( x ) = H_n \big( x, \tfrac{1}{4 \sqrt{A C}} \big) \, \exp \big( - 2 \sqrt{AC} \, x^2 \big) \, ,
		\end{align}
	with parameters
		\begin{align}
			&	A = \tfrac{D_{p p}}{4\gamma} \, ,
			&&	C = \tfrac{\gamma(m\omega)^2}{4(D_{p p} - 4 \gamma m D_{p x})}.
		\end{align}
	This means, that during the temporal evolution, the frequency $\omega$ is shifted from
		\begin{align}\label{eq:frequency}
			\omega \rightarrow \tilde{\omega} = \omega \, \sqrt{\tfrac{D_{p p}}{D_{p p} - 4 \gamma m D_{p x}}} \, , 
		\end{align}
	cf.\ \cref{sec:coord_space}, item 4.
	If $D_{p x} = 0$, \cref{eq:density_equi} agrees with the result from \reff\cite{BRE02}, and also in \cref{eq:frequency}, $\omega = \tilde{\omega}$. 

	In case of the harmonic oscillator
		\begin{align}\label{eq:condition_1}
			\tfrac{D_{p p}^2 - 4 \gamma m D_{p p} D_{p x}}{\gamma^2 m^2 \omega^2} \geq 1 \, ,
		\end{align}
	is another condition, which has to be satisfied, cf.\ \reff\cite{Homa2019}. 
	From \cref{eq:density_equi} one can immediately see, that the diagonal density matrix of the analytic result is given by
		\begin{align}\label{eq:rho_xx_equi}
			\rho_{\text{analyt}} ( x, x, \infty ) = \, & \tfrac{\sqrt{ m \omega^2} }{\sqrt{2\uppi T ( 1 + 2  \frac{\gamma}{\Omega} )}}  \exp \Big( - \tfrac{ m \omega^2 x ^2}{  2T(1 + 2\frac{\gamma}{\Omega} ) } \Big) \, .
		\end{align}
	Additionally, for the case $y=-x$, the orthogonal to the diagonal is given by
		\begin{align}\label{eq:rho_xy_equi}
			\rho_{\text{analyt}} ( x,- x, \infty ) = \, & \tfrac{\sqrt{ m \omega^2} }{\sqrt{2\uppi T ( 1 + 2  \frac{\gamma}{\Omega} )}} \exp (  - 2mT x^2 ) \, ,
		\end{align}
	which has the same width as expected from \cref{eq:cross_diag}. 

	Taking these two functions  \cref{eq:rho_xx_equi,eq:rho_xy_equi} as two fit functions and comparing them to the numerical results from the equilibrium solution given in \cref{fig:rho3d_ho_init1} will lead to a nearly perfect accordance, cf.\ \cref{fig:rho_xx_ho_1} and allows to extract a temperature using the orthogonal axis to the diagonal, taking $y=-x$ which agrees with the one of the heat  bath, namely $T_{\text{fit}} = 297,47$ MeV. 
	The same is valid for the case, where $n=10$, cf. \cref{fig:rho_xx_ho_10}, where the fit temperature is given by $T_{\text{fit}}=297,43$ MeV. 
	Therefore, we can state again, that the system indeed thermalizes in a physical sense.

	One other remark, we want to give at this point regarding thermalization, and why in \cref{fig:rho_xx_ho_1,fig:rho_xx_ho_10} indeed the thermalization can be seen already considering  the barometric formula of an ideal gas.
	The relation
		\begin{align}
			\braket{x_{\text{cl}}^2} = \tfrac{T}{m\omega^2} 
		\end{align}
	leads for the given parameters to $\sqrt{\braket{x_{\text{cl}}^2}} =  \frac{1}{2} \sqrt{T/m}$, which indeed approximately matches the standard deviation of \cref{eq:density_equi}, which is given by 
		\begin{align}
			\tfrac{1}{2 \braket{x_{\text{analyt}}}^2} = \tfrac{\gamma m^2 \omega^2}{D_{p p} - 4 \gamma m D_{p x}} 
		\end{align}
	and therefore
		\begin{align}
			\braket{x_{\text{analyt}}} =  \tfrac{1}{2} \sqrt{\tfrac{T}{m} \big(1 + \tfrac{2\gamma}{\Omega}\big)}\, ,
		\end{align}
	where the $\Omega-$dependency indicates a cutoff-effect.

	Having recapitulated the constraints for the Lindbald coefficients, we have to think of a suitable parameter set. 
	A frequency $\omega = 0.5$ c/fm being employed ensures that the potential is strong enough to centre the density matrix for all times at the given computational domain. 
	The temperature $T$ has to be high in comparison to energies of the systems itself.
	Therefore, we chose $T = 300$ MeV. 
	If we take the well known energy spectrum for the harmonic oscillator, $E_n = \omega ( n + \frac{1}{2} )$, this means, that we reach the bath temperature for an energy corresponding to $n \approx 600$, which is an astronomically large population. 
	Therefore, indeed, the temperature $T$ is comparatively high. For $n=10$, the energy is approximately $5.5$ MeV, which is very low in comparison to the bath temperature and definitely fulfils the constraints mentioned above.
	To satisfy the condition \labelcref{eq:condition_2}, $\Omega$ has to be at least $\sqrt{12T^2}$. 
	Therefore, we chose $\Omega = 4T$.
		\begin{figure}
			\begin{center}
				\includegraphics[width=1.0\columnwidth,clip=true]{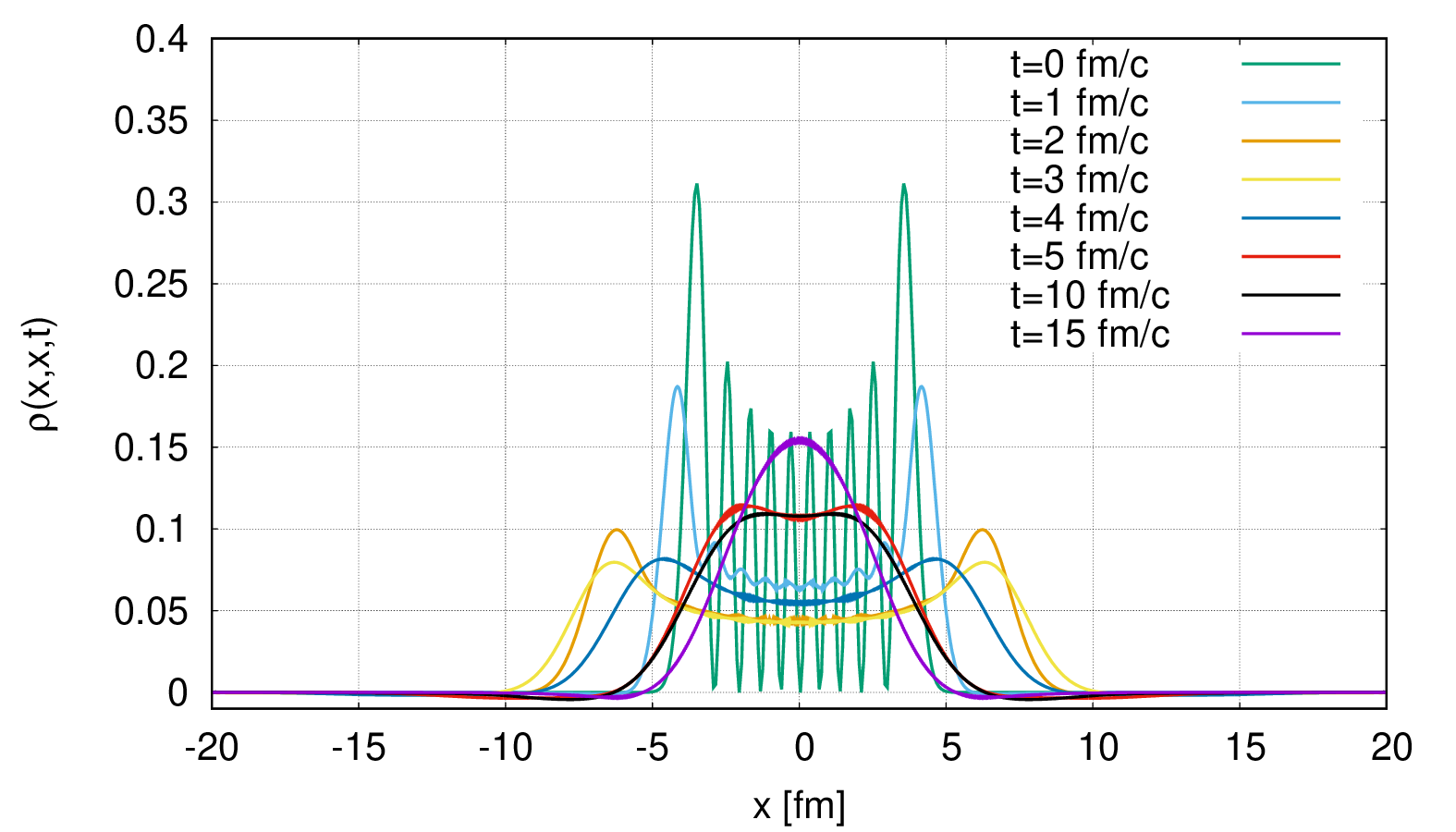}
				\caption{The diagonal of the density matrix $\rho ( x, x, t )$ in the harmonic oscillator potential at different times, until equilibrium is reached.
					Parameters are $n = 10$, $T = 300$ MeV, $\gamma = 0.5$ c/fm, $\omega = 0.5$ c/fm and the cutoff frequency $\Omega = 4 T$.
					We use $500 \times 500$ computational cells.
					Here the calculation includes the terms proportional to $D_{xx} = \frac{\gamma}{6mT}$.}
				\label{fig:rho_xx_ho_10_with_Dxx}
			\end{center} 
		\end{figure}
	Having discussed the necessary ingredients, we can turn to the results. 
	Remember that $D_{x x} = 0$, to allow a comparison with the analytical result.
	During the evolution, the real part of the density matrix broadens, cf.\ \cref{fig:rho_xx_ho_1}, conserving the norm, cf. \cref{fig:norm_ho} up to a maximal deviation of less then $1\%$ for a given amount of cells of $500^2$.
	This is of cause a small deviation of the norm, and therefore satisfying. 
	If one adds terms proportional to $D_{x x} = \gamma/6mT$, the norm can be handled by increasing the amount of cells up to $2000^2$, where the deviation of the norm, considering a equilibration time of around $5$ fm/c, is still in a satisfying regime for both initial conditions, $n = 1$ and $n = 10$, cf. \cref{fig:norm_ho}.
	In \cref{fig:rho_xx_ho_1} one can see, that the system equilibrates after approximately $t > 5$ fm/c.\footnote{The estimated equilibration time $\tau_R$ is $\tau_R = \frac{1}{\gamma} = 2$ fm/c, which means, that the actual equilibration time is of the same order but still higher \cite{Gao:1997}.}
	
	In the lower row of \cref{fig:rho3d_ho_init1} we show the imaginary part of the temporal evolution. 
	Note that the first panel shows the imaginary part at $t = 1$ fm/c, because the imaginary part is zero at $t = 0$.
	After a short time the imaginary part builds up to shrink down almost completely at $t = 10$ fm/c, where the system is expected to be thermalized.
	(Note the scales on the axis.)
	Again, the reason, why there is still some non-vanishing imaginary part is due to the resolution of the numerical scheme about the diagonal of the density matrix.
	Here, however, we can exclude boundary effects, because the density matrix does not spread all over the computational domain, but stays centred as a consequence of the confining harmonic potential.
 
	Next, let us turn to the scenario, where we start with an excited state with $n = 10$.
	The final ``(thermal)" distribution is expected to be the same as in the previous case.
	Indeed, this is also seen in our simulation in \cref{fig:rho3d_ho_init10} and \cref{fig:rho_xx_ho_10}.
	Especially \cref{fig:rho_xx_ho_10} helps to understand the non-trivial dynamics.
	The oscillations of the initial condition lead to a more and more smooth distribution while the overall width decreases towards equilibrium.
	After approximately $15$ fm/c the system equilibrates to the distribution given in \cref{eq:density_equi}.
	
	At this point, it is instructive to see, how the system behaves, if $D_{x x} = \frac{\gamma}{6mT}$, instead of $D_{x x} = 0$. 
	As we have already shown in \cref{eq:diffusion}, the $D_{xx}$-term leads to diffusion along the diagonal axis, and therefore spreads the diagonal distribution. 
	This can be seen in \cref{fig:rho_xx_ho_10_with_Dxx}, where the same parameters were used as in \cref{fig:rho_xx_ho_10}, but including the $D_{xx}$ term. 
	This clearly shows the large impact of a non-vanishing $D_{xx}$ term, which causes a spatial diffusion and also tends to a different equilibrium, which is not described by \cref{eq:density_equi}.
		
	The imaginary part, lower two rows of \cref{fig:rho3d_ho_init10}, shows similar behaviour as the real part in the beginning.
	It develops two maxima from the left and the right of the diagonal, which melt down to a negligibly small value as time passes.
	Regarding \cref{fig:norm_ho}, one can see, that the norm for times $t\geq 15$ fm/c in the case where $D_{x x} = 0$ is violated only less than $1$ \%.
	As we have already discussed, the case where $D_{x x}$ is finite leads to a significant violation of the norm, which is larger by one order of magnitude at this given time in comparison to the case where $D_{x x} = 0$.  
	
	To  take control of this problem in the case, where $D_{x x} = 0$, it is necessary to introduce even more cells, since the error of $N ( t )$ decreases, cf. \cref{fig:norm_ho}, for higher numbers of cells. 
	One has to see at this point, that the initial condition covers only around 1\% of the computational domain being highly oscillating in this area and therefore also $2000^2$ cells are a relatively low number of cells, especially in comparison to the initial condition of a free particle, where the relevant parts of the density matrix covered the whole computational domain.

\begin{figure*}
			\begin{center}
				\includegraphics[width=\linewidth]{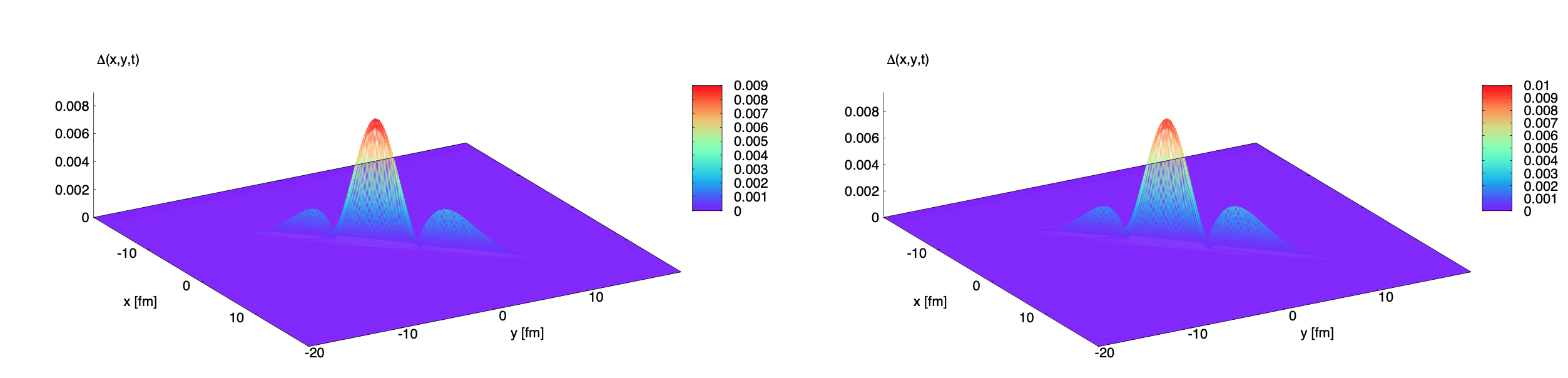}
				\caption{The deviation of the numerical from the analytical $( t \to \infty )$-solution measured in terms of $\vert N(x,y,t)\vert$, \cref{eq:nxy}, for the initial state with $n = 1$ (left panel) and $n = 10$ (right panel) at large time $t = 20$ fm/c, analytic result given in \cref{eq:density_equi}.
				Parameters of the test are $T = 300$ MeV, $\gamma = 0.5$ c/fm, $\omega = 0.5$ c/fm and the cutoff frequency $\Omega = 4 T$.  The amount of cells is chosen to be $500\times 500$.
				}
				\label{fig:norm_dev_init0}
			\end{center}
		\end{figure*}

	At the end of this chapter, let us comment on \cref{fig:norm_dev_init0}, where we show
		\begin{align}\label{eq:nxy}
			 \Delta( x, y, t ) = \bigg|\rho ( x, y, t ) -\rho_{\text{analyt}}(x,y,\infty )\bigg| \, ,
		\end{align}
	where the first term is the numerical result, for some large time, here $t=20$ fm/c, where we expected equilibration and the second term is the true analytical result for $t \to \infty$, \cref{eq:density_equi}.
	From the plots, we find only small deviations from the analytical result for the whole computational domain  for $n=1$ and  $n=10$, which are basically equivalent at the given resolution and final time, for the case that $D_{x x} = 0$.
	
	From \cref{fig:norm_ho} we can see that for both initial conditions, $n = 1$ and $n = 10$, the norm decreases in time, as in the case of the free particle, cf.\ \cref{fig:norm_free}.
	However, we have already discussed, that one can minimize this decrease by using more cells on the computational domain, which leads to a better resolution of the density matrix close to its diagonal entries.
	As already said, the problem emerges, because the finite volume method averages the real and imaginary parts of the density matrix over the cells and one always has small contributions in the cells on the diagonal from the density matrix close to the diagonal.
	Therefore, it would be desirable to take control of this problem for example by using a different computational geometry in future works. 
	
	However, we hope to be able to convince the reader, that, if thermalization takes place long before a notable violation of the norm, which means a small deviation from the initial (numerical) value, one can perform these calculations with good conscience.
	As one can see in \cref{fig:norm_dev_init0}, here the time is $t = 20$ fm/c, where \cref{fig:rho_xx_ho_1,fig:rho_xx_ho_10} already show equilibration.

	Finally, we conclude, that also for the harmonic oscillator the results are promising.
	Here, we do not only have the analytical result, to which we can compare our numerical method.
	Au contraire to the analytical result, we have full insight into the dynamical evolution towards the thermal state.
	This also includes predictions of the thermalization time and process, as was discussed before.
	Independently of the initial state, the same final thermal state is reached, while we find that the equilibration time varies due to the initial energy that dissipates into the environment.
	Hence, the \gls{kt} scheme seems to be a promising tool to analyse the temporal evolution of the harmonic oscillator potential with Lindblad dynamics and reproduces the analytical findings of \reffs\cite{Bernad2018,Homa2019}.

\section{Conclusions}

	Treating open quantum systems with the Lindblad approach is known to be a challenging task -- both from a theoretical and a numerical point of view.
	Not only, that the Lindblad equation is Markovian and valid only for small couplings to the bath, the question of  thermalization is not clearly answered.
	Besides this phenomenological questions, that have not been tackled in detail in this work, another question is about a powerful and reliable numerical treatment of this \gls{pde}.

	In this paper, we presented a numerical method, which is successfully used in computational fluid dynamics and is based on a \gls{fv} discretization -- the \gls{kt} scheme, which is based on a so-called \gls{muscl} reconstruction.
	It involves slope and flux regulators, and therefore avoids possible numerical instabilities emerging during the temporal evolution.
	As a ready-made spatial discretization it is optimal to solve \glspl{pde} that can be classified as advection-diffusion equations with sources/sinks in conservative form in their coordinate space representation.

	In order to apply this method, we demonstrated that the Lindblad equation in spatial representation can be reformulated as an advection-diffusion equation in conservative form, where real and imaginary parts of the density matrix take the role of the ``fluid''.
	It turned out, that most of the terms describing diffusion, advection as well as sources or/and sinks are mixtures of terms appearing in the original Lindblad equation, which of course depends on the choice of coordinates.
	To the best of our knowledge this approach and interpretation is new and may provide new insights in the structure and physical meaning of the Lindblad equation and dynamics.

	Of course, we did not only stay at the theoretical level, but also implemented the numerical method to demonstrate its practical use.
	For example, as one of the benchmark tests for all the computations we checked the norm conservation of the density matrix, which is a necessary condition for the validity of the Lindblad equation.

	A first minimal test case, was to solve the von-Neumann equation for different setups. This comprises stationary problems, where the density matrix should stay constant during time evolution.
	The initial values where composed by the analytical wave functions of the square-well potential, and two arbitrary initial conditions, which were chosen to be a Gaussian and a rectangular, box-like wave function. 
	It turned out that for all cases the norm was perfectly conserved and the output data enabled us to study patterns and stationary distributions of the density matrix for arbitrary initial conditions.
	We have studied these systems for different numbers of cells, because the numbers of cells dictate the precision of the computation.

	The second testing setup was for the full Lindblad dynamics. 
	Here two cases have been considered: the free particle and the harmonic oscillator with dissipation and therefore interaction with a heat bath.
	Besides the question of how many cells and which boundary conditions have to be used for accurate results, we discussed the question of approaching the correct equilibrium thermal distribution.
	The thermal state is dictated by the bath, and should therefore not depend on the initial state.
	Indeed, this was also observed in our numerical results, where it can be seen that for the free particle as well as for the harmonic oscillator the right (Boltzmann distributed) equilibrium state is achieved.
	We observed, that for the free particle the density matrix is populated along the full diagonal and slightly left and right of the diagonal because of the quantum nature of the system.
	All the other entries are zero at equilibrium. 
	For the harmonic oscillator the thermal state along the diagonal is distributed in a Gaussian shape and confined due to the harmonic potential, which is in agreement with the analytical result.
	
	We have compared both cases with analytic results the statistical distributions, considering, that the system thermalizes and proved, that this is the case for all setups, where $D_{xx} = 0$.
	Considering spatial diffusion leads to the necessity of increasing the amount of cells in the computational domain in order to preserve  the trace of the density matrix on a constant level. 
	However, we find, that the $D_{xx}$-proportional term is not norm-conserving and therefore the Lindblad equation does not lead to thermal distributed states if one considers finite spatial domains. 

	Another important test, which we performed for all computations is to check that the imaginary part of the density matrix has to vanish close to equilibrium.
	This is related to decoherence and turns out to be a good numerical test.
	Of course, we find, that the imaginary part does not vanish completely, which cannot be expected numerically, but gets very small compared to typical scales of the system.
	The violation of the vanishing imaginary part might therefore be considered as a measure for the numerical error.

	To summarize, we propose a new numerical method, which can be used as a black-box solver based on the well-known \gls{kt} scheme to solve Lindblad dynamics and tested this tool.
	The numerical error is controlled by the number of cells and can therefore be easily reduced by increasing the number of cells and computational time.

	Lastly, we want to stress, that further tests of the scheme are necessary, especially for more complex systems and that our approach should be seen as being complementary to other methods and not as a replacement.

\section{Outlook}

	In the literature, there is a variety of different Lindblad dynamical evolutions describing various systems.
	Usually they are formulated in the Fock representation.
	However, it is possible to evolve every system in a complete set of eigenfunctions, and therefore obtain wave functions, which allows to formulate the problem in coordinate space.
	It would be of major interest to apply the numerical schemes proposed in this work to other physical systems than those presented here.
	This might give new insights into the thermalization process, the equilibration time, and the equilibration process.
	A main advantage might be that our approach is agnostic towards the particular features of the systems, such as attractive or repulsive potentials, different $D$-coefficients, time-dependent dampings or even time-dependent potentials, which can all be implemented straightforwardly.
	The origin of the decreasing norm emerges from the geometry of the considered system, which is rotated with respect to the computational domain. 
	Therefore, there is a undesirable mixing of real-and imaginary part, which can, most probably be tackled by adjusting the geometry of the computational domain towards the geometry of the system itself.
	Furthermore, we believe that it is in general possible to extend the numerical framework to two or even three spatial dimensions (four/six computational ones), which would be a major step forward in the field of open quantum systems, but comes with tremendous computational costs.
	
\begin{acknowledgments}

	J.~R.\ and N.~Z.\ acknowledge support by the \textit{Helmholtz Graduate School for Hadron and Ion Research for the Facility for Antiproton and Ion Research} (HGS-HIRe for FAIR) and the \textit{Deutsche Forschungsgemeinschaft} (DFG, German Research Foundation) through the CRC-TR 211 ``Strong-interaction matter under extreme conditions'' -- project number 315477589 -- TRR 211.

	J.~R.\ and C.~G.\ thank H.~van Hees for creative and  thought-provoking impulses, profitable discussions and lecturing.

	J.~R.\ thanks D.~Schuch for sophisticated and profitable discussions.
	
	J.~R.\ thanks T.~Neidig  for lecturing and debating.
	
	A.~K.\ and N.~Z.\ thank S.~Floerchinger and J.~Braun for valuable discussions and their great support and encouragement at the TPI in Jena and the IKP in Darmstadt, respectively.

	A.~K.\ is further grateful to M.~Gärttner, E.~Grossi, and M.~J.~Steil for answering questions and for discussions on the numerical treatment of the Lindblad equation, fluid dynamics, open quantum systems and other related topics.

	All the numerical results in this work have been obtained by using \texttt{Python 3} \cite{10.5555/1593511} with various libraries, including {\it numpy} \cite{harris2020array} and {\it scipy} \cite{2020SciPy-NMeth}.

\end{acknowledgments}

\appendix

\section{Norm conservation of the Lindblad form}
\label{sec:normconservation}

	Usually, a density matrix in space satisfies the boundary conditions of a asymptotically free particle, namely 
	$\rho ( \pm \infty, \pm \infty, t ) = 0$ and $\partial_{x, y} \rho ( \pm \infty, \pm \infty, t ) = 0$.
	However, the boundary condition of a ``confined density matrix", as we use it for our box calculations is on the one hand still $\rho ( \pm L/2, \pm L/2, t ) = 0$, which on the other hand does not mean, that $\partial_{x, y} \rho ( \pm L/2, \pm L/2,t ) = 0$.
	
	Therefore, in order to calculate the spatial integral over the diagonal of the last term of \cref{eq:lindblad_spatial}, we consider the centre-of-mass coordinate $q = \frac{1}{2} \, ( x + y )$ to evaluate the integral in the direction of the diagonal of the density matrix $\tilde{\rho} ( r, q, t )$, re-expressed in the centre-of-mass coordinate and the relative coordinate $r = \frac{1}{2} \, (x - y)$.
	We obtain,
		\begin{align}
			& \ii D_{x x} \int_{-a}^{a} \dd q \, \tfrac{\partial^2}{\partial q^2} \, \tilde{\rho} ( r, q, t ) = 	\Vdistance
			\\	
			= \, & \ii D_{x x} \int_{-a}^{a} \dd q \, \tfrac{\partial}{\partial q} \, \big( \tfrac{\partial}{\partial q} \, \tilde{\rho} ( r, q, t ) \big) =	\Vdistance	\nonumber
			\\
			= \, & \ii D_{x x} \tfrac{\partial \tilde{\rho} ( r, q, t )}{\partial q} \Big|_{- a}^{a} \, ,	\Vdistance	\nonumber
		\end{align}
	which leads to a non-vanishing result, if $a$ is finite (e.g. $a = L/2$) and vanishes only if $a \rightarrow \infty$, because of the asymptotic boundary condition.
	
	From a physical point of view, the last term of \cref{eq:lindblad_spatial} is a diffusion equation, which diffuses the density matrix parallel to its diagonal, cf. \cref{sec:reinterpretation}.
	However, similar to the situation of ideal cooling at the boundaries for a standard heat equation, we enforce the density matrix to vanish at the boundaries, which ultimately always leads to a situation where the density matrix vanishes completely, if there is no source term, which compensates the outflow.
	Hence, without a source contribution on the diagonal, the Lindblad equation is not norm conserving on a finite domain.
	Interestingly, in numerical computations, this is also the case for situations where the domain is infinite and the density matrix is only localized by some external potential and therefore asymptotically vanishing:
	Here, the artificial restricition to a finite sized computational domain leads to a violation of the norm conservation.
	However, one can control this violation by increasing the computational domain, such that other numerical errors become dominant.

\section{Derivation of the conservative form of the Lindblad equation}\label{sec:kt_form}

	Starting from \cref{eq:lindblad_spatial}, we split the density matrix into its real and imaginary part, $\rho(x,y,t) = \rho_R ( x, y, t ) + \ii \rho_I ( x, y, t )$.
	This also allows to split the Lindblad equation into its real and imaginary part.
\begin{widetext}
		\begin{align}\label{eq:imag_rho}
			\tfrac{\partial}{\partial t} \, \rho_I = \, & \Big[ \tfrac{1}{2 m} \, \big( \tfrac{\partial^2}{\partial x^2} - \tfrac{\partial^2}{\partial y^2}\big) + ( V ( y ) - V ( x ) ) + 2 D_{p x} \, ( x - y ) \big( \tfrac{\partial}{\partial x} + \tfrac{ \partial}{\partial y} \big) \Big] \, \rho_R +	\vdistance	
			\\	
			& - \Big[ D_{p p} \, ( x - y )^2 + \gamma \, ( x - y ) \big( \tfrac{\partial}{\partial x} - \tfrac{ \partial}{\partial y} \big) + D_{x x} \, \big( \tfrac{\partial^2}{\partial x^2} + 2 \, \tfrac{\partial^2}{\partial x \, \partial y} + \tfrac{ \partial^2}{\partial y^2} \big) \Big] \, \rho_I \, ,	\vdistance	\nonumber
			\\
			\tfrac{\partial}{\partial t} \, \rho_R = \, & \Big[ \tfrac{1}{2 m} \, \big( \tfrac{\partial^2}{\partial y^2} - \tfrac{\partial^2}{\partial x^2}\big) + ( V ( x ) - V ( y ) ) - 2 D_{p x} \, ( x - y ) \, \big( \tfrac{\partial}{\partial x} + \tfrac{\partial}{\partial y} \big) \Big] \, \rho_I +	\vdistance	\label{eq:real_rho}
			\\	
			& - \Big[ D_{p p} \, ( x - y )^2 + \gamma \, ( x - y ) \, \big( \tfrac{\partial}{\partial x} - \tfrac{\partial}{\partial y} \big) - D_{x x} \, \big( \tfrac{\partial^2}{\partial x^2} + 2 \, \tfrac{\partial^2}{\partial x\partial y} + \tfrac{ \partial^2}{\partial y^2} \big) \Big] \, \rho_R \, .	\vdistance	\nonumber
		\end{align} 
	We use $\frac{\partial}{\partial \xi} \, [ \xi \, f ( \xi ) ] - f ( \xi ) = \xi \, \frac{\partial}{\partial \xi} \, f ( \xi )$ to write
		\begin{align}
			( x - y ) \, \big( \tfrac{\partial}{\partial x} - \tfrac{\partial}{\partial y} \big) \, \rho = \, & \big( \tfrac{\partial}{\partial x} \, x - \tfrac{\partial}{\partial y} \, x - \tfrac{\partial}{\partial x} \, y + \tfrac{\partial}{\partial y} \, y - 2 \big) \, \rho \, ,	\vdistance
			\\
			( x - y ) \, \big( \tfrac{\partial}{\partial x}+ \tfrac{\partial}{\partial y} \big) \, \rho = \, & \big( \tfrac{\partial}{\partial x} \, x + \tfrac{\partial}{\partial y} \, x - \tfrac{\partial}{\partial x} \, y - \tfrac{\partial}{\partial y} \, y \big) \, \rho \, ,	\vdistance
		\end{align}
	which can be applied to \cref{eq:imag_rho} to obtain
		\begin{align}
			\tfrac{\partial}{\partial t} \, \rho_I = \, & \tfrac{1}{2 m} \, \tfrac{\partial^2}{\partial x^2} \, \rho_R - \tfrac{1}{2 m} \, \tfrac{\partial^2}{\partial y^2} \, \rho_R + D_{x x} \, \tfrac{\partial^2 }{\partial x^2} \, \rho_I + D_{x x} \, \tfrac{\partial^2}{\partial y^2} \, \rho_I +	\vdistance
			\\
			& + ( V ( y ) - V ( x ) ) \, \rho_R + \big[ 2 \gamma - D_{p p} \, ( x - y )^2 \big] \, \rho_I + 2 D_{p x} \, \tfrac{\partial}{\partial x} \, \big[ ( x - y ) \, \rho_R \big] +	\vdistance	\nonumber
			\\
			& + 2 D_{p x} \, \tfrac{\partial}{\partial y} \, \big[ ( x - y ) \, \rho_R \big] + \gamma \, \tfrac{\partial}{\partial x} \, \big[ ( y - x ) \, \rho_I \big] + \gamma \, \tfrac{\partial}{\partial y} \, \big[ ( x - y ) \, \rho_I \big] + 2 D_{x x} \, \tfrac{\partial^2}{\partial x \partial y} \, \rho_I \, .	\vdistance 	\nonumber
		\end{align}
	After additional rearrangements, we find
		\begin{align}\label{eq:first}
			\tfrac{\partial}{\partial t} \, \rho_I = \, & \tfrac{1}{2 m} \, \tfrac{\partial^2}{\partial x^2} \, \rho_R - \tfrac{1}{2 m} \, \tfrac{\partial^2}{\partial y^2} \, \rho_R + D_{x x} \, \tfrac{\partial^2}{\partial x^2} \, \rho_I + D_{x x} \, \tfrac{\partial^2}{\partial y^2} \, \rho_I +	\vdistance
			\\
			& + ( V ( y ) - V ( x ) ) \, \rho_R + \big[ 2 \gamma - D_{p p} \, ( x - y )^2 \big] \, \rho_I + \vdistance	\nonumber
			\\
			& + \tfrac{\partial}{\partial x} \, \Big[ 2 D_{p x} \, ( x - y ) \, \rho_R + \gamma \, ( y - x ) \, \rho_I + D_{x x} \, \tfrac{\partial}{\partial y} \,\rho_I \Big] +	\vdistance	\nonumber
			\\
			& + \tfrac{\partial}{\partial y} \, \Big[ 2 D_{p x} \, ( x - y ) \, \rho_R + \gamma \, ( x - y ) \, \rho_I + D_{x x} \, \tfrac{\partial}{\partial x} \, \rho_I \Big] \, .	\vdistance	\nonumber
		\end{align}
	Performing the same steps for \cref{eq:real_rho} we obtain
		\begin{align}
			\tfrac{\partial}{\partial t} \, \rho_R = \, & - \tfrac{1}{2 m} \, \tfrac{\partial^2}{\partial x^2} \, \rho_I + \tfrac{1}{2 m} \, \tfrac{\partial^2}{\partial y^2} \, \rho_I + D_{x x} \, \tfrac{\partial^2}{\partial x^2} \, \rho_R + D_{x x} \, \tfrac{\partial^2}{\partial y^2} \, \rho_R +	\vdistance
			\\
			& + ( V ( x ) - V ( y ) ) \, \rho_I + \big[ 2 \gamma - D_{p p} \, ( x - y )^2 \big] \, \rho_R +	\vdistance	\nonumber
			\\
			& + \tfrac{\partial}{\partial x} \, \Big[ - 2 D_{p x} \, ( x - y ) \, \rho_I + \gamma \, ( y - x ) \, \rho_R + D_{x x} \, \tfrac{\partial}{\partial y} \, \rho_R \Big] +	\vdistance	\nonumber
			\\
			& + \tfrac{\partial}{\partial y} \, \Big[ - 2 D_{p x} \, ( x - y ) \, \rho_I + \gamma \, ( x - y ) \, \rho_R+ D_{x x} \, \tfrac{\partial}{\partial x} \, \rho_R \Big] \, .	\vdistance	\nonumber
		\end{align}
	Combining both equations in matrix-vector notation and reordering the terms, we arrive at our final result, which is the conservative form of an advection-diffusion equation with sources and sinks,
		\begin{align}\label{eq:final_form}
			& \tfrac{\partial}{\partial t}
			\begin{pmatrix}
				\rho_I
				\\
				\rho_R
			\end{pmatrix}
			+
			\tfrac{\partial}{\partial x}
			\begin{pmatrix}
				- 2 D_{p x} \, ( x - y ) \, \rho_R + \gamma \, ( x - y ) \, \rho_I
				\\
				+ 2 D_{p x} \, ( x - y ) \, \rho_I + \gamma \, ( x - y ) \, \rho_R
			\end{pmatrix}
			+ \tfrac{\partial}{\partial y}
			\begin{pmatrix}
				- 2 D_{p x} \, ( x - y ) \, \rho_R - \gamma \, ( x - y ) \, \rho_I
				\\
				+ 2 D_{p x} \, ( x - y ) \, \rho_I - \gamma \, ( x - y ) \, \rho_R
			\end{pmatrix} =
			\\
			= \, &
			\tfrac{\partial}{\partial x}
			\begin{pmatrix}
				\frac{\partial}{\partial x} \big[ \frac{1}{2 m} \, \rho_R + D_{x x} \, \rho_I \big] + D_{x x} \, \frac{\partial}{\partial y} \, \rho_I
				\\
				\frac{\partial}{\partial x} \, \big[ - \frac{1}{2 m} \, \rho_I + D_{x x} \, \rho_R \big] + D_{x x} \, \frac{\partial}{\partial y} \, \rho_R
			\end{pmatrix}
			+ \tfrac{\partial}{\partial y}
			\begin{pmatrix}
				\frac{\partial}{\partial y} \big[ - \frac{1}{2 m} \, \rho_R + D_{x x} \, \rho_I \big] + D_{x x} \, \frac{\partial}{\partial x} \, \rho_I
				\\
				\frac{\partial}{\partial y} \, \big[ \frac{1}{2 m} \,  \rho_I + D_{x x} \, \rho_R \big] + D_{x x} \, \frac{\partial}{\partial x} \, \rho_R
			\end{pmatrix}
			+	\nonumber
			\\
			& +
			\begin{pmatrix}
				( V ( y ) - V ( x ) ) \, \rho_R + \big[ 2 \gamma - D_{p p} \, ( x - y )^2 \big] \,\rho_I
				\\
				( V ( x ) - V ( y ) ) \, \rho_I + \big[ 2 \gamma - D_{p p} \, ( x - y )^2 \big]\rho_R
			\end{pmatrix} \, .	\nonumber
		\end{align}
		Therefore we obtain in accordance to \cref{sec:reinterpretation}
		\begin{align}\label{eq:fqs1}
			\vec{f}^x[\vec{x}, \vec{u}] &=\begin{pmatrix}
				- 2 D_{p x} \, ( x - y ) \, \rho_R + \gamma \, ( x - y ) \, \rho_I
				\\
				+ 2 D_{p x} \, ( x - y ) \, \rho_I + \gamma \, ( x - y ) \, \rho_R
			\end{pmatrix}\\
			\label{eq:fqs2}
			\vec{f}^y[\vec{x}, \vec{u}] &=\begin{pmatrix}
				- 2 D_{p x} \, ( x - y ) \, \rho_R - \gamma \, ( x - y ) \, \rho_I
				\\
				+ 2 D_{p x} \, ( x - y ) \, \rho_I - \gamma \, ( x - y ) \, \rho_R
			\end{pmatrix}\\
			\label{eq:fqs3}
			\vec{Q}^x[\partial_x \vec{u}, \partial_y \vec{u}] &=	\begin{pmatrix}
				\frac{\partial}{\partial x} \big[ \frac{1}{2 m} \, \rho_R + D_{x x} \, \rho_I \big] + D_{x x} \, \frac{\partial}{\partial y} \, \rho_I
				\\
				\frac{\partial}{\partial x} \, \big[ - \frac{1}{2 m} \, \rho_I + D_{x x} \, \rho_R \big] + D_{x x} \, \frac{\partial}{\partial y} \, \rho_R
			\end{pmatrix}\\
			\label{eq:fqs4}
			\vec{Q}^y[\partial_x \vec{u}, \partial_y \vec{u}] &=	\begin{pmatrix}
				\frac{\partial}{\partial y} \big[ - \frac{1}{2 m} \, \rho_R + D_{x x} \, \rho_I \big] + D_{x x} \, \frac{\partial}{\partial x} \, \rho_I
				\\
				\frac{\partial}{\partial y} \, \big[ \frac{1}{2 m} \,  \rho_I + D_{x x} \, \rho_R \big] + D_{x x} \, \frac{\partial}{\partial x} \, \rho_R
			\end{pmatrix}\\
			\label{eq:fqs5}
			\vec{S}[t, \vec{x},\vec{u}] &=	\begin{pmatrix}
				( V ( y ) - V ( x ) ) \, \rho_R + \big[ 2 \gamma - D_{p p} \, ( x - y )^2 \big] \,\rho_I
				\\
				( V ( x ) - V ( y ) ) \, \rho_I + \big[ 2 \gamma - D_{p p} \, ( x - y )^2 \big]\rho_R
			\end{pmatrix} 
		\end{align}
	where $\vec{u} = (\rho_I(x,y,t), \rho_R(x,y,t))^T$.
\end{widetext}

\section{Eigenvalues for KT method}
\label{app:eigenvalues}
	
	For the \gls{kt} scheme we also need the Jacobian of the advection fluxes, which is given by
		\begin{align}
			\frac{\partial \vec{f}^{\, x}}{\partial \vec{u}} =
			\begin{pmatrix}
				\gamma \, ( x - y )	&	2 D_{p x} \, ( x - y )
				\\
				- 2 D_{p x} \, ( x - y )	&	\gamma \, ( x - y )
			\end{pmatrix} \, ,
			\\
			\frac{\partial \vec{f}^{\, y}}{\partial \vec{u}} =
			\begin{pmatrix}
				- \gamma \, ( x - y )	&	2 D_{p x} \, ( x - y )
				\\
				- 2 D_{p x} \, ( x - y )	& 	- \gamma \, ( x - y )
			\end{pmatrix} \, .
		\end{align}
	The corresponding eigenvalues are given by
		\begin{align}
			\lambda_{1,2}^{x} = \, & ( x - y ) \, ( \gamma \mp 2 \ii D_{p x} ) \, ,	\vdistance
			\\
			\lambda_{1,2}^{y} = \, & ( x - y ) \, ( - \gamma \mp 2 \ii D_{p x} ) \, ,	\vdistance
		\end{align}
	which leads for both cases to the same spectral radius
		\begin{align}
			\rho = \max \vert \lambda_i \vert = \vert ( x - y ) \vert \, \sqrt{4 D_{p x}^2 + \gamma^2} \, .
		\end{align}
	Interestingly, the spectral radius and consequently the fluid velocity is independent of $\vec{u}$ itself and only depends on the distance from the diagonal of the density matrix as well as the Lindblad coefficients $D_{p x}$ and $\gamma$.

\section{Distribution of the equilibrium density matrix of a free particle in a square well potential}
\label{app:distribution}

	The equilibrium density matrix in coordinate space of a particle described by
		\begin{align}\label{free_part}
			& \bra{x} \hat{\rho}_\text{free} \ket{y} 	
			= \,  \tfrac{1}{Z} \bra{x} \ee^{- \beta ( \frac{\hat{p}^2}{2m}  )} \, \ket{y} =	\Vdistance	
			\\
			= \, & \tfrac{1}{Z} \int \tfrac{\mathrm{d} p}{(2 \uppi)} \,  \braket{x \vert p} \bra{p} \ee^{-\beta \frac{\hat{p}^2}{2m}} \ket{y} =	\Vdistance	\nonumber
			\\
			= \, & \tfrac{1}{Z} \int \tfrac{\mathrm{d} p}{(2 \uppi)} \, \ee^{- \beta \frac{p^{2}}{2 m} + \ii p ( x - y )} = \tfrac{1}{L} \, \ee^{- \frac{m}{2 \beta} ( x - y )^2} \, ,	\Vdistance	\nonumber
		\end{align}
	where $Z$ can be obtained, calculating the trace of $\hat{\rho}$, and considering, that the we want to confine the particle in a box of lengths $L$, we obtain
		\begin{align}
			Z = \, & \int_{- \frac{L}{2}}^{\frac{L}{2}} \mathrm{d} x \bra{x} \ee^{- \beta \frac{\hat{p}^{2}}{2 m}} \ket{x} =	\Vdistance
			\\
			= \, & \int_{- \frac{L}{2}}^{\frac{L}{2}} \mathrm{d} x \, \int \tfrac{\dd p}{(2 \uppi)} \, \braket{x \vert p} \bra{p} \ee^{- \beta \frac{\hat{p}^{2}}{2 m}} \ket{x} =	\Vdistance	\nonumber
			\\
			= \, & \int_{- \frac{L}{2}}^{\frac{L}{2}} \mathrm{d} x \, \int \tfrac{\dd p}{(2 \uppi)} \, \ee^{- \beta \frac{p^{2}}{2 m}} = \sqrt{\tfrac{m L^2}{2\uppi  \beta}} \, .	\Vdistance	\nonumber
		\end{align}
	If we set $x = y$ in \cref{free_part} we obtain the constant $\frac{1}{L}$ as the value for all the diagonal entries.
	For $y = - x$, the orthogonal axis, we obtain a Gaussian
		\begin{align}
			\rho_\text{free} ( x, - x ) = \tfrac{1}{L} \, \ee^{- 2 m T x^2} \, .	\label{eq:cross_diag}
		\end{align}

\bibliography{bibliography.bib}

\end{document}